\documentclass[12pt]{article}
\pdfoutput=1

\usepackage[left=2.8cm,right=2.8cm,top=2.8cm,bottom=2.8cm]{geometry}

\usepackage[colorlinks=true,linkcolor=black,citecolor=teal,urlcolor=MidnightBlue,filecolor=black]{hyperref}

\usepackage{amsmath}
\usepackage{amsfonts}
\usepackage{amssymb}
\usepackage{mathtools}
\usepackage{mathrsfs}
\usepackage{scalerel}
\usepackage[dvipsnames]{xcolor}
\definecolor{SchoolColor}{rgb}{0.6471, 0.1098, 0.1882} 
\usepackage{cite}
\usepackage[normalem]{ulem}
\usepackage{multirow,booktabs}
\usepackage[colorinlistoftodos,prependcaption,textsize=tiny]{todonotes}
\usepackage{filecontents}
\usepackage{ wasysym } 
\usepackage{eucal} 
\usepackage{bbold}
\usepackage{footmisc} 

\providecommand{\Lt}{{\tt L}}
\renewcommand{\Lt}{{\tt L}}
\providecommand{\Mt}{{\tt M}}
\renewcommand{\Mt}{{\tt M}}

\providecommand{\Xt}{{\tt X}}
\renewcommand{\Xt}{{\tt X}}

 \csname
@addtoreset\endcsname{equation}{section}

\definecolor{color1}{rgb}{0.22,0.45,0.70}

\DeclareMathOperator{\extdm}{d}
\newcommand{\extd}{\extdm \!}

\newcommand{\unity}{\mathbb{1}}

\newcommand{\cA}{{\cal A}}
\newcommand{\cB}{{\cal B}}

\newcommand{\cF}{{\cal F}}

\newcommand{\cJ}{{\cal J}}
\newcommand{\cK}{{\cal K}}
\newcommand{\cL}{{\cal L}}
\newcommand{\cM}{{\cal M}}
\newcommand{\cN}{{\cal N}}
\newcommand{\cO}{{\cal O}}
\newcommand{\cP}{{\cal P}}

\newcommand{\cW}{{\cal W}}

\newcommand{\bb}{\bar\beta}

\newcommand{\beq}{\begin{equation}}
\newcommand{\eeq}{\end{equation}}
\newcommand{\bi}{\begin{itemize}}
\newcommand{\ei}{\end{itemize}}
\newcommand{\bt}{\begin{tabular}}
\newcommand{\et}{\end{tabular}}
\newcommand{\bc}{\begin{center}}
\newcommand{\ec}{\end{center}}

\newcommand{\ket}[1]{|#1\rangle}
\newcommand{\bra}[1]{\langle#1|}

\newcommand{\vev}[1]{\langle#1\rangle}

\def\one{{\hbox{ 1\kern-.8mm l}}}
\newcommand{\Dslash}{\not{\hbox{\kern-4pt $D$}}}
\newcommand{\pdslash}{\not{\hbox{\kern-2pt $\partial$}}}

\newcommand{\be}{\begin{equation}}
\newcommand{\ee}{\end{equation}}
\newcommand{\bea}{\begin{eqnarray}}
\newcommand{\eea}{\end{eqnarray}}
\newcommand{\ba}{\begin{array}}
\newcommand{\ea}{\end{array}}

\def\bbox{{\,\lower0.9pt\vbox{\hrule \hbox{\vrule height 0.2 cm
\hskip 0.2 cm \vrule height 0.2 cm}\hrule}\,}}
\newcommand{\dsl}{\pa \kern-0.5em /}

\newcommand{\vp}{\varphi}

\font\mybb=msbm10 at 12pt
\def\bb#1{\hbox{\mybb#1}}

\def\bR {\bb{R}}

\begin{document}


\vspace{30pt}

\begin{center}


{\Large\sc Geometric actions and flat space holography}\\

---------------------------------------------------------------------------------------------------------


\vspace{25pt}

{\sc Wout Merbis$^{\,\includegraphics[height=10pt]{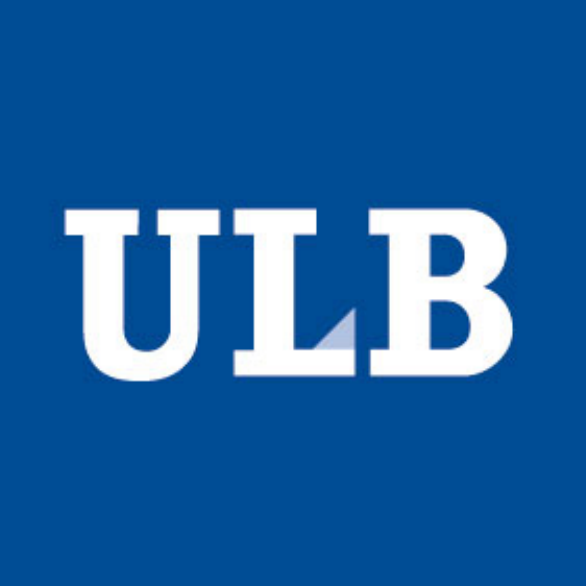}}$ and Max Riegler$^{\includegraphics[height=10pt]{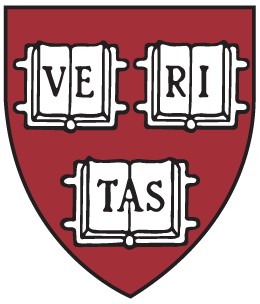}}$}

\vskip 25pt

{\em \includegraphics[height=14pt]{ulbnorm.pdf} \hskip -.1truecm  Universit\'e Libre de Bruxelles and International Solvay Institutes, Physique Th\'eorique et Math\'ematique, Campus Plaine - CP 231, B-1050 Bruxelles, Belgium\vskip 10pt }
{\em \includegraphics[height=15pt]{harvard_shield_cut.png} \hskip -.1truecm  Center for the Fundamental Laws of Nature, Harvard University, Cambridge, MA 02138, USA\vskip 5pt }
{email: {\tt wmerbis@ulb.ac.be}, {\tt mriegler@fas.harvard.edu}} \\

\vskip 10pt

\vspace{50pt} {\sc\large Abstract} \end{center}

\noindent
In this paper we perform the Hamiltonian reduction of the action for three-dimensional Einstein gravity with vanishing cosmological constant using the Chern-Simons formulation and Bondi-van der Burg-Metzner-Sachs (BMS) boundary conditions. 
An equivalent formulation of the boundary action is the geometric action on BMS$_3$ coadjoint orbits, where the orbit representative is identified as the bulk holonomy. 
We use this reduced action to compute one-loop contributions to the torus partition function of all BMS$_3$ descendants of Minkowski spacetime and cosmological solutions in flat space. We then consider Wilson lines in the ISO$(2,1)$ Chern-Simons theory with endpoints on the boundary, whose reduction to the boundary theory gives a bilocal operator. We use the expectation values and two-point correlation functions of these bilocal operators to compute quantum contributions to the entanglement entropy of a single interval for BMS$_3$ invariant field theories and BMS$_3$ blocks, respectively. 
While semi-classically the BMS$_3$ boundary theory has central charges $c_1 = 0$ and $c_2 = 3/G_N$, we find that quantum corrections in flat space do not renormalize $G_N$, but rather lead to a non-zero $c_1$.

\thispagestyle{empty}


\newpage


\pagenumbering{arabic}
\tableofcontents
\hypersetup{linkcolor=SchoolColor}
\newpage

\section{Introduction}

The holographic principle \cite{tHooft:1993dmi,Susskind:1995vu} plays a vital role in our current understanding of quantum gravity. It paved the way in which one can define a theory of quantum gravity using its dual quantum field theory. Particularly in anti-de Sitter (AdS) spacetimes much progress has been made in terms of connecting bulk physics to the dual boundary conformal field theory (CFT). The special case of two-dimensional CFTs has a rich history (see for instance \cite{Witten:2007kt,Maloney:2007ud,Yin:2007gv,Giombi:2008vd,Hellerman:2009bu,Hartman:2013mia,Fitzpatrick:2014vua,Hartman:2014oaa,Keller:2014xba}) and much ongoing interest \cite{Javadinezhad:2017jnv,Gonzalez:2018jgp,Hartman:2019pcd,Benjamin:2019stq,Gliozzi:2019ewk} in trying to understand pure AdS$_3$ quantum gravity from dual CFT arguments.

Similar attempts to understand pure quantum gravity in three-dimensional flat spacetimes from a holographic perspective are, despite much recent progress \cite{Bagchi:2010eg,Barnich:2012rz,Bagchi:2013lma,Afshar:2013vka,Gonzalez:2013oaa,Fareghbal:2013ifa,Detournay:2014fva,Bagchi:2014iea,Barnich:2015mui,Bagchi:2015wna,Hosseini:2015uba,Basu:2015evh,Campoleoni:2015qrh,Campoleoni:2016vsh,Bagchi:2016bcd,Riegler:2016hah,Oblak:2016eij,Oblak:2017ect,Grumiller:2017sjh,Basu:2017aqn,Jiang:2017ecm,Hijano:2019qmi,Godet:2019wje},  
less developed in comparison to the AdS/CFT case. In part, this is due to the fact that it is not immediately clear what the theory should be dual to. Without a top-down example -- such as the AdS/CFT-correspondence -- 3D flat space holography in its current incarnation mainly consists of matching asymptotic symmetries at null infinity to the symmetries of ultra-relativistic limits of CFT$_2$, statements about the kinematics and consequences of this symmetry \cite{Barnich:2006av,Barnich:2010eb,Bagchi:2010eg,Bagchi:2012cy,Barnich:2012aw,Bagchi:2012yk,Barnich:2012xq,Bagchi:2012xr}. Fortunately, like in the case of 2D CFTs, the asymptotic BMS$_3$ symmetries \cite{Bondi:1962,Sachs:1962} of flat spacetime at null infinity are infinite dimensional and hence very powerful. It is thus not unimaginable that the program of applying universal CFT$_2$ methods to constrain pure AdS$_3$ quantum gravity can be adapted to flat space in a similar fashion.

Alongside these developments there has been a parallel approach to holography in three dimensions. The crucial observation is that three-dimensional gravity can be formulated as a Chern-Simons theory of the gauge group corresponding to the isometries of the maximally symmetric background solution in question \cite{Achucarro:1987vz,Witten:1988hc}. Motivated by the `constrain first, quantize later' approach to the quantization of Chern-Simons theory \cite{Elitzur:1989nr}, Coussaert, Henneaux and van Driel \cite{Coussaert:1995zp} performed a Hamiltonian reduction of the action under Brown-Henneaux \cite{Brown:1986nw} boundary condition. They found that the action can be presented as two chiral Wess-Zumino-Witten (WZW) models, that combine into a single non-chiral WZW. Under the Brown-Henneaux boundary conditions, this action reduces further to the Liouville action on the asymptotic boundary. However, it would be problematic to then declare quantum Liouville theory to be AdS$_3$ quantum gravity, as Liouville theory contains non-normalizable modes, has a continuous spectrum and no normalizable ground state. Furthermore, in the reduction of \cite{Coussaert:1995zp}, bulk holonomies were not considered (and only partially considered in the appendix of \cite{Henneaux:1999ib}), excluding a large portion of the asymptotically AdS$_3$ solution space, including the Ba\~nados-Teitelboim-Zanelli (BTZ) black holes of \cite{Banados:1992wn}.

If one wants to consider also the BTZ black holes (and their descendants) in this framework one would have to deal with more generic topologies than the filled cylinder of global AdS$_3$ and include both asymptotic boundaries of the eternal black hole \cite{Henneaux:2019sjx}. One effective way to consider the dynamics on a single boundary is to `cut out' the spacetime of one of the two asymptotic regions and work with Chern-Simons theory where the constant time-slice is a disk with a puncture. Now the Wilson loop surrounding the puncture is no longer contractible and the resulting holonomy measures the black hole mass and angular momentum (or any other global charges of the region that we cut out). A repetition of the Hamiltonian reduction in this situation would forbid one to combine the two chiral WZW models as each now have independent zero modes. Instead, the two chiral WZW models reduce further to two chiral bosons \cite{NavarroSalas:1999sr,Barnich:2013yka}, where the zero modes of the bosons are related to the bulk holonomy \cite{Grumiller:2019tyl,Henneaux:2019sjx}. An equivalent formulation of this action arises from the Kirillov-Konstant coadjoint orbit method for the Virasoro group, worked out by Alekseev and Shatashvili in the eighties \cite{Alekseev:1988ce}.

The relation between the Alekseev-Shatashvili (AS) action, quantization of the coadjoint orbits and three-dimensional gravity follows from ingredients that have been known in the literature for around thirty years \cite{Brown:1986nw,Witten:1987ty} and has been revived and expanded upon recently for flat spacetimes \cite{Barnich:2017jgw}, AdS$_3$ \cite{Cotler:2018zff} and dS$_3$ \cite{Cotler:2019nbi}, see also \cite{Raeymaekers:2014kea,Campoleoni:2017xyl,Duval:2014uoa,Duval:2014lpa}. Originally, it was noted in \cite{Alekseev:1988ce} that the AS action can be obtained as a Drinfeld-Sokolov reduction of the $\widehat{\mathfrak{sl}(2,\bR)}$ WZW model, which is the CFT counterpart of choosing Brown-Henneaux boundary conditions.
Another way to view this is that asymptotically locally AdS$_3$ spacetimes are Ba\~nados geometries \cite{Banados:1998gg} parametrized by two functions that are the dual stress-tensor expectation values semi-classically. Since the stress-tensor transforms in the coadjoint representation of the Virasoro group \cite{Witten:1987ty}, the Ba\~nados geometries are intrinsically related to the coadjoint orbits of the Virasoro group \cite{Balog:1997zz,NavarroSalas:1999sr,Nakatsu:1999wt,Sheikh-Jabbari:2016unm} and the orbit representative $b_0$ corresponds to the global charges of the bulk Ba\~nados geometry. 

The Alekseev-Shatashvili geometric action on the coadjoint orbits of the Virasoro group captures all Virasoro descendants of a given bulk Ba\~nados geometry with global charges given by the orbit representatives. The action is one-loop exact \cite{Cotler:2018zff}, which can be shown by suitable adaptation of the Duistermaat-Heckman theorem \cite{Duistermaat:1982vw} (see also \cite{Stanford:2017thb}). A given generic Virasoro orbit can be related by a field redefinition to the $b_0 = 0$ orbit \cite{Barnich:2017jgw}. This is in essence the uniformizing transformation that proved useful in computing Virasoro blocks in the heavy-light limit from AdS$_3$ geometry \cite{Fitzpatrick:2015zha}. It is also possible to compute the identity Virasoro blocks directly using the AS action, as was demonstrated in \cite{Cotler:2018zff}. Hence this is a useful framework for AdS$_3$ holography, which is in our opinion still underexplored.

In this paper, we make progress on a similar framework for flat space holography in 2+1 dimensions by utilizing the geometric action on the coadjoint orbit of the BMS$_3$ group \cite{Barnich:2015uva,Barnich:2017jgw}. We first show explicitly in section \ref{sec:reduction} how an effective action of pure three-dimensional flat space gravity is obtained by reducing the classical gravity action with Barnich-Comp\`ere boundary conditions \cite{Barnich:2006av} to a two-dimensional boundary theory. This was done first in \cite{Barnich:2013yka} where a flat space version of Liouville theory was found \cite{Barnich:2012rz} for the case of vanishing bulk holonomies.
Here we repeat this analysis for generic holonomy of the CS connection and obtain
\begin{align} \label{IbmsIntro}
    I_{\textrm{CS}} = - \frac{k}{2\pi} \int \extd u \extd\vp \; \Big[&  \left( \cL_0 + \cM_0 \partial_f \alpha(f) - \partial_f^3\alpha(f) \right) \dot{f} f'  -  \frac12 \left( \cM_0 f'^2  - 2 \{f,\vp\} \right) \Big].
\end{align}
Here $\cM_0$ and $\cL_0$ are proportional to the mass and angular momentum of the gravitational saddle and $f$ and $\alpha$ are fields which generate BMS$_3$ superrotations and supertranslations, respectively. The boundary theory coincides with the geometric action on the coadjoint orbit of the BMS$_3$ group of \cite{Barnich:2017jgw}, where the orbit representatives are given by $\cM_0$ and $\cL_0$ and the BMS$_3$ central charges are $c_1 = 0$ and $c_2 = 3/G_N$. This provides a map between the bulk gravitational solutions and the coadjoint orbit of the BMS$_3$ group \cite{Barnich:2015uva} with constant representatives. The global Minkowski vacuum corresponds to the first exceptional BMS$_3$ orbit, with ISL$(2,\bR)$ stabilizer subgroup. The flat space cosmology solutions \cite{Cornalba:2002fi,Cornalba:2002nv} correspond to generic (massive) orbits with representatives $(j_0,p_0) > 0$ and the smaller, two dimensional Abelian stabilizer subgroup.  

We then compute the one-loop torus partition function of BMS$_3$ descendants around vacuum Minkowski spacetime and flat space cosmologies in section \ref{sec:partitionfun}. We find that these are given by the vacuum character and the massive characters of the BMS$_3$ group \cite{Oblak:2015sea,Garbarz:2015lua,Bagchi:2019unf}, respectively. This gives a map between the BMS$_3$ weights ($\xi$ and $\Delta$) and the orbit representatives that can in term be expressed as the mass and angular momentum of the flat space cosmology. For the global Minkowski spacetime our results match with the partition function computed in \cite{Barnich:2015mui} by heat kernel methods only if no regularization is performed. Using the zeta function regularization applied for the AdS$_3$ case in \cite{Cotler:2018zff}, we find agreement with the highest-weight characters of \cite{Bagchi:2019unf} and a quantum shift of $c_1$, while $c_2 \sim 1/G_N$ does not receive any corrections.

In section \ref{sec:EE} we construct the following bilinear operator in the geometric theory
\begin{equation}\label{bilocalintro}
\mathcal{B}_{\Delta, \xi}(\vp_1,u_1;\vp_2,u_2) =  \; e^{ \xi\left(-2\frac{W_{12}}{X_{12}}+\frac{W_1'}{X'_1}+\frac{W_2'}{X'_2} \right)}\left(\frac{X'_1 X'_2}{(X_{21})^2}\right)^\Delta,
\end{equation}
by reducing to the boundary theory a Wilson line with end-point attached at the boundary points $(\vp_1,u_1)$ and $(\vp_2,u_2)$. This expression is valid for superrotations $X$ and supertranslations $W$ around the null orbifold with vanishing mass and angular momentum (or equivalently: vanishing orbit representatives). To obtain this operator on the generic orbit of BMS$_3$ with non-zero $\cL_0$ and $\cM_0$, one can use the map
\begin{subequations}\label{BMSuniformization}
\begin{align}
	X(\vp,u) & = e^{-\sqrt{\cM_0} f(\vp,u)} \,, \\
	W(\vp,u) & = - \sqrt{\cM_0} e^{-\sqrt{\cM_0} f(\vp,u)} \left( \alpha(f(\vp,u),u) + \frac{\cL_0}{\cM_0} f(\vp,u) \right)\,,
\end{align}
\end{subequations}
where $f$ and $\alpha$ are the superrotations and supertranslation fields with action \eqref{IbmsIntro}. 
For the saddle point values $f = \vp$ and $\alpha = u$ this is the BMS$_3$ analogue of the uniformizing transformations used in two dimensional CFTs.

The expectation value of the bilocal operator \eqref{bilocalintro} corresponds to the two-point function of probe operators in a background set by the respective orbit representative. This construction allows us to generalize the results for holographic entanglement entropy to any asymptotically flat geometry of \cite{Barnich:2010eb} by using the uniformizing BMS$_3$ transformation \eqref{BMSuniformization} and computing the expectation value on the relevant orbit. Using this setup we find that the leading order in small Newtons constant $G_N$ reproduces known results for entanglement entropy and we extend the analysis by computing the subleading corrections in $G_N$. We find once again that the subleading corrections lead to a non-zero $c_1$ of order unity, while $G_N$ is not renormalized. We furthermore show that entanglement entropy is one-loop exact and does not receive any further perturbative corrections within the present setup. These results are particularly useful for the recently discussed novel quantum energy conditions in flat spacetimes \cite{Grumiller:2019xna}.

Next we proceed to compute the building blocks for any BMS$_3$ invariant quantum field theory -- the BMS$_3$ blocks \cite{Bagchi:2016geg,Bagchi:2017cpu} -- in section \ref{sec:BMSblocks}. The BMS$_3$ identity block in the light-light limit (with external weights $\xi, \Delta \sim \cO(1)$) is computed by evaluating the two-point correlator of bilocal operators on the vacuum Minkowski orbit. We perform this computation to first order in perturbation theory in $1/c_2$ and compare the results with a direct way of computing the BMS$_3$ blocks using highest-weight representations in appendix \ref{sec:BMSblockdirect}. Furthermore, we show that in the limit of large central charge $c_2$ with $\frac{\Delta \xi}{c_2}$ and $\frac{\xi^2}{c_2}$ fixed, the vacuum block exponentiates as
    \begin{equation}
        \cF_{\unity}=\exp{\left[\frac{2}{c_2} \Big(  ( \Delta_1 \xi_2 + \Delta_2 \xi_1-\frac{c_1}{c_2}\xi_1\xi_2)  \mathcal{F}(x) + t\,\xi_1 \xi_2  \partial_x \mathcal{F}(x) \Big)\right]} + \ldots \,,
    \end{equation}
where $\cF(x) = x^2\, {}_2F_1(2,2;4,x)$, with ${}_2F_1(a,b;c,z)$ the ordinary hypergeometric function and the dots denote terms subleading in $1/\sqrt{c_2}$.
    
To compute the identity block in the heavy-light limit we compute the two-point function of a light probe bilocal in a flat space cosmological background, corresponding to the massive orbit of BMS$_3$. The leading order result can equivalently be obtained from the plane by the uniformizing BMS$_3$ transformation \eqref{BMSuniformization}. We additionally compute the one-loop corrections to the heavy-light BMS$_3$ identity block, which we use to compute the subleading corrections to the entanglement entropy in BMS$_3$ invariant QFTs dual to flat space cosmological backgrounds. Concluding remarks can be found in section \ref{sec:discussion}. 

\section{Flat space gravity and BMS$_3$ orbits}\label{sec:review}

General relativity in three spacetime dimensions with vanishing cosmological constant can equivalently be formulated as an ISL$(2,\bR)$ Chern-Simons theory \cite{Witten:1988hc}. In the next section we work out the Hamiltonian reduction of the Chern-Simons action to a boundary theory, similar in spirit to \cite{Coussaert:1995zp,Barnich:2015mui}. First, we state our conventions and review some relevant statements about the three-dimensional Chern-Simons formulation of gravity and the coadjoint orbits of the BMS$_3$ group.

\subsection{Setting the scene}

For the purposes of this work it is convenient to use the isomorphism of the 3D Poincar\'e algebra with $\mathfrak{isl}(2,\bR)$ and work with those variables. We work in a basis where the $\mathfrak{isl}(2,\bR)$ algebra is given by
\begin{subequations}
\begin{align}
[\Lt_m, \Lt_n] & = (m-n) \Lt_{m+n} \,,\\
[\Lt_m, \Mt_n] & = (m-n) \Mt_{m+n} \,, \\
[\Mt_m, \Mt_n] & = 0\,,
\end{align}
\end{subequations}
for $m,n = -1,0,1$. A convenient representation of $\mathfrak{isl}(2,\bR)$ is given in terms of Grassmann valued matrices with a Grassmann odd parameter $\epsilon$ \cite{Krishnan:2013wta}
\begin{equation}
\Lt_{-1}  = \left( \begin{array}{cc}
0 & -1 \\
0 & 0 
\end{array}
\right)\,, \qquad 
\Lt_{0} = \frac12 \left( \begin{array}{cc}
1 & 0 \\
0 & -1 
\end{array}
\right)\,, \qquad  
\Lt_{1} = \left( \begin{array}{cc}
0 & 0 \\
1 & 0 
\end{array}
\right)\,, \qquad
\Mt_{n}  = \epsilon\, \Lt_{n}\,.
\end{equation}
The non-degenerate bilinear form on this algebra is $\vev{\Lt_m \Mt_n} = -2 \gamma_{mn}$, with $\gamma_{mn}$ given by
\begin{equation}\label{gamma}
    \gamma_{mn} = \left(
			\begin{array}{c|ccc}
				  &\Mt_1&\Mt_0&\Mt_{-1}\\
				\hline
				\Lt_1&0&0&1\\
				\Lt_0&0&-\frac{1}{2}&0\\
				\Lt_{-1}&1&0&0
			\end{array}\right) \,.
\end{equation}
In terms of the Grassmann valued representation this bilinear form is simply 
    \begin{equation}\label{eq:InvBilForm}
        \vev{\Lt_m \Mt_n}\equiv2 \partial_{\epsilon}\textrm{Tr}(\Lt_m \Mt_n).        
    \end{equation}
The BMS$_3$ boundary conditions have been discussed at length in the literature, (see for instance \cite{Barnich:2010eb,Barnich:2012aw,Barnich:2013yka,Barnich:2014cwa,Gary:2014ppa}). Asymptotically Minkowski metrics in three dimensions can be written as
\begin{equation}\label{BMS3metric}
\extd s^2 = \cM(u,\vp) \extd u^2 - 2 \extd r \extd u + 2 \cN(u,\vp) \extd u \extd\vp + r^2 \extd\vp^2,
\end{equation}
where $r \to \infty$ is the boundary at null infinity  $\mathscr{I}^+$. The spectrum of zero mode solutions (with constant $\cM=\cM_0$ and $\cN=\cN_0$) contains Minkowski spacetime for $\cM_0 = -1, \cN_0 = 0 $; angular deficit solutions for $-1 < \cM_0 < 1$ and the null orbifold at $\cM_0 = 0 = \cN_0$.

Cosmological solutions are obtained by taking $\cM_0 > 0$ and they can be parameterized as
\begin{equation}
\cM_0 = r_+^2 \,, \qquad \cN_0 = r_0\, r_+\,.
\end{equation}
The coordinate transformations
\begin{equation}
u = t + \frac{r - r_0 {\rm Tanh}^{-1} (r_0/r)}{r_+^2} \,, \qquad \vp = \theta + \frac{{\rm Tanh}^{-1} (r/r_0)}{r_+} \,,
\end{equation}
then make it more apparent that \eqref{BMS3metric} is, indeed, a flat space cosmology \cite{Cornalba:2002fi,Bagchi:2013lma}
\begin{equation}
\extd s^2 = r_+^2 \extd t^2 - \frac{r^2 \extd r^2}{r_+^2 (r^2 - r_0^2)} + r^2 \extd\theta^2 - 2 r_+ r_0 \extd t \extd\theta\,.
\end{equation}

The Einstein-Hilbert action can be represented in a first-order formulation by changing variables from the metric to a dreibein $e=e_\mu \extd x^\mu$ and an independent (dualized) spin-connection $\omega=\omega_\mu \extd x^\mu$. Linearly combining $e$ and $\omega$ into a $\mathfrak{isl}(2, \bR)$ gauge connection
\begin{equation}
	\mathcal{A} = e^m \Mt_m + \omega^m \Lt_m\,,
\end{equation}
one finds that the Chern-Simons action, defined on some manifold $\mathfrak{M}$
\begin{equation}\label{Scs}
I_{\textrm{CS}} = \frac{k}{4\pi} \int_{\mathfrak{M}} \vev{\mathcal{A}\wedge \extd \mathcal{A} + \frac23 \mathcal{A} \wedge \mathcal{A} \wedge \mathcal{A}}\,,
\end{equation}
is equivalent (up to boundary terms) to the Einstein-Hilbert-Palatini action with vanishing cosmological constant in three dimensions when $k=1/4G_N$, where $G_N$ denotes Newton's constant. The metric is recovered from  \mbox{$g_{\mu\nu}= - 2 \gamma_{mn} e^m_\mu e^n_\nu$}. 

Since Einstein gravity in three dimensions is a purely topological theory it is imperative to introduce suitable boundary conditions in order to obtain interesting physics. A particular popular choice in terms of the Chern-Simons gauge field $\mathcal{A}$ is given by
first introducing coordinates $(r,u,\vp)$ where $0\leq r<\infty$, $-\infty<u<\infty$ and $\vp\sim\vp+2\pi$ as well as fixing $\mathcal{A}_r = b^{-1} \partial_r b$ and taking the $\vp$ component of the Chern-Simons connection to be
\begin{equation}\label{BMSbcs}
\mathcal{A}_{\vp} = b^{-1}(\Lt_{1} - \frac{\cM}{4} \Lt_{-1} - \frac{\cN}{2} \Mt_{-1})b\,, \qquad b = e^{\frac{r}{2} \Mt_{-1}}.
\end{equation}
The $u$-component of the Chern-Simons connection is a Lagrange multiplier that can be taken to be proportional to an infinitesimal gauge transformation that preserves the form of $\mathcal{A}_\vp$. These are the gauge transformations satisfying $\delta \mathcal{A}_{\vp} = \Lambda' + [\mathcal{A}_{\vp},\Lambda]$ and are given in terms of two arbitrary functions of $(u,\vp)$ that we denote $\epsilon_L$ and $\epsilon_M$
\begin{align}\label{Lambda}
\Lambda[\epsilon_L, \epsilon_M] = & \; \epsilon_M \Mt_1  - \epsilon_M' \Mt_0 - \frac14 \left(2 \epsilon_L \cN + \epsilon_M \cM - 2 \epsilon_M''\right) \Mt_{-1} \\
& \; +\epsilon_L \Lt_1  - \epsilon_L' \Lt_0 - \frac14 \left(\epsilon_L \cM - 2 \epsilon_L''\right) \Lt_{-1}\,. \nonumber
\end{align}
Under these transformations the state dependent functions $\cM$ and $\cL$ transform as
\begin{subequations}\label{delsol}
\begin{align}
	\delta \cM & = \epsilon_L \cM' + 2 \epsilon_L' \cM - 2 \epsilon_L'''\,, \\
	\delta \cN & = \frac12 \epsilon_M \cM' + \epsilon_M' \cM + \epsilon_L \cN' + 2 \epsilon_L' \cN - \epsilon_M'''\,,
\end{align}
\end{subequations}
where a prime denotes a derivative with respect to $\varphi$. The relations \eqref{delsol} describe precisely the coadjoint action of the $\mathfrak{bms}_3$ algebra \cite{Barnich:2015uva}. 
The $u$-component of the connection can now be obtained as a gauge transformation compatible with \eqref{Lambda}, where we replace the gauge parameters $\epsilon_{L,M}$ by the `chemical potentials' $\mu_{L,M}$:
\begin{equation} \label{Au}
    \mathcal{A}_u = b^{-1} \Lambda[\mu_L,\mu_M] b \,.
\end{equation} 
Taking $\mu_M = 1$ and $\mu_L=0$ is equivalent to the metric \eqref{BMS3metric} in the second order formulation. For these values of the chemical potentials the state dependent functions $\cM$ and $\cN$ satisfy
\begin{equation}\label{CSeom}
    \partial_u \cM = 0 \,, \qquad \partial_u \cN = \frac12 \partial_\vp \cM \,,
\end{equation}
as a consequence of the Chern-Simons field equations $F=\extd \mathcal{A}+\mathcal{A}\wedge\mathcal{A}=0$. This allows one to parametrize the solutions by two functions on the boundary circle $\cM = \cM(\vp)$ and 
\begin{equation}
    \cN(u,\vp) = \cL(\vp) + \frac{u}{2} \partial_\vp \cM(\vp).
\end{equation}
Using these functions the asymptotic charges generating the $\mathfrak{bms}_3$ transformations are given by
\begin{equation}\label{CSQ}
    Q[\epsilon_L] = \frac{k}{2\pi} \oint \extd\vp \; \epsilon_L \cL  \,, \qquad Q[\epsilon_M] = \frac{k}{4\pi} \oint \extd\vp \;   \epsilon_M \cM \,.
\end{equation}
The Fourier modes of the charges $M_n = Q[\epsilon_M = e^{im\vp}]$ and  $L_n = Q[\epsilon_L = e^{im\vp}]$ span the $\mathfrak{bms}_3$ algebra by their Dirac brackets. After promoting the modes to operators via the commutators the resulting $\mathfrak{bms}_3$ algebra reads
\begin{subequations}
	\label{BMS3}
	\begin{align}
	[\Lt_n, \Lt_m] & = (n-m)\Lt_{m+n} + \frac{c_1}{12} n^3 \delta_{m+n,0}\,, \\
	[\Lt_n, \Mt_m] & = (n-m)\Mt_{m+n} + \frac{c_2}{12} n^3 \delta_{m+n,0}\,, \\
	[\Mt_n, \Mt_m] & = 0\,,
	\end{align}
\end{subequations}
with the central charges $c_1 = 0 $ and $c_2 = 12 k = \frac{3}{G_N}$\,. 
In order to obtain the more standard normalization of $n(n^2-1)$ for the central charge terms in the algebra, one would have to shift the zero mode of $M_0$ by $c_2/24$, which is equivalent to shifting $\cM$ by 1. We prefer to keep working with the current normalization and hence have the Minkowski vacuum correspond to $\cM = -1$ (and $\cL =0$). 

\subsection{Coadjoint orbits of the BMS$_3$ group}

The previous considerations show that the reduced phase space of three-dimensional asymptotically flat gravity at null infinity is parametrized by two functions $\cM$ and $\cL$ that transform in the coadjoint representation of BMS$_3$. Here we collect some relevant statements about the BMS$_3$ group, its coadjoint orbits and the relation to gravitational solutions. For more details we refer to \cite{Barnich:2015uva,Oblak:2016eij,Barnich:2017jgw} and references therein.

\subsubsection{Coadjoint action}

The centrally extended BMS$_3$ group is the semi-direct product of the (universal cover of the) Virasoro group $\widehat{\rm Diff}(S^1)$ and its algebra (seen as an Abelian vector space) under the adjoint action, or:
\begin{equation}
{\rm BMS}_3 = \widehat{\rm Diff}(S^1) \ltimes_{{\rm Ad}} {\rm Vec} (S^1)_{\rm ab}\,.
\end{equation}
Its elements are denoted by $(f,\lambda; \alpha,\mu)$, where $f$ is a diffeomorphism of the circle, satisfying
\begin{equation}
f(\vp + 2\pi) = f(\vp) + 2\pi\,, \qquad f'(\vp) > 0\,,
\end{equation}
and the constant $\lambda$ denotes the central extension of ${\rm Diff}(S^1)$. The function $f$ parametrizes a superrotation, while $\alpha$ corresponds to a supertranslation and $\mu$ is its corresponding central extension. The supertranslations are periodic functions on the circle
\begin{equation}
\alpha(\vp + 2\pi) = \alpha(\vp)\,.
\end{equation}
The space of coadjoint vectors of BMS$_3$ is the dual space to the $\mathfrak{bms}_3$ algebra and its elements are denoted by $(j,c_1;p,c_2)$ where $j=j(\vp)\extd\vp^2$ and $p=p(\vp)\extd\vp^2$ are quadratic densities on the circle. These densities are dual to infinitesimal superrotations and supertranslations and are sometimes referred to as angular supermomentum and supermomentum, respectively. The constants $c_1 $ and $c_2$ are the two $\mathfrak{bms}_3$ central charges.

Elements of the $\mathfrak{bms}_3$ algebra are denoted by $(X,a;\alpha,b)$ and they are paired with elements $(j,c_1; p,c_2)$ via
\begin{equation}
	\langle (j,c_1; p,c_2) , (X,a; \alpha,b) \rangle = \frac{1}{2\pi}\int_0^{2\pi} \extd\vp \; \left[ j(\vp) X(\vp) +p(\vp) \alpha(\vp) \right] + c_1 a + c_2 b \, .
\end{equation}

The coadjoint action for semi-direct product groups $H = G \ltimes_{\rm Ad} \mathfrak{g}_{\rm ab}$ with elements $(g, \alpha)$ on its dual space with elements $(j,p)$ can be derived in terms of the coadjoint action of $G$ with elements $g$ as
\begin{equation}
{\rm Ad}_{(g,\alpha)^{-1}}^*(j,p) =  \left( {\rm Ad}_{g^{-1}}^* j - {\rm Ad}_{g^{-1}}^* {\rm ad}_{\alpha}^* p, {\rm Ad}_{g^{-1}}^* p \right)\,,
\end{equation}
where ${\rm ad}_{\alpha}^*$ is the coadjoint action of the algebra $\mathfrak{g}$ of $G$. See \cite{Barnich:2015uva} for more details on the general construction. In our case, it implies that the coadjoint action of BMS$_3$ on the dual space elements can be derived from the coadjoint action of the Virasoro group, with elements $(f, \lambda)$ on the elements of its dual space $(b, c)$:
\begin{equation}
    {\rm Ad}^*_{(f,\lambda)^{-1}} (b ,c) = \left(b(f) f'(\vp)^2 - \frac{c}{24\pi} \left\{ f, \vp \right\} ,c \right)\,.
\end{equation}
Here $\{f,\vp\}$ is the Schwarzian derivative
\begin{equation}\label{Schwarzian}
\{f,\vp\} = \frac{f'''(\vp)}{f'(\vp)} - \frac32 \left(\frac{f''(\vp)}{f'(\vp)}\right)^2\,.
\end{equation}
Following the general construction for semi-direct product groups of \cite{Barnich:2015uva} this implies that the coadjoint action of BMS$_3$ is 
\begin{equation}
{\rm Ad}^*_{(f,\alpha)^{-1}} (j,c_1 ; p, c_2) = (\tilde{j}\,  \extd\vp^2 , c_1 ; \tilde{p} \, \extd\vp^2, c_2) \,,
\end{equation}
where
\begin{subequations}\label{eq:CoadjointMN}
\begin{align}
\tilde{p} & =  f'(\vp)^2 p(f) - \frac{c_2}{24\pi} \{f,\vp\} \,, \\
\tilde{j} & =  f'(\vp)^2 \left( \partial_f p(f) \alpha (f ) + 2 \partial_f \alpha(f) p(f) - \frac{c_2}{24\pi} \partial_f^3 \alpha(f) \right) \\
& \qquad + f'(\vp)^2 j(f) - \frac{c_1}{24\pi} \{f,\vp\}. \nonumber
\end{align}
\end{subequations}
The coadjoint action of the $\mathfrak{bms}_3$ algebra is obtained as the infinitesimal version of the above, expanding $f(\vp) = \vp + \epsilon_L(\vp)$ and $\alpha = \epsilon_M(\vp)$ one obtains 
\begin{equation}
    {\rm ad}_{(\epsilon_L; \epsilon_M)}^*(j,c_1; p,c_2) = (\delta j \, \extd\vp^2, 0;\delta p \, \extd\vp^2, 0),
\end{equation}
with
\begin{subequations}
\begin{align}
    \delta p & = \epsilon_L p' + 2 \epsilon_L' p - \frac{c_2}{24\pi} \epsilon_L'''\,, \\
	\delta j & = \epsilon_M p' + 2 \epsilon_M' p  - \frac{c_2}{24\pi} \epsilon_M''' + \epsilon_L j' + 2 \epsilon_L' j - \frac{c_1}{24\pi} \epsilon_L'''\,.
\end{align}
\end{subequations}
This exactly corresponds to the transformation laws \eqref{delsol} upon the identifications
\begin{equation}\label{orbitreps}
    c_1 = 0\,, \qquad c_2 = 12k = \frac{3}{G_N} \,, \qquad j = \frac{k}{2\pi} \cL \,, \qquad p = \frac{k}{4\pi} \cM \,.
\end{equation}

\subsubsection{Coadjoint orbits}

Just as the coadjoint action of BMS$_3$ is derived from the coadjoint action of the Virasoro group, so are the coadjoint orbits of BMS$_3$ classified in terms of Virasoro coadjoint orbits (and coadjoint orbits of the little groups $G_p$ of elements $p \in \mathfrak{g}_{\rm ab}^*$). To be more precise, following \cite{Barnich:2015uva}, each coadjoint orbit $\cW_{(j,p)}$ of a semi-direct product group $H = G \ltimes_{\rm Ad} \mathfrak{g}_{\rm ab}$ is a fibre bundle over the orbits $\cO_p$ for $p \in \mathfrak{g}_{\rm ab}^*$ under the coadjoint action of $G$. The fibre above $q \in \cO_p$ is a product of the cotangent bundle $T^*_q \cO_p$ with the coadjoint orbit of the corresponding little group $G_p$. To classify the coadjoint orbits of $H$ it is sufficient to know the set of all orbits $\cO_p$ and all coadjoint orbits of the corresponding little groups. 

A systematic approach to classify the coadjoint orbits of BMS$_3$ is to
\begin{enumerate}
    \item Pick an element $p \in {\rm Vec} (S^1)_{\rm ab}^*$ and compute its orbit under the coadjoint action of $\widehat{\rm Diff}(S^1)$. 
    \item Find the corresponding little group $G_p$ or, the stabilizer subgroup on the coadjoint orbit $\cO_p$.
    \item Pick an element $j_p \in \mathfrak{vir}^* $ and compute its coadjoint orbit under the action of $G_p$.
\end{enumerate}
The first two steps are equivalent to the classification of coadjoint orbits of the Virasoro group \cite{Witten:1987ty}. There are two types of Virasoro coadjoint orbits with constant representatives and non-vanishing central charge (to which we will restrict ourselves)
\begin{itemize}
    \item The {\bf exceptional orbits} $ \cO_{(p_n,c_2)}$ of the Virasoro group have representative $p = p_n = - \frac{c_2}{48\pi} n^2$ for positive integer $n$. For these orbits the little group is the $n$-fold cover of PSL$(2,\bR)$, hence these orbits are manifolds $\widehat{\rm Diff}(S^1)/\textrm{PSL}^{(n)}(2,\bR)$.
    \item The {\bf generic orbit} $ \cO_{(p_0,c_2)}$ have representatives $p = p_0 \neq - \frac{c_2}{48\pi} n^2$. The little group in this case is a one-dimensional Abelian group and the orbits are manifolds $\widehat{\rm Diff}(S^1)/S^1$.
\end{itemize}
In the latter case, the coadjoint representation of the little group is trivial and hence the generic BMS$_3$ orbit $\cW_{(j_0,c_1;p_0,c_2)}$ is diffeomorphic to the cotangent bundle $T^* \cO_{(p_0,c_2)}$. The exceptional orbits of BMS$_3$, $\cW_{(j_{p_n},c_1;p_n,c_2)}$ are fibre bundles over $T^* \cO_{(p_n,c_2)}$, with coadjoint orbits of PSL$^{(n)}(2, \bR)$ as its fibres. 

For the sake of this work we will restrict to orbits on which the energy is bounded from below. For this we first need an appropriate measure of energy. Asymptotically, time translations are generated by Chern-Simons gauge transformations with $\epsilon_L = 0$ and \mbox{$\epsilon_M =1$} and hence energy can be defined as a Chern-Simons charge \eqref{CSQ} with those values, or
\begin{equation}
    E = \frac{k}{4\pi} \oint \extd\vp \; \cM(\vp) = \oint \extd\vp\;  p (\vp) \,.
\end{equation}
Because $p(\vp)$ transforms as a coadjoint vector of the Virasoro group, the energy $E$ satisfies the same bounds as the energy on the Virasoro coadjoint orbits \cite{Barnich:2015uva}. This was analyzed in \cite{Witten:1987ty} and the result is that the energy is bounded for orbits with constant representative $p_0 \geq - \frac{c_2}{48\pi}$. This implies that all the flat space cosmologies (with $p_0 > 0$) and all conical deficit solutions ($- \frac{c_2}{48\pi} < p_0 < 0$) have energy bounded from below. We will refer to the orbits $\cW_{(j_0,0; p_0,c_2)}$ corresponding to these solutions as the massive BMS$_3$ orbits. Furthermore, the Minkowski vacuum (with $j = 0$ and $p = -\frac{c_2}{48\pi}$) also has its energy bounded from below and it does so for the lowest value of the energy $E = - \frac{c_2}{24} $. The corresponding orbit is the first exceptional orbit of BMS$_3$ $\cW_{(0,0;p_1,c_2)}$, to which we will refer as the vacuum orbit of BMS$_3$.

\subsubsection{Geometric action}\label{sec:coadjointaction}

The coadjoint orbit $\cO_p$ of any group $G$ is a homogeneous symplectic space $G/G_p$, where $G_p$ is the little group at $p$. To each of these orbits one can associate a geometric action which admits $G$ as global symmetry and $G_p$ as gauge symmetry. The kinetic term of this action is fixed by the Kirillov-Konstant symplectic form  \cite{Kirillov:1976}, which is the pullback to the coadjoint orbit of the pre-symplectic form on $G$. Reviews on the construction of the Kirillov-Konstant symplectic form and the associated geometric action are aplenty in the literature (see for instance \cite{Alekseev:1988ce,Alekseev:1990mp,Aratyn:1990dj,Aratyn:1989qq,Cotler:2018zff}). It will suffice here to say that the construction was generalized to infinite dimensional semi-direct product groups with central extensions in \cite{Barnich:2017jgw}. The result for the BMS$_3$ group is the action with kinetic term
\begin{align}\label{Ibms3}
I_{{\rm BMS}_3}[f,\alpha,j_0,p_0,c_1,c_2] = & - \int  \extd u \extd\vp\; \bigg[  j_0(f) \dot{f}f' + \frac{c_1}{48\pi} \frac{\dot{f}''}{f'} \\
& + \dot{f}f' \Big(\partial_f p_0(f) \alpha(f) + 2 p_0(f) \partial_f \alpha(f) - \frac{c_2}{24\pi} \partial_f^3\alpha(f) \Big) \bigg] \nonumber\,.
\end{align}
Where primes denote $\vp$ derivatives, dots denotes $u$ derivatives and $u$ parametrizes a path along the orbit. Here there is no restriction on the orbit representatives $j_0$ and $p_0$, they need not be constant, although the action simplifies if they are.

Another accomplishment of the work \cite{Barnich:2017jgw} was to show how the geometric action on the coadjoint orbits of any gauge group can be deformed by adding Hamiltonians that preserve the global symmetries of the theory. One can add to the kinetic term \eqref{Ibms3} as Hamiltonian the Noether charge $Q_{(\epsilon_L,\epsilon_M)}$ of a global symmetry (generated by $\mathfrak{bms}_3$ vector fields $(\epsilon_L(\vp),\epsilon_M(\vp))$) and the resulting action will by construction preserve the global symmetries of \eqref{Ibms3}. In the case at hand, global symmetries act as
\begin{equation}
    \delta_{(\epsilon_L,\epsilon_M)} (f, \alpha(f)) = ( \epsilon_L(\vp) \partial_\vp f, \epsilon_M(\vp) \partial_\vp f) \,,
\end{equation}
leading to the Noether charges:
\begin{align}\label{Noethercharge}
    Q_{(\epsilon_L,\epsilon_M)} & =  \oint \extd \vp \, \bigg[ \epsilon_M \left( f'{}^2 p_0(f) - \frac{c_2}{24\pi} \left\{ f, \vp \right\} \right)\\
    & \nonumber + \epsilon_L \left(  f'^2 \left( j_0(f) + \partial_f p_0(f) \alpha (f) + 2 p_0(f) \partial_f \alpha(f) - \frac{c_2}{24\pi} \partial_f^3 \alpha(f)  \right) - \frac{c_1}{24\pi} \left\{ f, \vp \right\} \right) \bigg]\,.
\end{align}
For constant $(\epsilon_L,\epsilon_M)$, the action \eqref{Ibms3} including the  Hamiltonian \mbox{$\int H \extd u = \int Q_{(\epsilon_L,\epsilon_M)} \extd u$} is invariant under global symmetries modified by $\delta_{(X,v)}(f, \alpha (f)) = (X(\vp,u) \partial_\vp f, v(\vp,u) \partial_\vp f)$ with
\begin{equation}
    X = X_0( \vp + \epsilon_L u) \,, \qquad v = v_0 (\vp + \epsilon_L u) + u \epsilon_M \partial_\vp X_0\,.
\end{equation}

The gauge symmetry of the action is related to the stabilizer subgroup on the orbit. For constant representatives, the stabilizer subalgebra is generated by $\mathfrak{bms}_3$ vectors $(\epsilon_L, \epsilon_M)$ satisfying
\begin{subequations}
\begin{align}
    \epsilon_L''' - P_0 \epsilon_L' & = 0 \,, \\
    \epsilon_M''' - P_0 \epsilon_M' & = J_0 \epsilon_L' - \frac{c_1}{c_2} \epsilon_L''' \,,
\end{align}
\end{subequations}
where we have defined $P_0 = \frac{48 \pi}{c_2} p_0 $ and $J_0 = \frac{48 \pi}{c_2} j_0 $. The generic solution to these equations is given by
\begin{subequations}
\begin{align}
    \epsilon_L & = \ell_0 + \ell_{+} e^{\sqrt{P_0} \vp } + \ell_{-} e^{-\sqrt{P_0} \vp}\,, \\
    \epsilon_M & = m_0 + \left( m_+ + \ell_+ \frac{c_1 P_0 - c_2 J_0}{2 c_2 \sqrt{P_0} } \vp \right) e^{\sqrt{P_0}\vp} + \left( m_-  - \ell_- \frac{c_1 P_0 - c_2 J_0}{2 c_2 \sqrt{P_0} } \vp \right) e^{-\sqrt{P_0}\vp} \,,
\end{align}
\end{subequations}
where $\ell_{0,\pm} $ and $m_{0, \pm}$ may be arbitrary functions of $u$.
It's easy to see that these solutions are only periodic in $\vp$ for $P_0 = - n^2$ and $J_0 = - \frac{c_1}{c_2} n^2$ with $n\in \mathbb{Z}$, or whenever
\begin{equation}
    p_0 = - \frac{c_2}{48\pi} n^2 \,, \qquad   j_0 = - \frac{c_1}{48\pi} n^2\,.
\end{equation}
For these values, the $\mathfrak{bms}_3$ vectors span an $n$-fold cover of $\mathfrak{isl}(2,\bR)$. Hence, the exceptional orbits $\cW_{(j_{p_n},c_1;p_n,c_2)}$ have stabilizer subgroup ISL$^{(n)}(2,\bR)$. For the generic orbits, one would have to set $\ell_{\pm} = 0 = m_{\pm}$ and the gauge symmetry on these orbits consists solely of shifts by an arbitrary function of $u$, with Abelian algebra.

\section{Reduction of the action}\label{sec:reduction}

In this section we perform the Hamiltonian reduction of the Chern-Simons action for three-dimensional flat Einstein gravity. By imposing the BMS$_3$ boundary conditions of \cite{Barnich:2006av} the bulk Chern-Simons theory reduces to a two-dimensional boundary theory. Allowing for non-trivial bulk holonomies the boundary theory is shown to be equivalent to the geometric action on the coadjoint orbit of the BMS$_3$ group that has been discussed in the last section, with the orbit representatives proportional to the bulk holonomies. We comment on the boundary Hamiltonian and the classical saddle points.

\subsection{Chern-Simons to Wess-Zumino-Witten}

The first step in the reduction is to write the Chern-Simons action as a chiral WZW model \cite{Elitzur:1989nr}. This part is generic for all Chern-Simons theories on manifolds with the topology of filled cylinder and has already appeared in various places in the literature (see \cite{Donnay:2016iyk} for a review and references). We follow here the presentation of \cite{Grumiller:2019tyl}, but this step in the reduction has first appeared for gravity with asymptotically flat spacetimes in \cite{Barnich:2013yka}. 

The starting point of the reduction is the Hamiltonian form of the Chern-Simons action \eqref{Scs} for ISL$(2,\bR)$ on a manifold $\cM$ with the topology of a filled cylinder and equipped with coordinates $u, r,  \vp$, supplemented by a boundary term $I_{\rm bdy}$ 
\begin{equation}\label{ScsHam}
I_{\textsc{cs}}[\mathcal{A}] = \frac{k}{4\pi} \int_{\cM} \extd u \extd r \extd\vp\, \vev{ \mathcal{A}_{r} \dot{\mathcal{A}}_{\vp} - \mathcal{A}_{\vp} \dot{\mathcal{A}}_{r} + 2 \mathcal{A}_{u} F_{\vp r} } + I_{\rm bdy} \,.
\end{equation}
The boundary term $I_{\rm bdy}$ should be fixed such that the variational principle is well-defined, i.e., the variation of the action should vanish exactly on-shell:
\begin{equation}\label{bdyterm}
\delta I_{\textsc{cs}}[\mathcal{A}]\big|_{\textrm{\tiny EOM}} = \delta I_{\rm bdy} -  \frac{k}{2\pi} \int_{\partial \cM} \extd u \extd\vp \, \vev{\mathcal{A}_u \delta \mathcal{A}_{\vp}} = 0\,.
\end{equation} 
We are going to fix $I_{\rm bdy}$ in the next section, which will give the boundary Hamiltonian. For now, we focus on the symplectic terms in \eqref{ScsHam}.

The $u$-component of the Chern-Simons connection imposes the constraint $F_{\vp r}=0$, which is solved locally by
\begin{equation}
\mathcal{A}_i = G^{-1}\partial_i G, \qquad\qquad i =\vp, r, \qquad\qquad G \in \textrm{ISL}(2,\bR)\,.
\end{equation}
In a gauge where $\cA_r^\prime = 0$ (prime denotes $\partial_\vp$) the group element $G$ can be factorized as
\begin{equation}\label{eq:nha7}
G(u,\,\vp,\,r) = g(u,\,\vp)\,b(u,\,r),
\end{equation}
implying
\begin{equation}
\mathcal{A}_{\vp} = b^{-1} a_{\vp}b =  b^{-1}g^{-1}g^\prime b, \qquad \qquad \mathcal{A}_{r} = b^{-1} \partial_{r} b\,.
\end{equation}
In the present work we always assume $u$-independence of $b$ at the boundary, $\dot b|_{\partial\cM}=0$.

For smooth and non-singular Chern-Simons connections on the disk $D$ the Wilson loop around the $\vp$-cycle is contractible and hence the holonomy is trivial. However, the solutions of interest here are strictly speaking not all non-singular everywhere (such as the conical defect solutions) and neither do they have the topology of a filled cylinder; there are two asymptotic regions, $\mathscr{I}^+$ and $\mathscr{I}^-$ and both are null cylinders. We should hence actually allow defects or other boundaries in the interior of the disk. In the case of multiple boundaries, there are independent actions at each boundary and the bulk holonomy need not vanish, but will couple the two boundaries non-locally.\footnote{We are also ignoring possible matching conditions to be imposed at $\mathscr{I}^-_+$ and $\mathscr{I}^+_-$.}
We prefer to keep working with one asymptotic region here, but we do not simply want to ignore the possibility to have non-trivial global charges. One effective way of doing so is to consider the dynamics on the outer boundary only, but to keep the holonomies along the $\vp$-cycle non-trivial to account for whatever has been ignored in the inside. This effectively replaces the annulus by a punctured disk and we do not impose boundary conditions or consider dynamics at the puncture. It is merely there to prevent loops around the $\vp$-cycle to contract to a point. In principle, a complete analysis should take into account the inner boundary with its own boundary conditions, a dynamical holonomy and matching conditions at the corners. This is, however, not necessary for our purposes and hence beyond the scope of this work. For a recent treatment of this in the AdS$_3$ case, see \cite{Henneaux:2019sjx}.

There are two ways to treat non-trivial holonomies. One may write the gauge connection as sum of a periodic group element $g$ plus a term representing the holonomy. Alternatively, the holonomies can be encoded in the periodicity properties of the group element $g$. We follow the latter approach and write
\begin{equation} \label{bcimpl}
a_{\vp} = g^{-1} g^\prime, \qquad \qquad 
g(u,\,\vp + 2\pi) = h g(u,\,\vp), 
\end{equation}
where $h\in$\,ISL$(2,\bR)$ such that $\vev{h} = H_\vp$, where the bracket was given in  \eqref{eq:InvBilForm} and $H_\vp=\vev{\cP  e^{\oint \mathcal{A}_\vp}}$ denotes the holonomy around the $\vp$-cycle. We assume in this work that $h$ is $u$-independent.\footnote{Dropping this assumption would be required only in the presence of (matter) sources that can change the holonomy in a time-dependent way and would imply additional boundary terms in the WZW action \eqref{WZW2} \cite{Grumiller:2019tyl}.} 

After choosing the above gauge we can write the Chern-Simons action on the punctured disk times $\bR$ as
\begin{align}\label{WZW2}
I_{\textrm{CS}}[G] = - \frac{k}{4\pi} \int_{\partial \cM} \extd u\extd\varphi\, \vev{\partial_\varphi {g} {g}^{-1}  \partial_u {g} {g}^{-1} } - I_{\rm WZ}[{G}] + I_{\rm bdy},
\end{align}
where
\begin{equation}\label{WZterm2}
I_{\rm WZ}[G] =  \frac{k}{12\pi} \int_{\cM} \vev{ G^{-1} \extd G\wedge G^{-1} \extd G\wedge G^{-1} \extd G }\,.
\end{equation}
This is the WZW model for affine $\widehat{\mathfrak{isl}}(2,\bR)$.

\subsection{WZW to the geometric action of BMS$_3$}

We now decompose the group elements $G(u,\vp,r)$ into different 
ISL$(2,\bR)$ components by writing
\begin{equation}
	G(u,\vp,r) = e^{X \Lt_+} e^{W \Mt_+} e^{\Phi \Lt_0} e^{\zeta \Mt_0} e^{Y \Lt_-}  e^{V \Mt_-},
\end{equation}
where $X, \Phi, Y, W, \zeta, V$ are functions of $u,\vp,r$ and their pull-back to the boundary depends only on $(u,\vp)$. Using this decomposition, the first term in \eqref{WZW2} becomes
\begin{equation}
	 \frac{k}{2\pi}\int \extd u\extd\vp \; \left[ e^{\Phi}\left(\dot{V}X'+\dot{X}V'+\dot{Y}W'+\dot{W}Y'+\zeta\left(\dot{X}Y'+\dot{Y}X'\right) \right) -\frac{1}{2}\left(\dot{\Phi}\zeta'+\dot{\zeta}\Phi'\right)\right].
\end{equation}
The Wess-Zumino term \eqref{WZterm2} can be conveniently written as a total derivative
\begin{equation}
	I_{\text{WZ}}[G] = -\frac{k}{2\pi}\int_{\cM} \extd^3x \epsilon^{\mu\nu\rho} \partial_{\mu} \bigg[e^{\Phi}\left(\partial_\nu Y\partial_\rho W+\partial_\nu V\partial_\rho X+\zeta\partial_\nu Y\partial_\rho X\right)
	\bigg].
\end{equation}
This brings the total action \eqref{WZW2} to the form:
\begin{equation}\label{WZW3}
    I_{\rm CS}[G] =  \frac{k}{2\pi}\int \extd u\extd\vp \;\left[ 2 e^{\Phi}\left(\dot{Y}W' +X' \left(\dot V + \zeta\dot{Y} \right) \right) -\frac{1}{2}\left(\dot{\Phi}\zeta'+\dot{\zeta}\Phi'\right)	\right] + I_{\rm bdy}\,.
\end{equation}
Due to \eqref{bcimpl} the fields appearing here are not periodic in $\vp$, so in principle one should take total derivatives in $\vp$ into account. However, these terms do not contribute here or below due to our assumption that $h$ is $u$ independent.

The conditions \eqref{BMSbcs} impose constraints on the set of 6 fields. They are:
\begin{align}\label{constraintsflat}
e^{\Phi} X' & =  1\,, &  Y & = - \frac{\Phi'}{2}\,, &
e^{\Phi}W' & = - \zeta ,\,  & V & = - \frac{\zeta'}{2}\,.
\end{align}
In addition, they define $\cM$ and $\cN$ in terms of the fields $X,W$ as:
\begin{subequations} \label{MNdef}
\begin{align}
    \cM & = 4(Y^2-Y') = - 2 \{X,\vp\},\\ 
    \mathcal{N} & = - X' \partial_\vp \left( \frac{1}{X'} \partial_\vp \left( \frac{1}{X'} \partial_\vp W \right)  \right) 
\equiv -(X')^2\partial_X^3W \,.
\end{align}
\end{subequations}
Here $\{X,\vp\}$ is again the Schwarzian derivative \eqref{Schwarzian}. Implementing these constraints in the action \eqref{WZW3} gives
\begin{equation}\label{redaction}
	I_{\textrm{CS}}[W,X] =  \frac{k}{2\pi}\int \extd u \extd\vp\; \dot{X} X' \partial_X^3W + I_{\rm bdy} \,.
\end{equation}
This action is the geometric action on the BMS$_3$ coadjoint orbits \eqref{Ibms3} with vanishing orbit representatives. If the bulk Chern-Simons theory would have trivial holonomy then the fields $W$ and $X$ are periodic. In that case the bulk corresponds to the null orbifold and the action \eqref{redaction} is boundary action describing BMS$_3$ transformations of the null orbifold.  

Note that this action is equivalent to the BMS$_3$ Liouville theory of \cite{Barnich:2012rz,Barnich:2013yka} upon using the constraints \eqref{constraintsflat} to write the action in terms of $\Phi$ and $\zeta$. Including the boundary Hamiltonian, this action is
\begin{equation}\label{BMSLiouville}
	I[\Phi,\zeta] = \frac{k}{2\pi}\int \extd u \extd\vp\; \left(\Phi'{}^2 - \dot \zeta  \Phi'\right)\,,
\end{equation}
up to total derivatives.

In order to understand how the orbit representative enters the action we need to study the periodicity properties of the fields that are inherited from the holonomy of the bulk Chern-Simons connection $\mathcal{A}_{\vp}$. The holonomy is given by the path-ordered exponential of the Chern-Simons connection integrated over a closed $\vp$ loop. In general the path ordered exponential is quite difficult to compute for arbitrary $\cM(u,\vp)$ and $\cN(u,\vp)$ and we will therefore take the following point of view. We compute the holonomy for the classical saddle points of interest, with constant $\cM = \cM_0$ and $\cN = \cL_0$. The reduction procedure will then lead to effective action of gauge transformations around these classical saddles, consistent with the BMS$_3$  boundary conditions \eqref{BMSbcs}.

For constant $\cM = \cM_0$ and $\cN = \cL_0$, the holonomy is 
\begin{equation}
	H_{\vp} = \vev{\cP  e^{\oint \mathcal{A}_\vp}} = \frac{4 \pi \cL_0 \sinh (\pi \sqrt{\cM_0})}{\sqrt{\cM_0}},
\end{equation}
We can parametrize the holonomy by including a non-standard periodicity in the group elements $g$ in the reduction. We take
\begin{equation}
g(\vp +2\pi) = h g(\vp)\,, \qquad \text{with:} \;\; h = e^{2\pi \sqrt{\cM_0}\Lt_0} e^{\frac{2\pi \cL_0}{\sqrt{\cM_0}}\Mt_0}.
\end{equation}
This implies the above fields have the following $\vp$-periodicities
\begin{subequations}\label{period1}
\begin{align}
	X(\vp+2\pi) & = e^{-2\pi \sqrt{\cM_0}}X(\vp)	, & \Phi(\vp+2\pi) & = \Phi(\vp) + 2\pi \sqrt{\cM_0} \,, \\
	W(\vp+2\pi) & = e^{-2\pi \sqrt{\cM_0}} \left( W(\vp) - \frac{2\pi \cL_0  X(\vp)}{\sqrt{\cM_0}} \right),  & \zeta(\vp+2\pi) & = \zeta(\vp) + \frac{2\pi \cL_0}{\sqrt{\cM_0}}, 
\end{align}
\end{subequations}
and
\begin{align}\label{period2}
	Y(\vp+2\pi) & = Y(\vp) , &&
	V(\vp+2\pi) = V(\vp).
\end{align}
To connect with the geometric action \eqref{Ibms3}, we need a field redefinition describing the action in terms of fields $f(\vp,u)$ and $\alpha(f,u)$ with the periodicities
\begin{equation}
f(\vp+2\pi,u) = f(\vp,u) + 2\pi\,, \qquad \qquad \alpha(f+2\pi, u) = \alpha(f,u).
\end{equation}
A field redefinition that achieves precisely this is 
\begin{subequations}\label{redef}
\begin{align}
	X(\vp,u) & = e^{-\sqrt{\cM_0} f(\vp,u)} \,, \\
	W(\vp,u) & = - \sqrt{\cM_0} e^{-\sqrt{\cM_0} f(\vp,u)} \left( \alpha(f(\vp,u),u) + \frac{\cL_0}{\cM_0} f(\vp,u) \right).
\end{align}
\end{subequations}
It is easy to check that this satisfies the periodicity conditions \eqref{period1}-\eqref{period2} by using the constraints \eqref{constraintsflat}. In terms of the new variables \eqref{MNdef} becomes:
\begin{subequations}\label{Tdef}
\begin{align}
    \cM & = \cM_0 f'^2 - 2 \{f,\vp\},\\ 
    \mathcal{N} & = f'^2 \left(\cL_0 + \cM_0 \partial_f\alpha(f)  - \partial_f^3 \alpha(f) \right) \,.
\end{align}
\end{subequations}
This is exactly the coadjoint action of BMS$_3$ \eqref{eq:CoadjointMN} generated by $f$ and $\alpha$, starting from the orbits of constant representative.
Plugging \eqref{redef} into the action \eqref{redaction} we find
\begin{equation}
I_{\textrm{CS}}[f,\alpha,\cL_0,\cM_0] = - \frac{k}{2\pi} \int \extd u \extd\vp \; \left( \cL_0 + \cM_0 \partial_f\alpha(f) - \partial_f^3\alpha(f) \right) \dot{f} f'.
\end{equation}
Comparing this to the geometric action on the BMS$_3$ coadjoint orbit \eqref{Ibms3}.
We find that it matches for constant $j_0$ and $p_0$ when
\begin{equation}\label{identifications}
	j_0 = \frac{k }{2\pi}\cL_0\,, \qquad p_0 = \frac{k }{4\pi} \cM_0 \,, \qquad c_1 = 0 \,, \qquad c_2 = 12k = \frac{3}{G_N}\,.
\end{equation}
This agrees with the values obtained before in \eqref{orbitreps}.

\subsection{Hamiltonian and classical saddles}

The Hamiltonian of the boundary theory comes from the boundary term added to the Chern-Simons action in order to ensure a well-defined variational principle. The variation of this term is
\begin{equation}
\delta I_{\rm bdy}= \frac{k}{2\pi} \int \extd u \extd\vp \vev{\mathcal{A}_u \delta \mathcal{A}_\vp}\,.
\end{equation}
Using that $\mathcal{A}_u$ is given by \eqref{Au} we obtain that the boundary term is
\begin{equation}
I_{\rm bdy} = \frac{k}{4\pi} \int \extd u  \extd\vp \left(2\mu_L\mathcal{N} + \mu_M \cM \right).
\end{equation}
Using  \eqref{Tdef} gives a boundary Hamiltonian for constant $\mu_L$ and $\mu_M$
\begin{align}\label{genHam}
	I_{\rm bdy} = \frac{k}{2\pi} \int \extd u \extd\vp \Big[& \mu_L\left( \cL_0 + \cM_0 \partial_f\alpha(f) - \partial_f^3\alpha(f) \right)  f'^2 \\
	& \nonumber \qquad \qquad + \frac12 \mu_M \left( \cM_0 f'^2  - 2 \{f,\vp\} \right) \Big]\,.
\end{align}
This exactly corresponds to the Hamiltonian added in section \ref{sec:coadjointaction} as the Noether charge for global symmetries \eqref{Noethercharge} upon making the identifications \eqref{identifications}.

When $\mu_M= 1$ and $\mu_L=0$ this is the Hamiltonian suggested in \cite{Barnich:2017jgw} to be relevant for three-dimensional gravity in asymptotically flat spacetimes. We continue with this choice of chemical potentials. The final action is then
\begin{align} \label{Sfinal}
    I_{\textrm{CS}}[f,\alpha,\cL_0,\cM_0] = - \frac{k}{2\pi} \int \extd u \extd\vp \; \Big[& \left( \cL_0 + \cM_0 \partial_f\alpha(f) - \partial_f^3\alpha(f) \right) \dot{f} f' \\
    & \nonumber \qquad \qquad  -  \frac12 \left( \cM_0 f'^2  - 2 \{f,\vp\} \right) \Big].
\end{align}
To verify whether or not the reduction has been consistent\footnote{Inconsistencies in the reduction can arise where the reduced action has equations of motion differing from original field equations with the constraints implemented. This is the case in the reduction to Liouville theory and can be remedied by changing from Dirichlet to Neumann boundary conditions of the reduced variables \cite{Coussaert:1995zp}.}, we see if the final action \eqref{Sfinal} has the same equations of motion as the original Chern-Simons theory. The equation obtained by varying with respect to $\alpha(f)$ is
\begin{equation}\label{alphaeqn}
    \frac{1}{f'} \partial_u \left( \{f,\vp\} - \frac12 \cM_0 f'^2  \right)  = 0\,.
 \end{equation}
Varying the action with respect to $f$ gives
\begin{equation}\label{feqn}
    \frac{1}{f'} \partial_u f'^2 \left(\cL_0 + \cM_0 \partial_f\alpha(f)  - \partial_f^3 \alpha(f) \right)  = \frac{1}{f'} \partial_\vp \left( \frac12 \cM_0 f'^2  - \{f,\vp\} \right) \,.
 \end{equation}
Fortunately, using \eqref{Tdef}, we find that for $f' \neq 0$ these equations are equivalent to the original field equations \eqref{CSeom}. 

Equation \eqref{feqn} can be solved by taking
\begin{equation}\label{alphasol}
    \alpha(f,u) = g(f,u) + u f' \,,
\end{equation}
where $g(f,u)$ is a solution to the homogeneous equation
\begin{equation}
    \frac{1}{f'} \partial_u \left( f'^2 \left(\cL_0 + \cM_0 \partial_f g(f,u)  - \partial_f^3 g(f,u) \right) \right) = 0\,.
\end{equation}

Solutions corresponding to the gravitational saddle points of interest have constant $\cM$ and $\cN$ with their values given by the zero-modes $\cM_0$ and $\cL_0$. For flat space cosmologies we have $\cM_0 >0$ and $\cL_0 \neq0$, leading to 
\begin{equation}\label{saddle}
f = \vp \,, \qquad \alpha(f,u) = u\,,
\end{equation}
as the unique saddle points respecting the periodicity conditions $f(\vp + 2\pi) = f( \vp) + 2 \pi$ and modulo the gauge redundancy that we can use to shift $f$ and $\alpha$ by arbitrary functions of $u$. For the Minkowski ground state, where $\cM_0 = -1$ and $\cL_0 = 0$ and for the null orbifold where $\cM_0=0$ and $\cL_0=0$ there are more solutions consistent with the periodicity conditions, but also the gauge symmetry is enlarged to the global ISL$(2,\bR)$. Modulo this gauge redundancy the unique saddle is still given by \eqref{saddle}. 

\subsubsection*{What have we learned from all this?}

In this section we have shown that the bulk Chern-Simons action for ISL$(2,\bR)$ can be reduced to a boundary action for fields generating BMS$_3$ transformations around a given classical saddle. This action coincides with the geometric action on the coadjoint orbits of BMS$_3$, where the orbit representatives are the charges of the saddles of interest.  For the null orbifold, with vanishing boundary charges, the action becomes \eqref{redaction} and we have seen that the orbit representatives can be instated from this by the transformations \eqref{redef}. This also implies that the inverse of these transformations can be used to map the theory for arbitrary constant, but non-zero representatives to the null orbifold. 

Now that we have found the effective action for BMS$_3$ transformations around a given background, we can use it to compute quantum ($\cO(1/c_2)$) corrections to a given classical quantity. In the remainder of the paper we will do so in several examples. First we show how the geometric action on the coadjoint orbit of BMS$_3$ can be used to compute the one-loop contribution to the partition function of flat space gravity for both Minkowski spacetime and flat space cosmologies. Then we discuss how to compute boundary correlators from  Wilson lines ending on the boundary and we will use them to compute entanglement entropy and its leading order quantum correction. Finally, in section \ref{sec:BMSblocks}, we show how the Wilson lines can be used to compute the BMS$_3$ identity block and its subleading terms, both for light operators and in the heavy-light limit.

\section{Flat space torus partition function}\label{sec:partitionfun}

In this section we compute the flat space torus partition function from the geometric action on the BMS$_3$ coadjoint orbit. The approach taken here follows the one taken in \cite{Cotler:2018zff} for AdS$_3$ gravity and extends their analysis to flat space. We will see that the result matches BMS$_3$ characters obtained in \cite{Oblak:2015sea} (see also \cite{Garbarz:2015lua,Bagchi:2019unf}) both for the one-loop contribution around the Minkoswki vacuum and for the flat space cosmologies. 

The first thing to do is to analytically continue $u \to - i y$ and periodically identify the new Euclidean ``time'' coordinate. Here we can choose to include the chemical potentials $\mu_L$ and $\mu_M$ into the periodicity and twist of the torus, or keep them explicitly in the action and use canonical periodicities $y \sim y +1, \vp \sim \vp+2\pi$ on the contractible cycle. Here we choose to set $\mu_L =0 $ and $\mu_M =1$ and use fields with periodicity conditions
\begin{subequations}\label{falphabc}
\begin{align}
f( \vp + \Omega \beta,y+  \beta) & = f(\vp,y)\,, & \alpha(f+ \Omega \beta,y +  \beta) & = \alpha(f,y)\,, \\
f(\vp+2\pi,y) & = f(\vp,y) + 2\pi \,, & \alpha(f+2\pi,y) & = \alpha(f,y)\,.
\end{align}
\end{subequations}
The Euclidean action under consideration is then
\begin{equation}
I_{\textrm{E}} = - \frac{k}{2\pi} \int \extd y \extd\vp \left[ i \left(\cL_0 + \cM_0 \alpha'(f) - \alpha'''(f)\right)f' \partial_y f + \{f,\vp\} -  \frac{\cM_0}{2} f'{}^2 \right].
\end{equation}
The real part of the Euclidean action is $\int \extd y H$ with
\begin{equation}
H = - \frac{c_2}{24\pi} \int \extd\vp \left[ \{f,\vp\} - \frac{\cM_0}{2}f'{}^2\right],
\end{equation}
which is bounded from below for $c_2>0, \cM_0 \geq -1$ \cite{Witten:1987ty}. In this case this bound is saturated by the Minkowski vacuum that has $\cM_0 =-1$ and $\cL_0 = 0$. 

Due to the periodicity conditions on the thermal $y$-cycle, we have to use the gauge ambiguity to make the $y$ dependence of the saddles consistent with \eqref{falphabc}. The solution is to take
\begin{equation}
	f_0 = \vp - \Omega y\,, \qquad \alpha (f_0) = 0 \,.
\end{equation}
The action on the saddle point is
\begin{equation}
	I_{\textrm{E}}^{(0)} = \frac{c_2}{24} \beta ( \cM_0 + 2 i \Omega \cL_0)\,.
\end{equation}
In the gravitational theory the parameters $\Omega, \beta$ are related to $\cL_0,\cM_0$ to ensure the regularity of the solution at the cosmological horizon \cite{Bagchi:2013lma}. In the Chern-Simons language this means that the connection has trivial holonomy along the thermal cycle, which is the contractible cycle of the torus. In our conventions this gives the conditions
\begin{equation}
	\exp \left[ - i \int_{0}^{\beta} a_y \extd y + \int_{0}^{\beta \Omega} a_\vp \extd\vp \right] = - \unity.
\end{equation}
These conditions are solved by taking
\begin{equation}\label{musol}
\Omega = \frac{i \cM_0}{\cL_0}\,, \qquad \qquad \beta = \frac{2\pi \cL_0}{\cM_0^{3/2}}\,.
\end{equation}
This implies that the on-shell Euclidean action is
\begin{equation}
I_{\textrm{E}}^{(0)} = - \frac{\pi r_0}{4 G_N}\,,
\end{equation}
which is consistent with the gravitational computation performed in \cite{Bagchi:2013lma}.

Next we expand the fields $f(\vp,y)$ and $\alpha(f,y)$ around their saddle points. Note that we allow $\alpha$ to depend on Euclidean time both implicitly through $f(\vp,y)$ and explicitly.
We write
\begin{subequations}
\begin{align}
f(\vp,y) & = f_0 + \sum_{m,n} \frac{\epsilon_{m,n}}{(2\pi)^2} e^{-\frac{2\pi i m y}{\beta} }e^{-i n f_0}\,, \\
\alpha(f,y) & = \sum_{m,n} \frac{\alpha_{m,n}}{(2\pi)^2} e^{-\frac{2\pi i m y}{\beta} }e^{-i n f}\,, \\
& = \sum_{m,n} \frac{\alpha_{m,n}}{(2\pi)^2} e^{-\frac{2\pi i m y}{\beta} }e^{-i n f_0} \left(1 - i n \sum_{m',n'} \frac{\epsilon_{m',n'}}{(2\pi)^2} e^{-\frac{2\pi i m' y}{\beta} }e^{-i n' f_0} + \cO(\epsilon^2) \right)\,. \nonumber 
\end{align}
\end{subequations}
The action then becomes
\begin{align}
I_{\textrm{E}} = I_{\textrm{E}}^{(0)} - \frac{ik}{(2\pi)^3} \sum_{m=-\infty}^{\infty} \sum_{n} \bigg\{ & \left(\cL_0 n (m- \theta n) +  \frac{i\beta}{4\pi}(n^4 + \cM_0 n^2)\right) |\epsilon_{m,n}|^2  \\ 
& \nonumber + (m- \theta  n)(n^3+\cM_0 n)\epsilon^*_{m,n} \alpha_{m,n} \bigg\} + \ldots\,,
\end{align}
where $2\pi \theta = \beta \Omega$, $\epsilon_{m,n}^* = \epsilon_{-m,-n}$ and the sum over $n$ excludes $n=0$ for generic values of $\cM_0$, and excludes $n = -1,0,+1$ when $\cM_0 = -1$.

The one-loop partition function is then found to be
\begin{equation}
	Z_{\rm 1-loop}[\beta,\theta] = N e^{-I_E^{(0)}} \prod_{m,n} (m-\theta n)^{-1} (n^3 + \cM_0 n)^{-1},
\end{equation}
where $N$ is a normalization constant independent of $\beta$ and $ \Omega$.

The $\beta$-dependence of the partition function is captured entirely by the saddle point contribution. To perform the product over $m$ we consider
\begin{subequations}
\begin{align}
\partial_{\theta} \log Z_{\rm 1-loop} & = - \frac{\pi i c_2}{6} \cL_0 + \sum_{n \neq 0} \sum_{m=-\infty}^{\infty} \frac{n}{m-\theta n} \\
& = -  \frac{\pi i c_2}{6} \cL_0 + \sum_{n \neq 0} \left\{ - \frac{1}{\theta} + \sum_{m=1}^{\infty} \frac{2\theta n^2}{m^2 - \theta^2 n^2} \right\} \\
& = - \frac{\pi i c_2}{6} \cL_0 - 2\pi\sum_{n = 1}^{\infty} n \cot (\pi n \theta )\,. \label{cotsum}
\end{align}
\end{subequations}
The above sum diverges and so there are several ways how one can deal with this divergence. In the following we describe first an approach that was followed by Barnich et al. in \cite{Barnich:2015mui} that can be roughly described as ``integrate first, regularize later''. After that we describe an alternative approach that switches the order of these operations around i.e. ``regularize first, integrate later''. Both approaches yield sensible results, albeit with different quantum shifts of the central charges involved.

The first approach, that was followed by Barnich et al. in \cite{Barnich:2015mui} is to immediately integrate the divergent sum split it into two parts and then perform a very specific analytic continuation of $\theta$ (for one part $\theta\to\theta+i\epsilon$ and the other $\theta\to\theta-i\epsilon$) in order to obtain a regular expression. The result after exponentiation is (up to a normalization constant)
\begin{equation}\label{ZFSC}
Z_{\rm 1-loop} = e^{-I_E^{(0)}} \prod_{n=1}^{\infty} \frac{1}{|1-q^n|^2}\,,  \qquad \text{with:} \;\; q = e^{2\pi i (\theta+i\epsilon)},
\end{equation}
for the flat space cosmology solutions. For the Minkowski ground state $\cM_0 = -1$ and $\cL_0 = 0$ and we obtain
\begin{equation}\label{ZMink}
Z^{\rm Mink}_{\rm 1-loop} = e^{\frac{\beta}{8G_N} } \prod_{n=2}^{\infty} \frac{1}{|1-q^n|^2}\,,
\end{equation}
consistent with the result obtained in \cite{Barnich:2015mui}. 
The answer \eqref{ZFSC} obtained by not immediately taking care of the divergence of the sum \eqref{cotsum}, actually agrees with the massive BMS$_3$ character obtained in \cite{Oblak:2015sea}. 
\begin{equation}
	\chi_{p_0,j_0} [(f,\alpha)] = e^{2\pi i j_0 \theta} e^{- \beta (p_0 - c_2/24)} \prod_{n=1}^{\infty} \frac{1}{|1-q^n|^2}\,,
\end{equation} 
for $j_0 = - \frac{c_2}{12} \cL_0$, $p_0 = \frac{c_2}{24}(\cM_0 +1)$. 

Note that in order to avoid poles in the partition function one needs a non-vanishing imaginary part in $\theta$. From \eqref{musol} we see that the regularity conditions imply for $\theta$ that
\begin{equation}
    \theta = \frac{i}{\sqrt{\cM_0}} \,.
\end{equation}
Hence we see that for positive $\cM_0$ (corresponding to the flat space cosmologies), $\theta$ is purely imaginary. For negative  and real $\cM_0$, one would have to analytically continue $\theta$ to have a (positive) imaginary part. Recent evidence from a discretized Ponzano-Regge model of three-dimensional flat space \cite{Goeller:2019zpz} indicates that $\theta$ indeed obtains a finite imaginary shift. 

Alternatively, one can also immediately deal with the divergent sum \eqref{cotsum} and then integrate in order to obtain the logarithm of the partition function. In \cite{Cotler:2018zff} the same sum as in \eqref{cotsum} appeared and they used zeta function regularization. Following this approach to regularize the partition function, one can write the sum as
\begin{equation}
 \sum_{n=1}^{\infty} n  \cot(\pi n \theta) = \sum_{n=1}^{\infty} n ( \cot(\pi n \theta) + i) - i \sum_{n=1}^{\infty} n \,.
\end{equation}
The first sum on the left hand side converges for $\text{Im}(\theta)> 0$, so also now we should analytically continue $\theta$ to have a positive imaginary part. The second sum can be regularized by zeta function regularization. Now the answer is
\begin{equation}\label{ZFSC2}
Z_{\rm 1-loop} = e^{-I_E^{(0)}} q^{-\frac{1}{12}} \prod_{n=1}^{\infty} \frac{1}{(1-q^{n})^2},
\end{equation}
for the flat space cosmology solutions. For the Minkowski ground state $\cM_0 = -1$ and $\cL_0 = 0$ and we obtain
\begin{equation}\label{ZMink2}
Z^{\rm Mink}_{\rm 1-loop} = e^{\frac{\beta}{8G_N}} q^{- \frac{13}{12}} \prod_{n=2}^{\infty} \frac{1}{(1-q^n)^2}\,.
\end{equation}
We compare now to the results for the BMS$_3$ characters obtained using the highest-weight representations in \cite{Bagchi:2019unf}:
\begin{subequations}
\begin{align}
\chi^{\rm vacuum}_{(c_1,c_2,0,0)} (\theta, \beta) & = e^{- 2 \pi i \theta  \frac{c_1}{24}  + \beta  \frac{c_2}{24} } \prod_{n=2}^{\infty} \frac{1}{(1 - e^{2\pi i n \theta})^2 }\,, \\
\chi^{\rm massive}_{(c_1,c_2,\Delta,\xi)} (\theta, \beta)& = e^{2 \pi i \theta \left(\Delta - \frac{c_1}{24}  \right) - \beta \left(\xi - \frac{c_2}{24}\right) } \prod_{n=1}^{\infty} \frac{1}{(1 - e^{2\pi i n \theta})^2 }\,.
\end{align}
\end{subequations}
We find that
\begin{subequations}\label{eq:FSQuantumShifts}
\begin{align}
	{\rm vacuum}- \chi && c_1 & = 26\,, & c_2 & = \frac{3}{G_N}\,, \\
	{\rm massive}-\chi && c_1 & = 2 \,, & c_2 & = \frac{3}{G_N}\,, \label{weightsrel} \\
	\nonumber && \Delta & = - \frac{c_2}{12} \cL_0, & \xi & = \frac{c_2}{24}(\cM_0 + 1) \,.
\end{align}
\end{subequations}
Here $\Delta$ and $\xi$ are the  $\Lt_0$ and $\Mt_0$ weights of BMS$_3$ primary states (for a brief review of BMS$_3$ invariant quantum field theories see appendix \ref{sec:BMSblockdirect}).
We see that zeta function regularization of the one-loop factor introduces a quantum shift of the $c_1$ central charge by 26 in the Minkowski vacuum and by 2 in the case of flat space cosmologies.  The value of $c_2$ is robust under quantum corrections (it cannot shift by a number as it has a physical dimension, but it could have received corrections of the form $c_2 \to c_2 + \# \xi$). 

There is another way of looking at possible quantum shifts of $c_1$ and $c_2$. The authors in \cite{Cotler:2018zff} used arguments from heat kernel computations in AdS$_3$ performed in \cite{Giombi:2008vd} in order to argue that the zeta function regularization they used is consistent with the one-loop result of \cite{Giombi:2008vd}. This basically boils down to keeping track of the divergences caused by the infinite volume near the conformal boundary. One can try to look at the results we obtained in this section along similar lines.

The one-loop partition function for 3D flat space Einstein gravity has been first computed in \cite{Barnich:2015mui}. The one-loop contributions to the partition function are computed via
    \begin{equation}\label{eq:FSHeatKernelOneLoop}
        S^{(1)}=-\frac{1}{2}\ln\det\Delta^{(2)}+\ln\det\Delta^{(1)}-\frac{1}{2}\ln\det\Delta^{(0)},
    \end{equation}
where $\Delta^{(2)}$, $\Delta^{(1)}$, $\Delta^{(0)}$ are the kinetic operators that correspond to Laplacian operators for a massless, traceless symmetric tensor, a vector, and a scalar, respectively. Computing all these determinants one finds that they all have divergences due to the infinite volume of $\mathbb{R}^3/\mathbb{Z}$. However, these terms precisely cancel when adding them up like in \eqref{eq:FSHeatKernelOneLoop}. Using the same reasoning as in the AdS$_3$/CFT$_2$ case we can interpret this as further evidence that Newton's constant does not receive one-loop corrections in 3D flat space i.e. the value of $c_2$ is robust under quantum corrections.
 
Now this also immediately raises the question precisely how universal the shifts of $c_1$ are that we obtained here using the zeta function regularization. At this point it seems that one gets different shifts depending on how and at what point in the computation one deals with the divergences that show up i.e. the result is regulator dependent. A putative positive answer to this question has some quite far reaching consequences. If the shifts \eqref{eq:FSQuantumShifts} turn out to be universal this would also mean that quantum gravity in 3D flat spacetime is a theory with both non-zero $c_1$ and $c_2$ and as such might exhibit features such as parity breaking at the quantum level. 

The results for the partition function computed in this section are likely to be one-loop exact. The reason for this is a theorem due to Duistermaat and Heckman \cite{Duistermaat:1982vw} that was adapted to the 1D Schwarzian action relevant for the Sachdev-Ye-Kitaev (SYK) model \cite{Sachdev:1992fk,Kitaev:2015talk} in \cite{Stanford:2017thb} and also invoked in \cite{Cotler:2018zff} for the Alekseev-Shatashvili action. The argument heuristically goes as follows. For geometric actions defined from Kirillov-Konstant (KK) symplectic forms, the path integral measure includes the volume element of phase space, which is the Pfaffian of the KK symplectic form. One can formally write this Pfaffian as a path integral over a set of Grassmann odd variables. One can think of the Grassmann odd fields as ghost fields. The action in the path integral now includes a term quadratic in the ghost fields and one can define a Grassmann odd $Q$ generator that leaves this extended action invariant. This $Q$ generates a supersymmetry of the geometric action plus ghosts, and one can use this supersymmetry to localize the partition function by adding the appropriate $Q$ exact term. It would be interesting to work out the details of this proof for geometric action of centrally extended semi-direct product groups such as the one we are dealing with here, but for the moment we will be satisfied by stating that the partition functions computed here are also derivable as a limit of the one-loop exact partition functions computed in \cite{Cotler:2018zff}, so we expect our result to also be one-loop exact. 

\section{Wilson lines and entanglement entropy} \label{sec:EE}

One of the big advantages of using geometric actions and the language of coadjoint orbits is that it allows one to very efficiently compute a variety of things. In the remaining sections we focus on two particularly interesting physical quantities: entanglement entropy and BMS$_3$ blocks. The reason for this is that both computations require knowing the exact form of certain bilocal operators i.e. two point functions that can be very efficiently computed using BMS$_3$ coadjoint orbits.

To make this statement more precise let us quickly recall an efficient strategy called the \emph{replica trick} (see e.g. \cite{Calabrese:2004eu,Ryu:2006ef}) that is usually employed for computing entanglement entropy in 2D CFTs as well as BMS$_3$ invariant QFTs. Assuming one has a quantum system with multiple degrees of freedom that can be divided into two subsystems $A$ and $B$ one can write the resulting Hilbert space $\mathcal{H}$ as a direct product $\mathcal{H}=\mathcal{H}_A\otimes\mathcal{H}_B$. The reduced density matrix $\rho_A$ of the subsystem $\mathcal{H}_A$ can then be defined by tracing out the degrees of freedom of $\mathcal{H}_B$ from the density matrix $\rho$ of the total system $\mathcal{H}$, i.e. $\rho_A=\textrm{Tr}_B\rho$. In order to quantify the ``amount" of entanglement between system $A$ and $B$ one can introduce the entanglement entropy between the two subsystems that is given by the von Neumann entropy
\begin{equation}\label{eq:EEvonNeumann}
S_A=-\textrm{Tr}[\rho_A\ln\rho_A].
\end{equation}
Computing the logarithm of a density matrix can be quite complicated depending on the quantum system in question. This problem can be circumvented by using the replica trick where one considers $n$ copies of the system that are glued together along the entangling interval $A$ in a certain fashion such that the resulting manifold is an $n$-sheeted Riemann surface with a partition function $Z_n(A)$. This partition function can be used to compute $\textrm{Tr}\rho^n_A=\frac{Z_n(A)}{Z_1(A)^n}$ and in turn the associated Renyi entropies $S^{(n)}_A=-\partial_n\textrm{Tr}\rho^n_A$. The entanglement entropy $S_A$ is related to the first Renyi entropy $S^{(1)}_A$ by the limit $n\rightarrow1$, or more explicitly
\begin{equation}\label{eq:EERenyiLimit}
S_A=-\lim_{n\rightarrow1}\partial_n\textrm{Tr}\rho^n_A.
\end{equation}
A key observation for both 2D CFTs \cite{Calabrese:2009qy} as well as 2D BMS$_3$ invariant QFTs \cite{Bagchi:2014iea} is that the quantity $\textrm{Tr}\rho^n_A$ transforms as a two-point function of two primaries $\Phi_{\Delta,\xi}(x,u)$ with certain weights under conformal or BMS$_3$ transformations, respectively. Focusing on the entanglement entropy of a single interval in an infinitely long (1+1)-dimensional quantum system invariant under BMS$_3$ symmetries at zero temperature one can relate $\textrm{Tr}\rho^n_A$ to the expectation value of said two-point function on the complex plane \cite{Bagchi:2014iea}
\begin{equation}\label{eq:EEReplicaTrace}
\textrm{Tr}\rho^n_A=k_n\langle \Phi_{\Delta,\xi}(x_1,u_1)\Phi_{\Delta,\xi}(x_2,u_2)\rangle_{\mathbb{C}}^n \equiv k_n\langle\mathcal{B}_{\Delta,\xi}(x_1,u_1;x_2,u_2)\rangle^n,
\end{equation}
where $k_n$ are some constants and the BMS weights are related to the central charges $c_1$ and $c_2$ as 
\begin{equation}\label{eq:BMSWeights}
\Delta=\frac{c_1}{24}(1-\frac{1}{n^2}),\qquad\xi=\frac{c_2}{24}(1-\frac{1}{n^2}).
\end{equation}
In a nutshell this means that knowing the precise form of $\mathcal{B}_{\Delta,\xi}(x_1,u_1;x_2,u_2)$ also means knowing the precise form of the entanglement entropy of a bipartite system. 

The Renyi entropies have been calculated using Galilean conformal field theory techniques \cite{Bagchi:2014iea} by explicitly evaluating the expectation value of the two-point function \eqref{eq:EEReplicaTrace}. In this section we take a slightly different approach by computing the bilocal operators from a Wilson line along a curve $\mathcal{C}$ with end points $(x_1,u_1)\rightarrow(x_2,u_2)$ at the boundary
\begin{equation}\label{eq:WilsonEE}
\mathcal{B}_{\Delta, \xi}(x_1,u_1;x_2,u_2) = \langle {\rm out}|  \cP \exp\left[\int_{\mathcal{C}} \mathcal{A}\right]| {\rm in} \rangle\,.
\end{equation}
The Wilson line is computed in some suitably defined representation of $\mathfrak{isl}(2,\bR)$ and then reduced to the boundary theory using the methods of section \ref{sec:reduction}. 

This computation is reminiscent of holographic computations of entanglement entropy using Wilson lines in 3D Chern-Simons theories of gravity \cite{Ammon:2013hba,deBoer:2013vca,Castro:2014tta,Bagchi:2014iea,Basu:2015evh}, but it differs in essential details. The difference with the approach followed here is that the aforementioned papers use a Wilson line construction to compute (a generalized notion of) bulk geodesic length of an extremal surface anchored at the boundary interval of interest. The entanglement entropy is then obtained from this by invoking (the appropriate generalization of) the Ryu-Takayanagi formula \cite{Ryu:2006ef}.

Here we use the bulk Wilson line to define a bilocal operator whose expectation value is a two-point function of BMS$_3$ primaries in the boundary theory. Since we now have an effective boundary field theory for BMS$_3$ transformations around a given bulk saddle, we can not only compute the large $c_2$ semi-classical result, but also the leading order $1/c_2$ corrections that correspond to stress tensor exchanges between the BMS$_3$ primaries. This is achieved in a perturbative expansion of the geometric action on the BMS$_3$ coadjoint orbits. 
The same techniques will be applied in the next section to compute the BMS$_3$ identity block, both for light operators and in the heavy-light limit, but first we will compute the Wilson line \eqref{eq:WilsonEE}, generalizing to flat space the construction of \cite{Kraus:2018zrn,Besken:2018zro}.

\subsection{Bilocal operators from Wilson lines} \label{sec:bilocal}

In order to make contact with \eqref{eq:EEReplicaTrace} we first focus on computing \eqref{eq:WilsonEE} on the null orbifold with $\cM_0 = 0 = \cL_0$. The results on the plane are easily derivable from this by decompactifying the $\vp$ circle. We have seen from the results in section \ref{sec:reduction} that $\cM_0$ and $\cL_0$ can be reinstated by the transformations \eqref{redef}. The two operator insertions that make up the bilocal are placed at $(\vp_1,u_1)$ and $(\vp_2,u_2)$, respectively. The connection $\mathcal{A}$ takes the form
\begin{equation}\label{eq:EEConnection}
\mathcal{A}= \left(\Lt_+ - \frac{1}{4} \cM(\vp) \Lt_- - \frac12 \mathcal{N}(\vp,u) \Mt_-\right)\extd \vp + \left(\Mt_+- \frac{1}{4} \cM(\vp) \Mt_-\right)\extd u.
\end{equation}
The function $\cM$ and $\cN$ are given by equation \eqref{MNdef} in terms of the fields $X$ and $W$ on the null orbifold. 
To compute this Wilson line, we use the transformation property
\begin{equation}
U^{-1}(\vp_2,u_2) \cP \exp\left[ \int_{\mathcal{C}} a\right] U(\vp_1,u_1) = \cP \exp\left[ \int_{\mathcal{C}} a_U\right],
\end{equation}
where $a_U = U^{-1} a U - U^{-1} \extd U$ to bring the path-ordered exponential in a computable form. When one takes
\begin{align}
a = \Lt_+ \extd x+ \Mt_+ \extd u\,, && U = e^{\lambda^+ \Lt_+ + \mu^+ \Mt_+} e^{\lambda^0 \Lt_0 + \mu^0 \Mt_0} e^{\lambda^- \Lt_- + \mu^- \Mt_-},
\end{align}
with
\begin{align}
\lambda^+ & = \vp - X(\vp) ,  & \mu^+ & = u - W(\vp,u), \\
\lambda^0 & = - \log(X') , & \mu^0 & = - \frac{W'}{X'}, \\
\lambda^- & = -\frac{X''}{2 X'}, & \mu^- & = \frac{2X''W'-X'W''}{2(X')^2},
\end{align}
one may verify that $a_U$ is exactly given by \eqref{eq:EEConnection}. Hence the Wilson line \eqref{eq:WilsonEE} can be computed as\footnote{Here we made a slight abuse of notation in order to emphasize the beginning and endpoints of the interval. More correctly one would have to write the argument of the path ordered exponential as $\int_0^1\extd \tau \left(\frac{\partial \vp}{\partial\tau}\Lt_++\frac{\partial u}{\partial\tau}\Mt_+\right)$, where $\vp(0)=\vp_1$, $\vp(1)=\vp_2$, $u(0)=u_1$ and $u(1)=u_2$ and for some parametrization of the coordinates $x^\mu=(\vp(\tau),u(\tau))$.}
\begin{equation}\label{eq:EEWilson1}
\mathcal{B}_{\Delta, \xi}(\vp_1,u_1;\vp_2,u_2)  = \langle {\rm out}|U^{-1}(\vp_2,u_2) \cP \exp\left[ \int_{\vp_1}^{\vp_2} \Lt_+ \extd \vp + \int_{u_1}^{u_2} \Mt_+ \extd u\right] U(\vp_1,u_1)|{\rm in} \rangle.
\end{equation}

Next, one has to use a suitable representation for $\ket{{\rm in}}$ and $\ket{{\rm out}}$. There are at least two possible choices: ``ordinary" highest-weight representations and induced representations of $\mathfrak{isl}(2,\bR)$. Highest-weight representations are very straightforward to work with and have been used heavily in previous works, however, they exhibit negative norm states in general (i.e. for $c_2\neq0$). On the other hand, induced representations by construction do not suffer from negative norm states, but are not as straightforward to work with in comparison. Since previous successful computations of entanglement entropy in BMS$_3$ invariant QFTs \cite{Bagchi:2014iea,Basu:2015evh} made use of highest-weight representations we will also employ them in the following computations.

In line with the AdS$_3$ computation of the Wilson line in \cite{Besken:2018zro,Kraus:2018zrn} we take the following choice of in- and out-state in a highest-weight representations of $\mathfrak{isl}(2,\bR)$
\begin{align}
\Lt_{-1} \ket{{\rm in}} & = 0 = \Mt_{-1}\ket{ {\rm in}}\,,& 
\Lt_0 \ket{{\rm in}}  &= - \Delta \ket{ {\rm in}}\,, & \Mt_0 \ket{ {\rm in}} & = - \xi \ket{{\rm in}}, \\
\Lt_{+1} \ket{{\rm out}} & = 0 = \Mt_{+1}\ket{{\rm out}}\,,&
\Lt_0 \ket{{\rm out}} & =  \Delta \ket{ {\rm out}}\,, & \Mt_0 \ket{ {\rm out}} & =  \xi \ket{ {\rm out}}. 
\end{align}
Thus, our in- and out-states are highest-weight states with weights $\ket{{\rm in}}\equiv\ket{-\Delta,-\xi}$ and $\ket{{\rm out}}\equiv\ket{\Delta,\xi}$.
Hermitian conjugation is defined by taking $\Lt_{n}^\dagger = \Lt_{-n}$ and likewise for $\Mt_n$. The Wilson line \eqref{eq:EEWilson1} then becomes
\begin{equation}\label{eq:EEWilson2}
\mathcal{B}_{\Delta, \xi}(\vp_1,u_1;\vp_2,u_2) =  \; e^{ \xi\left(\frac{W_1'}{X'_1}+\frac{W_2'}{X'_2} \right)} \left(X'_1 X'_2\right)^\Delta \langle {\rm out}| \cP \exp\left[ X_{21} \Lt_+ + W_{21} \Mt_+ \right]| {\rm in} \rangle\,,
\end{equation}
where here $X_i = X(\vp_i)$ and $X_{12} = X_1 - X_2$ and likewise for $W$.
By the Baker-Campbell-Hausdorff formula it is possible to prove the identity
\begin{equation}
\exp{[\alpha \Lt_++\beta \Mt_+]}=\exp{\left[-\frac{\beta}{\alpha}\Mt_0\right]}\exp{[\alpha \Lt_+]}\exp{\left[\frac{\beta}{\alpha}\Mt_0\right]}\,.
\end{equation}
Using this, equation \eqref{eq:EEWilson2} becomes
\begin{equation}\label{eq:EEWilson3}
\mathcal{B}_{\Delta, \xi}(\vp_1,u_1;\vp_2,u_2) =  \; e^{ \xi\left(-2\frac{W_{12}}{X_{12}}+\frac{W_1'}{X'_1}+\frac{W_2'}{X'_2} \right)} \left(X'_1 X'_2\right)^\Delta \langle {\rm out}| \cP \exp\left[ X_{21} L_+ \right]| {\rm in} \rangle\,.
\end{equation}
In order to compute the last remaining term one can first analytically continue $\Delta\rightarrow-j$ so that one effectively ends up with a finite-dimensional representation of the $\mathfrak{sl}(2,\bR)$ subalgebra spanned by $\Lt_n$, where\footnote{The most straightforward way to see this is to expand the exponential and to realize that all inner products between the occurring states generated by repeated application of $L_+$ are zero except the state that is generated by $(L_+)^{2j}|\textrm{in}\rangle$.} $\bra{-j,\xi}\mathcal{P}\exp{[\alpha L_+]}\ket{j,-\xi}=\alpha^{2 j}$. Thus we obtain as the final result for our bilocal field
\begin{equation}\label{eq:EEWilson4}
\mathcal{B}_{\Delta, \xi}(\vp_1,u_1;\vp_2,u_2) =  \; e^{ \xi\left(-2\frac{W_{12}}{X_{12}}+\frac{W_1'}{X'_1}+\frac{W_2'}{X'_2} \right)}\left(\frac{X'_1 X'_2}{(X_{21})^2}\right)^\Delta.
\end{equation}
This could have been obtained in an easier way by considering that under a finite BMS$_3$ transformation $(\vp,u) \to (X(\vp),W(\vp,u))$ the BMS$_3$ primaries transform as \cite{Hijano:2018nhq}
\begin{equation}
\cO_{\Delta,\xi}(\vp,u) \to (X')^\Delta e^{\xi \frac{W'}{X'} } \cO_{\Delta,\xi}(X,W).
\end{equation}
The bilocal operator  \eqref{eq:EEWilson4} is exactly the finite transformation of a two-point function of BMS$_3$ primaries on the plane, given in equation \eqref{twopt}.

From this result and the map \eqref{redef}, we can find the expression for the bilocal operator on the vacuum orbit of BMS$_3$ or on the massive orbits with non-zero $\cM_0$ and $\cL_0$. For instance, to map this result to the bilocal to the vacuum orbit, we take $\cM_0 = -1$ and $\cL_0 = 0$ in \eqref{redef}, that is, we take
\begin{equation}\label{maptoMink}
X = e^{- i f(u,\vp)} \,, \qquad 
W = - i e^{- i f(u,\vp)} \alpha(f,u)\,,
\end{equation}
to obtain
\begin{align}\label{bilocalcyl}
\cB^{\rm vac}_{\Delta, \xi}(\vp_1,u_1; \vp_2,u_2) = \left( \frac{f_1' f_2'}{4 \sin^2( \frac{f_{12}}{2}) } \right)^\Delta \exp \left[\xi \left(-\frac{\alpha_{12}}{\tan\left(\frac{f_{12}}{2}\right)} + \frac{\alpha_1'}{f_1'} + \frac{\alpha_2'}{f_2'} \right) \right]\,. 
\end{align}
On the massive BMS$_3$ orbits, the map \eqref{redef} for generic (non-zero) $\cM_0$ and $\cL_0$ gives
\begin{align}\label{bilocalFSC}
\cB^{\rm m}_{\Delta,\xi} (\vp_1, u_1; \vp_2, u_2)  = & \left( \frac{\gamma^2 f_1' f_2' }{2^2 \sin^2 \left(\frac{\gamma}{2}f_{12} \right)} \right)^{\Delta} \exp\left[ \xi\left(  \frac{\alpha_1'}{f_1'} + \frac{\alpha_2'}{f_2'}  - \frac{2 \cL_0}{\gamma^2} \right)\right] \nonumber \\
& \times  \exp \left[- \xi \gamma \left(\alpha_{12} - \frac{\cL_0}{\gamma^2} f_{12} \right) \cot \left( \frac{\gamma}{2}f_{12}\right) \right]\,,
\end{align}
where we have defined
\begin{equation}\label{gammaM0}
\gamma = \sqrt{- \cM_0} = \sqrt{1 - \frac{24 \xi_H}{c_2} }\,.
\end{equation}
The last equality follows from \eqref{weightsrel} and $\xi_H$ is the $M_0$ weight of the flat space cosmology. Note that the parameter $\gamma$ is related to the fugacity $\theta = \frac{\Omega \beta}{2\pi}$ of the last section as
\begin{equation}
\gamma = - \frac{1}{\theta}\,.
\end{equation}

For the saddle point solutions \eqref{saddle} the map to the vacuum orbit \eqref{maptoMink} exactly correspond to the coordinate transformation from the plane (with coordinates $(X,W)  = (x,t)$) to the null cylinder
\begin{equation}
\label{planetocyl}
x = e^{-i\varphi}\,, \qquad  t = - i u e^{- i \varphi}\,.
\end{equation}
A similar map for the saddles of the massive orbit
\begin{subequations}\label{planetoFSC}
	\begin{align}
	x & = e^{-i \gamma \vp} \,,& 	t & = - i \gamma e^{-i \gamma \vp} \left( u - \frac{\cL_0}{\gamma^2} \vp \right),
	\end{align}
\end{subequations}
defines the BMS$_3$ analogue of the uniformizing transformations in AdS$_3$/CFT$_2$ of \cite{Fitzpatrick:2015zha}, where these transformations were used to compute expectation values for light operators in a heavy (BTZ) background. It is now apparent that in the CFT case this is equivalent to the map from the zero representative orbit of the Virasoro group to the generic positive representative orbits corresponding to the BTZ black holes. Here we found the flat space analogues to these transformations.

\subsection{Computing entanglement entropy using the bilocal}

Now that we have found the bilocal operators and understood how to map this to the different coadjoint orbits of BMS$_3$, we can proceed to compute the entanglement entropy by evaluating \eqref{eq:EEReplicaTrace} and taking the limit \eqref{eq:EERenyiLimit}. Here we first compute the leading order result and we discuss quantum corrections in the next subsection. The leading order contribution to the EE comes from simply plugging in the saddle point values \eqref{saddle} in the bilocal operator on the relevant orbit. In this way, we can recover known results of \cite{Bagchi:2014iea,Hosseini:2015uba,Basu:2015evh} for the entanglement entropy on the plane (by taking \eqref{eq:EEWilson3} with $(X,W) = (x,t)$):
\begin{equation}\label{eq:EEplane}
S_{\textrm{EE}}^{\rm plane}=\frac{c_1}{6}\,\log \frac{x_{12}}{\epsilon_x}+\frac{c_2}{6}\,\bigg(\frac{t_{12}}{x_{12}}-\frac{\epsilon_u}{\epsilon_x}\bigg)\,,
\end{equation}
and for the cylinder (from the bilocal on the vacuum orbit \eqref{bilocalcyl})
\begin{equation}\label{eq:EEMink}
S_{\textrm{EE}}^{\rm cyl}=\frac{c_1}{6}\,\log \left( \frac{2}{\epsilon_\vp} \sin \frac{\vp_{12}}{2} \right) + \frac{c_2}{6}\,\bigg(\frac{u_{12}}{2\tan \frac{\vp_{12}}{2} }-\frac{\epsilon_u}{\epsilon_\vp}
	\bigg)\,,
\end{equation}
where we have introduced the UV cut-offs $\epsilon_x, \epsilon_\vp$ and $\epsilon_u$. The result for flat space cosmologies follows from \eqref{bilocalFSC} and reads
\begin{equation}\label{eq:EEFSC}
S_{\textrm{EE}}^{\rm FSC}=\frac{c_1}{6}\,\log \left( \frac{2}{\gamma \epsilon_\vp} \sin  \frac{\gamma \vp_{12}}{2} \right) + \frac{c_2}{6}\,\bigg(\frac{ \cL_0}{\gamma^2} + \frac{\gamma(u_{12}- \frac{\cL_0}{\gamma^2} \vp_{12})}{2  \tan \frac{\gamma \vp_{12}}{2} }-\frac{\epsilon_u}{\epsilon_\vp}
\bigg).
\end{equation}
These results agree exactly with those obtained in \cite{Bagchi:2014iea,Hosseini:2015uba,Basu:2015evh}.

In \cite{Grumiller:2019xna} holographic methods were used to compute the entanglement entropy of the flat space analogue of Ba\~nados geometries \cite{Banados:1998gg} whose holographic duals are generic excited states in a 2D BMS$_3$ invariant QFT. The term excited state in this context means that the expectation values of the energy-momentum operators $\mathcal{T}_M$ and $\mathcal{T}_L$ in a BMS$_3$ invariant QFT depend on two arbitrary functions $\mathcal{M}$ and $\mathcal{N}$ and can be written as (see e.g. \cite{Fareghbal:2013ifa})
\begin{equation}\label{eq:EMTensorAndCharges}
2\pi\langle\mathcal{T}_{M}\rangle=\frac{c_2}{24}\,\mathcal{M}(x),\qquad 2\pi\langle\mathcal{T}_{L}\rangle=\frac{c_1}{24}\, \mathcal{M}(x)+\frac{c_2}{12}\, \mathcal{N}(x,t)\,,
\end{equation}
where $2\partial_t\mathcal{N}=\partial_x \mathcal{M}$. Equivalently one can also look at these states as being generated by finite BMS$_3$ transformations $x \rightarrow f(x)$ and $t \rightarrow g(f(x),u)$ from a given reference state such as e.g. the null orbifold.
The entanglement entropy for such excited states found in \cite{Grumiller:2019xna} reads
\begin{equation}\label{eq:GenericEE}
S_{\textrm{EE}}=\frac{c_1}{6}\,\log\frac{(f_2-f_1)}{\epsilon_x\sqrt{f'_1f'_2}}+\frac{c_2}{6}\,\bigg(\frac{g_2-g_1}{f_2-f_1}-\frac{\epsilon_u}{\epsilon_x}-\frac{g'_2}{2f'_2}-\frac{g'_1}{2f'_1}\bigg).
\end{equation}
Having computed the bilocal \eqref{eq:EEWilson4} we now have a very simple way of reproducing this from a BMS$_3$ QFT point of view. The first thing to do is to regulate the interval sizes by introducing the UV cutoffs $\epsilon_x$ and $\epsilon_u$. This leads to
\begin{equation}\label{eq:EEWilson5}
\mathcal{B}_{\Delta, \xi}(x_1,u_1;x_2,u_2) =  \; e^{ \xi\left(-2\left(\frac{W_{21}}{X_{21}} - \frac{\epsilon_u}{\epsilon_x} \right) + \frac{W_1'}{X'_1} + \frac{W_2'}{X'_2} \right)} \left(\frac{\epsilon_x^2 X'_1 X'_2}{(X_{21})^2}\right)^\Delta.
\end{equation}
Equation \eqref{eq:EEReplicaTrace} instructs us to compute the expectation value of this bilocal. To leading order in $c_2$, we can do so by simply replacing the expectation values of fields $X$ and $W$ by the finite BMS$_3$ transformations $f(x)$ and $g(f(x),u)$ respectively. 
The Renyi limit $n\rightarrow1$ in \eqref{eq:EERenyiLimit}, together with the weights \eqref{eq:BMSWeights} gives precisely the expression \eqref{eq:GenericEE}. This provides a very nice and simple cross check for the holographic (large $c_2$) results obtained in \cite{Grumiller:2019xna}.

\subsection{Quantum corrections} \label{sec:quantum}

We have seen that the tree level expectation value for the bilocal operators 
exactly reproduces the known results for the entanglement entropy of BMS$_3$ invariant fields theories semi-classically. But our current setup allows us to do better and we can compute the subleading contributions to the entanglement entropy in a perturbative expansion in $1/c_2$. Since pure gravity in 3D flat space can be reduced to the geometric action on the coadjoint orbits of BMS$_3$, we can use this action to compute the one-loop contributions to the expectation values of the bilocals by a perturbative expansion around the classical saddle points \eqref{saddle}. These contributions are coming from stress tensor exchanges between the two BMS$_3$ primary fields, or in other words, from descendents of the BMS$_3$ primaries. 

Note that since $c_2$ is inversely proportional to Newton's constant it is a parameter that has a physical dimension of a mass. Thus, whenever we are referring to large $c_2$ we implicitly mean large compared to the typical masses of the problem at hand which are set by $\xi$. I.e. for light operators, (small masses), we mean $\xi/c_2 \ll 1$. In section \ref{sec:BMSblocks} we will also consider heavy operators, that have masses such that $\xi_H/c_2 \sim 1$. In that case there will still be a set of light operators which have small masses compared to $c_2$, validating a large $c_2$ expansion. 

To proceed we need two ingredients. We first need the quadratic action for perturbations around the saddle point \eqref{saddle} to compute the propagators on the relevant BMS$_3$ orbits. Then we expand the bilocal \eqref{bilocalcyl} to quadratic order and use the propagators to obtain the $\cO(1/c_2)$ corrections to \eqref{eq:EEMink}. In this section we restrict ourselves to the subleading corrections to the entanglement entropy on the cylinder, corresponding to the vacuum BMS$_3$ orbits. The results for the massive orbits of BMS$_3$ can be derived from the $\cO(1/c_2)$ corrections to the heavy-light BMS$_3$ identity block, which we compute in the next section.

\subsubsection{BMS$_3$ propagators}\label{sec:BMS3Propagators}

In section \ref{sec:reduction} we reduced the gravitational action for flat space to the boundary and found the geometric action for BMS$_3$ coadjoint orbits. Here we expand this action around the classical saddle points \eqref{saddle} and compute the quadratic action for fluctuations around these saddles.
To be more general, we include the Hamiltonian \eqref{genHam} for generic constant $\mu_{L/M}$ and set $\mu_M = 1$ and $\mu_L = 0$ afterwards. After analytically continuing $u \to - i y$ the action reads
\begin{align} \label{Sfinal2}
    I[f,\alpha,\cL_0,\cM_0] =  - \frac{c_2}{24\pi} \int \extd y \extd\vp \; \Big[ & \left( \cL_0  + \cM_0 \alpha'(f) - \alpha'''(f) \right) (i \partial_y f - \mu_L f') f'  \\
    & \nonumber  
    -  \frac12 \mu_M 
    \left( \cM_0 f'^2  - 2 \{f,\vp\} \right) \Big].
\end{align}
The propagators for fluctuations $\epsilon(\vp,y)$ and $\tilde \alpha (\vp,y)$ around the classical saddle points are obtained by expanding this action to quadratic order, i.e. we take
\begin{equation}\label{fluc}
    f(\vp, y) = \vp + \epsilon(\vp,y)\,, \qquad \alpha(\vp, y) = y  + \tilde{\alpha}(\vp, y)\,.
\end{equation}
Plugging \eqref{fluc} into \eqref{Sfinal2} we obtain $I_{\rm CS} = I^{(0)} + I^{(2)} + \ldots$, where the dots denote higher order terms and
\begin{align}
I^{(2)} = - \frac{c_2}{24\pi} \int \extd y \extd \vp \Big( & \left(\cL_0 \epsilon'  + \cM_0 \tilde \alpha'   + \tilde \alpha'''\right)\tilde \partial_-  \epsilon  - \frac{\mu_M}{2}(\epsilon''{}^2 + \cM_0 \epsilon'{}^2) \Big),
\end{align}
with $\tilde \partial_- = i \partial_y - \mu_L \partial_\vp$
In terms of the Fourier modes $\hat{\epsilon}_n(\omega)$ and $\hat{\alpha}_n(\omega)$ defined as
\begin{subequations}
\begin{align}
	\epsilon(\vp,y) = \frac{1}{(2\pi)^2} \int_{-\infty}^{\infty} \extd\omega \sum_{n = -\infty }^{\infty} e^{i n \vp + i \omega y} \hat{\epsilon}_n(\omega) \,, \\
	\tilde \alpha(\vp,y) = \frac{1}{(2\pi)^2} \int_{-\infty}^{\infty} \extd\omega \sum_{n = -\infty }^{\infty} e^{i n \vp + i \omega y} \hat{\alpha}_n(\omega) \,,
\end{align}
\end{subequations}
the quadratic action reads
\begin{equation}
	I^{(2)} = \frac{1}{2} \int_{-\infty}^{\infty} \extd \omega \sum_{n} A_{n}(\omega)_{ij} \hat{\chi}_n^i(\omega) \hat{\chi}^j_{-n}(-\omega) \,.
\end{equation}
Here $\hat{\chi}^i = \{\hat{\epsilon}, \hat{\alpha}\}$ and the sum over $n$ excludes $n=0$ for the massive orbits and \mbox{$n = 0, \pm1$} for the global Minkowski orbit. The matrix elements $A_n(\omega)_{ij}$ are given by
\begin{align}\label{Amat}
A_n(\omega) = & \, \frac{1}{(2\pi)^3} \frac{c_2}{12}\left( \begin{array}{cc}
(A_n(\omega))_{11}  & in(n^2+ \cM_0)(\omega - i \mu_L n)\\
 in(n^2+ \cM_0)(\omega - i \mu_L n) & 0
\end{array} \right)\,, \\
(A_n(\omega))_{11} = & \, \mu_M n^2 (n^2 + \cM_0) + 2 n i (\omega - i \mu_L n) \cL_0. 
\end{align}
The propagators in Fourier space are found by inverting the matrix $A_n(\omega)$. This gives
\begin{subequations}
\begin{align}
    \vev{\hat \epsilon_{n_1}(\omega_1) \hat{\epsilon}_{n_2}(\omega_2) } & = 0 \,,\\
    \vev{\hat \alpha_{n_1}(\omega_1) \hat{\epsilon}_{n_2}(\omega_2) } & =  \frac{24 \pi}{c_2} \frac{(2\pi)^2 \delta_{n_1 + n_2} \delta(\omega_1 + \omega_2)}{i n_1(n_1^2 + \cM_0) (\omega_1 - i \mu_L n_1)}  \,,\\
    \vev{\hat \alpha_{n_1}(\omega_1) \hat{\alpha}_{n_2}(\omega_2) } & =  \frac{24 \pi}{c_2} \Big[ \frac{\mu_M}{(n_1^2 + \cM_0)(\omega_1 - i \mu_L n_1)^2}  \\
    & \quad \quad 
     + \frac{2 i \cL_0}{n_1(n_1^2 + \cM_0)^2 (\omega_1 - i \mu_L n_1)}\Big]
    (2\pi)^2 \delta_{n_1 + n_2} \delta(\omega_1 + \omega_2) \,. \nonumber
\end{align}
\end{subequations}
The position space propagators are then obtained by Fourier transforming back as
\begin{equation}
    \vev{ \chi^i(\vp,y) \chi^j(0,0)} = \frac{1}{(2\pi)^4}\int_{-\infty}^{\infty} \extd\omega_1 \extd \omega_2  \sum_{n_1, n_2} e^{i n_1 \vp} e^{i \omega_1 y} \vev{ \chi^i_{n_1}(\omega_1) \chi^i_{n_2}(\omega_2) }\,,
\end{equation}
where the sum excludes $n_1 = 0 = n_2$ for the massive orbits and it excludes $n_1 = -1, 0 , +1 =  n_2$ whenever $\cM_0 = -1$ (in these cases the matrix \eqref{Amat} is not invertible). These modes are excluded due to the ISL$(2,\bR)$ gauge invariance on the vacuum orbit. We perform the Fourier transform assuming $\mu_L$ is real and non-negative.
For the vacuum orbit (with $\cM_0=-1$, $\cL_0 = 0$) the result is after continuing back to $y \to i u$:
\begin{subequations}\label{vacpropagators}
\begin{align}
\vev{\epsilon(\vp, u) \epsilon(0,0)} & = 0, \\
\vev{\tilde \alpha(\vp, u) \epsilon(0,0)} & = \, \frac{3}{c_2} \bigg(  3 \zeta -2 - 2 \frac{(1-\zeta)^2}{\zeta}\log\left(1-\zeta \right)  \bigg),  \\
\vev{\tilde \alpha (\vp, u) \tilde{\alpha}(0,0)} &  =  \frac{3 i u \mu_M}{c_2} \left(  2 + \zeta - 2 \frac{(\zeta^2 -1)}{\zeta} \log (1-\zeta)  \right)\nonumber\\
& =  i\mu_M  u \; \zeta \partial_\zeta  \vev{\tilde \alpha(\vp, u) \epsilon(0,0)},
\end{align}
\end{subequations}
with $\zeta = e^{i {\rm sign}(u) (\vp- \mu_L u)}$. For the massive orbits, the propagators are
\begin{subequations}\label{eq:Fluctuations2PointFunctions}
\begin{align}
\vev{\epsilon(\vp, u) \epsilon(0,0)} = & 0, \\
\vev{\tilde \alpha(\vp, u) \epsilon(0,0)}  = &\,  \frac{6}{c_2\gamma^2} \bigg( 2 \log (1-\zeta) + \Phi(\zeta,1,\gamma) + \Phi(\zeta, 1, - \gamma)  \bigg),  \\
\vev{\tilde \alpha(\vp, u) \tilde{\alpha}(0,0)}  = & \frac{6 \cL_0}{c_2 \gamma^3}\bigg(\Phi(\zeta,2,\gamma)-\Phi(\zeta,2,-\gamma)\bigg)\\
&+\left( \frac{2\cL_0}{\gamma^2}+i\mu_M u\, \zeta \partial_\zeta\right)\vev{\tilde \alpha(\vp, u) \epsilon(0,0)},\nonumber 
\end{align}
\end{subequations}
where $\gamma$ was given in \eqref{gammaM0}
and $\Phi(\zeta,s,a)$ is the Lerch transcendent
\begin{equation}
	\Phi(\zeta, s, a) = \sum_{n=0}^{\infty} \frac{\zeta^n}{(n+a)^s}\,.
\end{equation}
In what follows, we sometimes need to consider the coincidence point limit of these propagators. We regularize this by introducing a cutoff in imaginary $\vp$ and $u$, such that for the vacuum orbit:
\begin{align}
    \vev{\tilde \alpha (\vp + i \delta_\vp, u + i \delta_u) \epsilon(\vp ,u) } & = \frac{3}{c_2}+\ldots \,, \\
    \vev{\tilde \alpha (\vp + i \delta_\vp, u + i \delta_u) \tilde \alpha(\vp ,u) } & =  \ldots\,, \\
    \vev{\tilde \alpha '(\vp + i \delta_\vp, u + i \delta_u) \epsilon'(\vp ,u) } & = - \frac{9 + 12 \log (\delta_\vp-\mu_L\delta_u)}{c_2} +\ldots\,, \\
    \vev{\tilde \alpha '(\vp + i \delta_\vp, u + i \delta_u) \tilde\alpha'(\vp ,u) } & = -\frac{12 \delta_u}{c_2 (\delta_\vp-\mu_L\delta_u)}+\ldots \,,
\end{align}
where the dots denote subleading terms in $\delta_\varphi$ and $\delta_u$. The analogous expressions for the massive orbits are listed when we need them in \eqref{coincidencemassive} below.

In the following we set $\mu_M = 1$ and take the limit $\mu_L \to 0_\downarrow$ such that $\zeta = e^{{\rm sign}(u) i\vp}$. 

\subsubsection{Quantum corrections to the entanglement entropy}
We now proceed to compute the $\cO(1/c_2)$ corrections to the entanglement entropy in flat space Einstein gravity (with $\mu_L = 0, \mu_M = 1$ and $c_1 \sim \cO(1)$). To this end, we expand the bilocal operators $\cB_{\Delta,\xi}(\vp_1,u_1;\vp_2, u_2)$ around the saddle point as in \eqref{fluc}. We focus here on the result dual to the Minkowski vacuum, by expanding the bilocal on the vacuum orbit \eqref{bilocalcyl}. To obtain the quantum corrections to the entanglement entropy for field theories dual to the flat space cosmologies, we use the corrections to the heavy-light identity BMS$_3$ block which we compute in the next section.

The expansion of the bilocal \eqref{bilocalcyl} gives
\begin{align}\label{bilocalexp}
    \cB_{\Delta,\xi}(\vp_1,u_1; \vp_2,u_2) = & \vev{\cO_1 \cO_2}_{\rm cyl} \bigg(1 + \cF_{12}^{(1)} \cdot \chi  + \frac12 (\cF_{12}^{(1)} \cdot \chi)^2 + \ldots \\
    & + \Delta (\cJ_{12}^{(2)} \cdot \epsilon^2) + \xi \left( (\cK_{12}^{(2)} \cdot \epsilon^2)  + (\cF_{12}^{(2)} \cdot  \tilde \alpha \epsilon) \right) +  \ldots \bigg)\,, \nonumber
\end{align}
where $\vev{\cO_1\cO_2}_{\rm cyl}$ is the two-point function of two BMS$_3$ primary fields on the cylinder
\begin{equation}\label{twoptcyl}
    \vev{\cO_1 \cO_2}_{\rm cyl} =  \frac{1}{\left(2 \sin \left(\frac{\vp_{12}}{2}  \right) \right)^{2\Delta}}  e^{-  u_{12} \xi \cot \frac{\vp_{12}}{2} }\,.
\end{equation}
The first order contributions are
\begin{subequations}\label{firstorder}
\begin{align}
\cF_{12}^{(1)} \cdot \chi & = \Delta (\cJ_{12}^{(1)} \cdot \epsilon) + \xi (\cK_{12}^{(1)}\cdot \epsilon) + \xi (\cJ_{12}^{(1)} \cdot \tilde\alpha), \\
\cJ_{12}^{(1)} \cdot \epsilon & = \epsilon_1' + \epsilon_2' - \cot \left( \tfrac{\vp_{12}}{2}\right) \epsilon_{12}, \\
\cK_{12}^{(1)} \cdot \epsilon & = \frac{u_{12} \epsilon_{12}}{2 \sin^2 \left( \frac{\vp_{12}}{2} \right)}. 
\end{align}
\end{subequations}
At second order, the contributions are characterized by two separate types. One is the square of the first order contributions, quadratic in the weights $\Delta, \xi$. The second type of terms are linear
in $\Delta, \xi$ and read:
\begin{subequations}
\begin{align}
    \cJ_{12}^{(2)} \cdot \epsilon^2 & = \frac14 \left( \frac{\epsilon_{12}^2}{\sin^2\left( \tfrac{\vp_{12}}{2}\right)} - 2 (\epsilon_1'{}^2 + \epsilon_2'{}^2 )\right), \\
    \cK_{12}^{(2)} \cdot \epsilon^2 & = - \frac14 \frac{\cot \left( \tfrac{\vp_{12}}{2}\right) }{\sin^2  \left(\tfrac{\vp_{12}}{2}\right)} u_{12} \epsilon_{12}^2 = \frac{1}{2}\left(\cK_{12}^{(1)} \cdot \epsilon\right) \left(\cJ_{12}^{(1)} \cdot \epsilon - \epsilon_1' - \epsilon_2'\right), \\
    \cF_{12}^{(2)} \cdot \tilde \alpha \epsilon  & = \frac{\tilde \alpha_{12} \epsilon_{12}}{2 \sin^2  \left( \tfrac{\vp_{12}}{2}\right)} - \epsilon_1' \tilde \alpha_1' - \epsilon_2' \tilde \alpha_2'.
\end{align}
\end{subequations}
Next we compute the expectation value of \eqref{bilocalexp}. The first order terms $\vev{\cF^{(1)}_{12} \cdot \chi}$ vanish. Since $\vev{\epsilon \epsilon} = 0$, there are only three terms contributing at order $\cO(1/c_2)$. They are
\begin{equation}
    \vev{\cB_{\Delta,\xi} (\vp_1,u_1; \vp_2,u_2)} = \vev{\cO_1 \cO_2}_{\rm cyl} \left( 1 + \xi V_{\xi} +  \xi \Delta V_{\xi \Delta} + \xi^2 V_{\xi\xi} + \ldots \right),
\end{equation}
with
\begin{subequations}
\begin{align}
    V_{\xi} & = \vev{ \cF_{12}^{(2)} \cdot \tilde \alpha \epsilon  } =
    \frac{12}{c_2} \left( 3 -  2 \log \left( \frac{2}{\delta_\vp} \sin \left( \frac{\vp_{12}}{2} \right) \right) \right), \\
    V_{\xi \Delta} & 
    = \vev{ (\cJ^{(1)}_{12} \cdot \tilde \alpha)( \cJ^{(1)}_{12} \cdot \epsilon) } = \frac{12}{c_2} \left(  2 \log \left( \frac{2}{\delta_\vp} \sin \left( \frac{\vp_{12}}{2} \right) \right) - 2 \right),
\end{align}
and
\begin{align}
    V_{\xi\xi} & = \vev{ (\cJ^{(1)}_{12} \cdot \tilde \alpha)( \cK^{(1)}_{12} \cdot \epsilon) +\frac12 \vev{ (\cJ^{(1)}_{12} \cdot \tilde \alpha)( \cJ^{(1)}_{12} \cdot \tilde \alpha) }  } \nonumber \\
    & = \frac{12}{c_2} \left( \frac{1}{2} u_{12} \cot \left(\frac{\vp_{12}}{2} \right)  -  \frac{\delta_u}{\delta_\vp} \right).
\end{align}
\end{subequations}
We see that the UV cutoffs $\delta_\vp$ and $\delta_u$ correspond to the cutoffs $\epsilon_\vp$ and $\epsilon_u$ introduced in \eqref{eq:EEMink}.
Finally, we are ready to compute \eqref{eq:EEReplicaTrace} and use this to take the limit $n\to 1$ in \eqref{eq:EEvonNeumann}. Due to the scaling of the weights of the operators \eqref{eq:BMSWeights}, only the term $V_{\xi}$ contributes and the result is\footnote{It should be noted that we used a slightly different normalization for the constant $k_1$ in \eqref{eq:EEReplicaTrace} than in the previous section.}
\begin{equation}\label{eq:CylEEQuantumShift}
    S_{\rm EE}^{\rm cyl} = \frac{c_1+12}{6} \log \left( \frac{2}{\epsilon_\vp} \sin \frac{\vp_{12}}{2} \right) + \frac{c_2}{6} \left( \frac{u_{12}}{2 \tan \frac{\vp_{12}}{2} }  - \frac{\epsilon_u}{\epsilon_{\vp} } \right).
\end{equation}
We see that once again $c_1$ is shifted, but now by 12, not by 26 in contrast to the shift of the central charge for the one-loop partition function. We would like to stress at this point that this shift of $c_1$ is exact in the sense that the entanglement entropy does not receive any further perturbative corrections in $\cO(1/c_2)$. This can be easily seen by rewriting the expansion of the bilocal as
    \begin{equation}
        \cB_{\Delta,\xi}(\vp_1,u_1; \vp_2,u_2) = \vev{\cO_1 \cO_2}_{\rm cyl}\sum_{i=0}^\infty\sum_{j=0}^\infty a_{ij}\Delta^i\xi^j,
    \end{equation}
where $a_{00}=1$ and all the other coefficients $a_{ij}$ are linear combinations of multi-point functions of $\epsilon$ and $\tilde{\alpha}$ and then computing
\begin{equation}
S_{\textrm{EE}}=-\lim_{n\rightarrow1}\partial_n(\cB_{\Delta,\xi}(\vp_1,u_1; \vp_2,u_2))^n.
\end{equation}
It is straightforward to see that using the weights \eqref{eq:BMSWeights} the resulting entanglement entropy looks like
    \begin{equation}
        S_{\textrm{EE}}= S_{\textrm{EE}}^{\textrm{cyl}(0)}-\frac{1}{12}(c_1 a_{10}+c_2 a_{01}),
    \end{equation}
where $S_{\textrm{EE}}^{\textrm{cyl}(0)}$ is given by \eqref{eq:EEMink}. The coefficient $a_{10}$ is an infinite sum of propagators of arbitrary powers of $\epsilon$ that by virtue of \eqref{vacpropagators} all vanish thus yielding $a_{10}=0$. The coefficient $a_{01}$ on the other hand also contains -- in addition to arbitrary powers of $\epsilon$ -- terms of the form $\vev{\tilde{\alpha}\epsilon^m}$. For $m\neq1$ all these terms vanish as well, leaving only the terms that are proportional to $\vev{\tilde{\alpha}\epsilon}$ (and derivatives thereof) as possible contributions to corrections of the entanglement entropy. These terms are precisely what we computed in this section. This shows explicitly that the contribution from stress-tensor descendants to the entanglement entropy of a BMS$_3$ invariant quantum field theory is one-loop exact. This gives another indication that the same contributions to the partition function are also one-loop exact, as the entanglement entropy can equivalently be computed from the replica partition function.

We want to close this section with a brief discussion regarding the quantum corrections to entanglement entropy that we worked out here. The first thing we would like to point out is that the functional form of entanglement entropy is completely fixed by symmetry. The only thing that can change due to quantum corrections is the interpretation of the central terms $c_1$ and $c_2$ and their relation to Newton's constant (or other parameters). It might look curious at first sight that we find a shift of $c_1$ instead of $c_2=\frac{3}{G_N}$. However, this is in agreement with the partition function computation that we performed previously and further reinforces the interpretation that there seems to be no renormalization of Newton's constant due to quantum effect in 3D asymptotically flat Einstein gravity. This in turn raises the question on how universal the shift of 12 actually is that we found. While we have no conclusive answer to this we will argue in the following that it is very likely that the shift we computed here for entanglement entropy is \emph{not} universal.

Recent work \cite{Belin:2019mlt} in the context of AdS/CFT suggests that the interpretation of quantum corrections computed holographically using the Faulkner, Lewkowycz and Maldacena (FLM) prescription \cite{Faulkner:2013ana} depends on the bulk regulator used. To be more specific, the bulk entanglement entropy in the vacuum for both gauge fields and gravitons including quantum corrections for an entangling interval at constant time and an angular separation $\theta_{12}$ is given by \cite{Belin:2019mlt}
    \begin{equation}\label{eq:GaugeGravitonsQuantumEEShift}
        S_{\textrm{CFT}}=\left(\frac{c}{3}+\frac{c_{\textrm{top}}}{3}\frac{\ell}{\epsilon_{\textrm{bulk}}}+\frac{c_{\textrm{top}}}{3}\right)\log\left[\frac{2}{\epsilon_{\textrm{CFT}}}\sin\left(\frac{\theta_{12}}{2}\right)\right],
    \end{equation}
where $c=\frac{3\ell}{2G_N}$, $\ell$ is the AdS radius and $c_{\textrm{top}}$ counts the number of (boundary) degrees of freedom of the bulk effective field
theory. One can already see from this expression it is not completely clear on how to separate this shift into a contribution coming from a renormalization of Netwon's constant and a reinterpretation of the relation between the central charge and Newton's constant. This also suggests that interpretations regarding the universality of the quantum shifts are highly dependant on the regulator that is used in the bulk. Depending on how one chooses the bulk regulator one might have different interpretations for the quantum shift of the holographic central charge such as e.g. 1 in \cite{Belin:2019mlt}, 13 in \cite{Cotler:2018zff} and 26 in \cite{Huang:2019nfm}.

Even though as of yet it is not understood how the FLM prescription translates to flat space holography, the results and discussions in \cite{Belin:2019mlt} are very useful to discuss the universality of the shift of $c_1$ by 12 that we found. Based on previous results in flat space holography and BMS invariant quantum field theories it is known that in basically all the results obtained so far there is always a part that looks like a chiral half of a CFT$_2$ that is associated to $c_1$. Based on this it is very likely that a (bulk) BMS version of FLM will functionally look almost exactly like \eqref{eq:GaugeGravitonsQuantumEEShift} (modulo the AdS radius) and contain terms that count the number of boundary degrees of freedom of the bulk effective theory and some bulk cutoff. 

One possible way to make this discussion a little bit more explicit is to think about these quantum corrections in terms of a flat space limit of AdS$_3$/CFT$_2$ results. On the level of the dual quantum field theories this corresponds to an \.In\"on\"u--Wigner contraction of a parent 2D CFT. This procedure typically consists of assuming two chiral halves of a CFT with different generators and central charges, linearly combining the physical quantities of interest, properly introducing a contraction parameter and then to perform the contraction. By doing so one is able to get a glimpse of a putative BMS version of FLM for the entanglement entropy of boundary gravitons.

In this particular case this would amount to assuming two expressions\footnote{Note that for $c^+=c^-$, $c^+_{\textrm{top}}=c^-_{\textrm{top}}$, $\epsilon^+_{\textrm{bulk}}=\epsilon^-_{\textrm{bulk}}$, $\epsilon^+_{\textrm{CFT}}=\epsilon^-_{\textrm{CFT}}$ and $x^+=x^-=\vp_{12}$ one recovers precisely \eqref{eq:GaugeGravitonsQuantumEEShift} for $S_{\textrm{CFT}}=S^+_{\textrm{CFT}}+S^-_{\textrm{CFT}}$.} for the entanglement entropy $S^\pm_{\textrm{CFT}}$ of two chiral copies of a CFT with coordinates $x^\pm=\vp_{12}\pm\frac{u_{12}}{\ell}$ as
    \begin{equation}
        S^\pm_{\textrm{CFT}}=\left(\frac{c^\pm}{6}+\frac{c^\pm_{\textrm{top}}}{6}\frac{\ell}{\epsilon^\pm_{\textrm{bulk}}}+\frac{c^\pm_{\textrm{top}}}{6}\right)\log\left[\frac{2}{\epsilon^\pm_{\textrm{CFT}}}\sin\left(\frac{x_{12}^\pm}{2}\right)\right].
    \end{equation}
After defining the quantities
    \begin{equation}
        c^\pm=\frac{1}{2}(\ell c_2\pm c_1),\quad c^\pm_{\textrm{top}}=\frac{1}{2}(\ell c^{\textrm{top}}_2\pm c^{\textrm{top}}_1),\quad \epsilon^\pm_{\textrm{bulk}}=\ell \epsilon^{\textrm{bulk}}_\vp\pm \epsilon^{\textrm{bulk}}_u,\quad \epsilon^\pm_{\textrm{CFT}}=\epsilon_\vp\pm\frac{\epsilon_u}{\ell},
    \end{equation}
and taking the limit\footnote{One important aspect of performing the correct contraction that corresponds on the gravity side to a limit of vanishing cosmological constant is that instead of taking $S^+_{\textrm{CFT}}+S^-_{\textrm{CFT}}$ one has to consider $S^+_{\textrm{CFT}}-S^-_{\textrm{CFT}}$ in order to get a finte result. See for example \cite{Riegler:2014bia,Riegler:2016hah} and references therein or footnote~\ref{ref:FootnoteContraction} for more details.} $\lim\limits_{\ell\to\infty}(S^+_{\textrm{CFT}}-S^-_{\textrm{CFT}})$ one obtains a finite expression of the form \mbox{$S_{\textrm{BMS}}=S^{c_1}_{\textrm{BMS}}+S^{c_2}_{\textrm{BMS}}$} with
    \begin{subequations}
        \begin{align}
            S^{c_1}_{\textrm{BMS}}& =\left(\frac{c_1}{6}+\frac{c^{\textrm{top}}_1}{6\epsilon^{\textrm{bulk}}_\vp}+\frac{c^{\textrm{top}}_1}{6}-\frac{c^{\textrm{top}}_2\epsilon^{\textrm{bulk}}_u}{6(\epsilon^{\textrm{bulk}}_\vp)^2}\right)\log\left[\frac{2}{\epsilon_\vp}\sin\left(\frac{\vp_{12}^\pm}{2}\right)\right],\\
            S^{c_2}_{\textrm{BMS}}& =\left(\frac{c_2}{6}+\frac{c^{\textrm{top}}_2}{6\epsilon^{\textrm{bulk}}_\vp}+\frac{c^{\textrm{top}}_2}{6}\right) \left(\frac{u_{12}}{2 \tan \frac{\vp_{12}}{2} }  - \frac{\epsilon_u}{\epsilon_{\vp} }\right) .
        \end{align}
    \end{subequations}
There are a couple of interesting points about these expressions. First of all, it is noteworthy, that it seems in principle to be possible to have bulk corrections to entanglement entropy also for $c_2$. However, it is by no means clear what this parameter $c^{\textrm{top}}_2$ could mean physically (if it even exists in the first place) since, similar to $c_2$, it would have a physical dimension. Second, it is clear from e.g. the one-loop partition function computations that have been done in Section~\ref{sec:partitionfun} or the heat kernel method used in \cite{Barnich:2015mui} that  in this case $c^{\textrm{top}}_2$ has to be zero. Taking this into account the above expressions simplify to
    \begin{subequations}
        \begin{align}
            S^{c_1}_{\textrm{BMS}}& =\left(\frac{c_1}{6}+\frac{c^{\textrm{top}}_1}{6\epsilon^{\textrm{bulk}}_\vp}+\frac{c^{\textrm{top}}_1}{6}\right)\log\left[\frac{2}{\epsilon_\vp}\sin\left(\frac{\vp_{12}^\pm}{2}\right)\right],\\
            S^{c_2}_{\textrm{BMS}}& =\frac{c_2}{6}\left(\frac{u_{12}}{2 \tan \frac{\vp_{12}}{2} }  - \frac{\epsilon_u}{\epsilon_{\vp} }\right) .
        \end{align}
    \end{subequations}
This expression qualitatively agrees with \eqref{eq:CylEEQuantumShift} and is consistent with what we argued in the previous paragraphs. In particular, it is suggestive to interpret the quantum shift of 12 that we obtained previously as
    \begin{equation}
        12=\frac{c^{\textrm{top}}_1}{\epsilon^{\textrm{bulk}}_\vp}+c^{\textrm{top}}_1.
    \end{equation}
Since as of yet we have no universal way of splitting the number 12 that we get into contributions coming from $c^{\textrm{top}}_1$ and $\epsilon^{\textrm{bulk}}_\vp$ it seems likely that in the BMS case the quantum shifts of $c_1$ is not universal, but depends on the specific bulk regulator used. From this perspective it might also be more plausible why the quantum shift for the one-loop partition function differs from the one that we found for the entanglement entropy because both use different regulators. It is certainly not easy to see how the regularization used in this section compares with the zeta-function regularization of section \ref{sec:partitionfun}. This point will require more clarification and in particular a better understanding of the FLM proposal applied to BMS invariant quantum field theories. We will leave that for future work.

\section{BMS$_3$ blocks from the coadjoint orbit}\label{sec:BMSblocks}

In this section we use the methods developed in the last section to compute the BMS$_3$ identity block. BMS$_3$ blocks are elementary building blocks of correlation functions in BMS$_3$ invariant field theories. Whereas the two- and three-point functions of BMS$_3$ primary operators are completely fixed by symmetry, the four-point functions can be decomposed into BMS$_3$ blocks, defined in analogy to the conformal blocks in \cite{Bagchi:2016geg,Bagchi:2017cpu}. The correlator of four BMS$_3$ primary fields $\phi_i$ with $\Lt_0$ weights $\Delta_i$ and $\Mt_0$ weights $\xi_i$ can be written as a sum over BMS$_3$ blocks $\cF_p$, labeled by the exchanged primary fields $p$.  
\begin{equation}
    \label{BMSblockdef}
    \frac{\vev{\phi_1 \phi_1 \phi_2 \phi_2}}{\vev{\phi_1\phi_1 } \vev{ \phi_2 \phi_2}}  = \sum_p c_{11p} c_{22p} \cF_p (x,t; \Delta_i, \xi_i)\,.
\end{equation}
Here $c_{ijp}$ are the three-point function coefficients. The BMS$_3$ blocks depend on the cross ratios $x = \frac{x_{12}x_{34}}{x_{13}x_{24}}$ and $t/x = \frac{t_{12}}{x_{12}} + \frac{t_{34}}{x_{34}} - \frac{t_{13}}{x_{13}} -\frac{t_{24}}{x_{24}}$, the external weights $\Delta_i, \xi_i$ and the weights of the exchanged primary $\Delta_p ,\xi_p$. Some elementary BMS$_3$ field theory is reviewed in appendix \ref{sec:BMSblockdirect}.

Like conformal blocks, it is quite challeging to compute BMS$_3$ blocks in full generality, as they contain a sum over all BMS$_3$ descendents of the exchanged operator. The global BMS$_3$ blocks, that give the leading order contribution in large $c_2$ to the BMS$_3$ blocks, have been computed for light operators (with weights $\Delta, \xi \sim \cO(1)$ in \cite{Bagchi:2016geg,Bagchi:2017cpu} and in the heavy-light limit (with two external weights of order $c_2$ and two light operators) in \cite{Hijano:2017eii,Hijano:2018nhq} using monodromy methods. 

Here we compute for the first time the identity BMS$_3$ block in the large $c_2$ limit that corresponds to the exchange of the identity operator and all its descendents. In the main text we use the expansion of the bilocal operators and its expectation value in the geometric theory on the BMS$_3$ coadjoint orbits. We check our results for light external operators with those of a direct computation performed in appendix~\ref{sec:BMSblockdirect}, where we explicitly sum over descendents at the relevant order in $1/c_2$. In the heavy-light limit we use the expectation value of the bilocal operators on the massive BMS$_3$ coadjoint orbits to compute the leading and subleading order in $1/c_2$.

\subsection{BMS$_3$ identity blocks}

A BMS$_3$ four-point function can be decomposed into so-called BMS-blocks \cite{Bagchi:2016geg}. In the Chern-Simons theory, the blocks are computed by an open Wilson line network \cite{Bhatta:2016hpz} with end points at the boundary operator insertions. The representation of the external legs of the Wilson lines determine the weights of the external operators, while the Wilson lines are joined in bulk vertices by gluing them together using the appropriate Clebsch-Gordan coefficients. For the exchange of the identity operator, the Clebsch-Gordan coefficients are trivial and we do not need to worry about the bulk vertices. In that case, we can simply compute the expectation value of two bilocal operators \eqref{bilocalcyl}, evaluated on the global Minkowski orbit of BMS$_3$. The exchange of $\epsilon $ and $\tilde \alpha$ fields between the two bilocals will then account for the BMS$_3$ stress-tensor descendants of the identity operator in the exchange channel. 
When two of the external operators have common weights $\Delta_1, \xi_1$ and the other two have weights $\Delta_2 ,\xi_2$, the full identity block is given by the normalized two-point function of the bilocal operator:
\begin{equation}\label{idblock}
\cF_{\unity}(\vp_i,u_i) = \frac{\vev{\cB_{\Delta_1, \xi_1}(\vp_1, u_1; \vp_2, u_2) \cB_{\Delta_2, \xi_2}(\vp_3, u_3; \vp_4, u_4)}}{\vev{\cB_{\Delta_1, \xi_1}(\vp_1, u_1; \vp_2, u_2)} \vev{\cB_{\Delta_2,\xi_2}(\vp_3, u_3; \vp_4, u_4)}} \,.
\end{equation}
We now compute this four-point function in perturbation theory in $1/c_2$, using the same techniques as last section. The bilocal operators in \eqref{idblock} are expanded around the saddle point as in \eqref{bilocalexp}, which gives to leading order
\begin{align}\label{idblockexp}
\cF_{\unity}(\vp_i,u_i) & =  1  + \Delta_1 \xi_2 \vev{(\cJ_{12}^{(1)} \cdot \epsilon)( \cJ_{34}^{(1)} \cdot \tilde \alpha )} +   \xi_1 \Delta_2 \vev{(\cJ_{12}^{(1)} \cdot \tilde \alpha )(\cJ_{34}^{(1)} \cdot \epsilon)}  \\
\nonumber & + \xi_1 \xi_2 \left( \vev{(\cK_{12}^{(1)} \cdot \epsilon )(\cJ_{34}^{(1)} \cdot \tilde \alpha)}  +  \vev{(\cJ_{12}^{(1)} \cdot\tilde \alpha)(\cK_{34}^{(1)} \cdot  \epsilon )} + \vev{(\cJ_{12}^{(1)} \cdot\tilde \alpha)(\cJ_{34}^{(1)} \cdot \tilde \alpha )} \right) + \ldots 
\end{align} 
The dots denote higher order terms neglected here. 
We can now use the two-point correlators \eqref{vacpropagators} worked out in Section~\ref{sec:BMS3Propagators}  to compute the bilocal two-point function to first order in $\epsilon$ and $\tilde{\alpha}$.
The answer can most easily be expressed by moving back to the plane (by the inverse transformations $u_i = i \frac{t_i}{x_i}$ and $\vp_i = i \log(x_i)$ and using a global BMS$_3$ transformation to fix the points at:
\begin{equation}
	t_1 = t_2 = t_4 = 0\,, \quad t_3= t\,, \qquad x_1 = \frac{1}{\lambda}\,, \quad x_2 = 1 \,, \quad x_3 = x \,, \quad x_4 = \lambda, 
\end{equation}
and then taking $\lambda \to 0$. This results in the expression for the identity BMS$_3$ block to first order in $\frac{1}{c_2}$
\begin{align}\label{eq:VacuumBlockExpansion}
\cF_{\unity} (x,t) = & \, 1 - \frac{12}{c_2} \Big[  ( \Delta_1 \xi_2 + \Delta_2 \xi_1 
) \left(2+ (1-2/x) \log(1-x) \right) \nonumber\\
& \qquad \qquad + \frac{\xi_1 \xi_2 t}{x^2(1-x)} \left( (x-2)x + 2(x-1) \log(1-x) \right)   \Big] + \cO\left(\frac{1}{c_2^2}\right) \,  \\
 = & \, 1 + \frac{2}{c_2} \Big[  ( \Delta_1 \xi_2 + \Delta_2 \xi_1 
 ) \mathcal{F}(x) + t\,\xi_1 \xi_2\partial_x \mathcal{F}(x) \Big]+ \cO\left(\frac{1}{c_2^2}\right), \nonumber
\end{align}
where we have used the abbreviation $\mathcal{F}(x)= x^2 \; {}_2F_1(2,2;4,x)$. This result can also be computed by using the highest-weight representation of the BMS$_3$ algebra and summing all contributions at order $1/c_2$, as we show explicitly in appendix \ref{sec:BMSblockdirect}.

It is well known that for 2d CFTs the identity block exponentiates in the limit where the central charge $c \to \infty$ with $h^2 /c$ kept fixed. One may ask whether a similar limit leads to an exponentiation of the BMS$_3$ block as well i.e. can one write, for an appropriate scaling of $\Delta_i$ and $\xi$ 
\begin{equation}\label{eq:BMSVacuumBlockExp}
    \cF_{\unity}  = \exp{\left[\frac{2}{c_2} \Big(  ( \Delta_1 \xi_2 + \Delta_2 \xi_1 
    )  \mathcal{F}(x) + t\,\xi_1 \xi_2  \partial_x \mathcal{F}(x) \Big)\right]} + \cO\left(\frac{1}{\sqrt{c_2}} \right).
\end{equation}
One easy way to argue that this is indeed the case is by using a limiting procedure that can be interpreted as an $\ell\to\infty$ limit of the AdS radius by virtue of the AdS$_3$/CFT$_2$ correspondence.

The expression for the identity holomorphic block in a 2D CFT with conformal weights $h$ and central charge $c$ in the limit of large central charge (keeping $\frac{h_i}{\sqrt{c}}$ finite) reads \cite{Fitzpatrick:2014vua}
    \begin{equation}\label{eq:HolomorphicVacuumBlock}
        \mathcal{V}_0(z)=\exp{\left[\frac{2h_1h_2}{c}\mathcal{F}(z)\right]},
    \end{equation}
and similar for the anti-holomorphic block $\bar{\mathcal{V}}_0(\bar{z})$ where all quantities in \eqref{eq:HolomorphicVacuumBlock} are simply replaced by their barred counterparts. The limit of $\frac{1}{\ell}=\epsilon\to\infty$ in AdS$_3$ corresponds to a particular \.In\"on\"u--Wigner contraction in a 2D CFT. For the case at hand one has to consider the following quantity\footnote{One might wonder why there is a minus sign between the two term in \eqref{eq:BlocksBeforeContraction} instead of a plus sign. This change in sign is related to an automorphism of the Virasoro algebra of the form $\bar{L}_n\to-\bar{L}_{-n}$ and $\bar{c}\to-\bar{c}$ that is necessary for the \.In\"on\"u--Wigner contraction to correspond to the limit $\ell\to\infty$.\label{ref:FootnoteContraction}}
\begin{equation}\label{eq:BlocksBeforeContraction}
    \cF_{\unity} = \lim_{\epsilon\to0}\exp{\left[\frac{2h_1h_2}{c}\mathcal{F}(z)-\frac{2\bar{h}_1\bar{h}_2}{\bar{c}}\mathcal{F}(\bar{z})\right]},
\end{equation}
where
    \begin{subequations}
        \begin{align}
            h_i & = \frac{1}{2}\left(\frac{\xi_i}{\epsilon}+\Delta_i\right), & \bar{h}_i & = \frac{1}{2}\left(\frac{\xi_i}{\epsilon}-\Delta_i\right),\\
            c & = \frac{1}{2}\left(\frac{c_2}{\epsilon}+c_1\right), & \bar{c} & = \frac{1}{2}\left(\frac{c_2}{\epsilon}-c_1\right),\\
            z & = x + \epsilon t, & \bar{z} & = x - \epsilon t.
        \end{align}
    \end{subequations}
In the limit $\epsilon\to0$ this expression reduces to
    \begin{equation}\label{eq:VacuumBlockAdSLimitResult}
        \cF_{\unity} = \exp{\left[\frac{2}{c_2} \Big(  ( \Delta_1 \xi_2 + \Delta_2 \xi_1-\frac{c_1}{c_2}\xi_1\xi_2)  \mathcal{F}(x) + t\,\xi_1 \xi_2\partial_x \mathcal{F}(x) \Big)\right]}  + \cO\left(\frac{1}{\sqrt{c_2}} \right),
    \end{equation}
and matches \eqref{eq:BMSVacuumBlockExp} up to the term proportional to $c_1$. In our case we have considered $c_1$ to be of $\cO(1)$ due to quantum corrections and hence this term is subleading in $1/c_2$. There is, however, a simple way to see how this term would appear from the point of view of the geometric theory on which we now comment briefly.

There are several ways of obtaining $c_1\neq 0$ from the onset. One could look at the reduction of the gravitational sector of parity violating theories of gravity, such as e.g. topologically massive gravity \cite{Deser:1981wh,Deser:1982vy} in flat space or ``reloaded" \cite{Giacomini:2006dr,Barnich:2014cwa} versions of Einstein gravity. A simple way to achieve this in the context of Chern-Simons theory is to perform the reduction of section \ref{sec:reduction} including a non-zero trace for the $\mathfrak{sl}(2,\bR)$ generators in $\mathfrak{isl}(2,\bR)$, i.e. take $\langle \Lt_m \Lt_n \rangle = - 2\tilde k \gamma_{mn}$ in addition to the non-zero trace elements defined in \eqref{eq:InvBilForm}. The reduction for the terms proportional to $\tilde k$ then proceeds exactly as (one chiral half of) the AdS$_3$ case worked out in \cite{Cotler:2018zff} and the result of this will be given by the term proportional to $c_1$ in \eqref{Ibms3} for the kinetic part. In addition, the Schwarzian action proportional to the first line of \eqref{Noethercharge} with coefficient $c_1$ should be added to the Hamiltonian.

In terms of the propagators of section \ref{sec:BMS3Propagators} the addition of a non-zero $c_1$ part in the quadratic action would lead to adding the $\vev{\tilde \alpha \epsilon}$ propagator to the $\vev{\tilde \alpha \tilde \alpha}$ propagator with coefficient $- c_1/c_2$. Due to the expansion \eqref{idblockexp}, this contributes a factor of \mbox{$- \frac{c_1}{c_2} \xi_1 \xi_2 \vev{\cJ_{12} \cdot \tilde\alpha \, \cJ_{34} \cdot \epsilon}$} to the final result, which gives exactly the term proportional to $c_1$ in the exponent of \eqref{eq:VacuumBlockAdSLimitResult}. For the interested reader we collect the full expressions of the propagators with $c_1\neq0$ in Appendix~\ref{sec:NonZerocOne}.

There is also an intrinsic argument due to \cite{Cotler:2018zff} to see that \eqref{eq:VacuumBlockExpansion} exponentiates in the limit $c_2\to\infty$ but keeping $\frac{\xi_i}{\sqrt{c_2}}$ and $\frac{\Delta_i}{\sqrt{c_2}}$ constant.\footnote{Here we consider again that $c_1 \sim \cO(1)$, otherwise the appropriate scaling limit would be to keep $\frac{1}{\sqrt{c_2}} \left( \Delta_i + \frac{c_1}{c_2} \xi_i \right)$ constant.} Since all propagators \eqref{vacpropagators} are of order $1/c_2$, we rescale the fields as
\begin{equation}
    \epsilon\rightarrow\frac{\epsilon}{\sqrt{c_2}}\,, \qquad 
    \tilde{\alpha}\rightarrow\frac{\tilde \alpha}{\sqrt{c_2}} \,,
\end{equation}
and we take
\begin{equation}
    \xi_i=\sqrt{c_2} \, \mathfrak{X}_i\,, \qquad \Delta_i= \sqrt{c_2} \, \mathfrak{D}_i 
    \,.
\end{equation}
In the limit of large $c_2$ with $\mathfrak{X}$ and $\mathfrak{D}$ fixed the bilocal \eqref{bilocalcyl} exponentiates as
    \begin{equation}
        \frac{\cB_{\Delta_i,\xi_i} (u_1,\vp_1; u_2, \vp_2)}{\vev{\cO_1\cO_2}_{\rm cyl}} = \exp{\left[ (\mathfrak{F}_{12}^{(1)})_i \cdot \chi + \cO\left(\frac{1}{\sqrt{c_2}}\right)\right]},
    \end{equation}
with
\begin{equation}
    (\mathfrak{F}_{12}^{(1)} \cdot \chi)_i = \mathfrak{D}_i (\cJ_{12}^{(1)} \cdot \epsilon) + \mathfrak{X}_i (\cK_{12}^{(1)}\cdot \epsilon) + \mathfrak{X}_i (\cJ_{12}^{(1)} \cdot \tilde\alpha)\,.
\end{equation}
The operators $\cJ_{12}^{(1)}$ and $\cK_{12}^{(1)}$ are defined in \eqref{firstorder}.
This means that the vacuum block is given by
    \begin{equation}\label{eq:ApproxVacBlock}
        \cF_{\unity} = \langle e^{ (\mathfrak{F}_{12}^{(1)} \cdot \chi)_1 } e^{ (\mathfrak{F}_{34}^{(1)} \cdot \chi)_2 } \rangle\left(1+\cO\left(\frac{1}{\sqrt{c_2}}\right)\right).
    \end{equation}
In the limit $c_2\to\infty$ we can neglect self interactions of the fields $\epsilon$ and $\alpha$ that means e.g. terms of the form $\langle(\cJ_{12} \cdot \epsilon)(\cJ_{12} \cdot \epsilon)\rangle$ or $\langle(\cK_{34} \cdot \epsilon)(\cJ_{34} \cdot \tilde\alpha)(\cJ_{34} \cdot \epsilon)\rangle$. Thus, the remaining terms in \eqref{eq:ApproxVacBlock} that contribute have $n$ powers of $(\mathfrak{F}_{12}^{(1)} \cdot \chi)_1$ that are contracted with $n$ powers of $(\mathfrak{F}_{34}^{(1)} \cdot \chi)_2$. There are $n!$ possible contractions that are weighted with a factor of $\frac{1}{n!^2}$ coming from the expansion of the two exponentials. After performing the contractions one finds that the $n$-exchange contribution of the $\epsilon$ and $\alpha$ fields is given by $n!$ times the $n^{\rm th}$ power of the single-exchange i.e. one has
    \begin{align}
        \langle (\mathfrak{F}_{12}^{(1)} \cdot \chi)_1^n (\mathfrak{F}_{34}^{(1)} \cdot \chi)_2^n\rangle = n!\langle(\mathfrak{F}_{12}^{(1)} \cdot \chi)_1(\mathfrak{F}_{34}^{(1)} \cdot \chi)_2\rangle^n. 
    \end{align}
That means that the vacuum block (up to terms of order $\cO\left(\frac{1}{\sqrt{c_2}}\right)$) in this particular limit can be written as
    \begin{align}
        \cF_{\unity} & \approx\sum_{n=0}^{\infty}\frac{1}{n!}\langle(\mathfrak{F}_{12}^{(1)} \cdot \chi)_1(\mathfrak{F}_{34}^{(1)} \cdot \chi)_2\rangle^n \nonumber \\
        & =\exp{\left[\frac{2}{c_2} \Big(  ( \Delta_1 \xi_2 + \Delta_2 \xi_1 
        )  \mathcal{F}(x) + t\,\xi_1 \xi_2  \partial_x \mathcal{F}(x) \Big)\right]},
    \end{align}
which is precisely the expected expression \eqref{eq:BMSVacuumBlockExp} and consistent with \eqref{eq:VacuumBlockExpansion}. 

\subsection{Heavy-light identity block}

The results of the last subsection could be obtained by explicitly summing over BMS$_3$ descendants in the exchange channel, as shown in section \ref{sec:BMSblockdirect}. Those results are valid for primary operator weights small compared to $c_2$. When $\Delta_H, \xi_H \sim c_2$ the true power of the geometric theory comes to light and we are able to produce results not easily computable by explicitly summing over BMS$_3$ descendants at the relevant order of $c_2$.

Operators with weights of order $c_2$ are denoted as heavy operators. Their holographic interpretation is that they source flat space cosmologies \cite{Cornalba:2002fi,Cornalba:2003kd} with mass $\cM_0$ and angular momentum $\cL_0$. By comparing the 1-loop partition function on the massive orbit to the characters of BMS$_3$ in \eqref{weightsrel} we have found that the weights of the operators creating a flat space cosmology are:
\begin{equation}
\xi_H = \frac{c_2}{24} (\cM_0 + 1)  \,, \qquad \Delta_H = - \frac{c_2}{12} \cL_0 \, .
\end{equation}
The heavy-light limit of the BMS$_3$ identity blocks can be computed by considering the expectation value for the bilocal operator \eqref{bilocalcyl} in the geometric theory on a massive BMS$_3$ orbit, corresponding to a flat space cosmology with mass and angular momentum $\cM_0$ and $\cL_0$ of order unity. To this end we set out to compute
\begin{equation}
\vev{\cO_H | \cO_L(\vp_1,u_1) \cO_L(\vp_2,u_2) |\cO_H } =
 \vev{\cB_{\Delta_L, \xi_L}(\vp_1, u_1; \vp_2, u_2) }_{\rm FSC} + \ldots
\end{equation} 
Here the dots denote terms corresponding to other operator exchanges in the intermediate channel, since the bilocal only captures the identity operator exchange plus all its BMS$_3$ descendants.

The machinery of the last section now needs to be adapted to the massive BMS$_3$ orbits, starting with appropriate bilocal on the massive orbit. We have already encountered this operator in section \ref{sec:bilocal} and we reinstate the result here for convenience
\begin{align}\label{bilocalHL}
\cB^{\rm m}_{\Delta,\xi} (\vp_1, u_1; \vp_2, u_2)  = & \left( \frac{\gamma^2 f_1' f_2' }{2^2 \sin^2 \left(\frac{\gamma}{2}f_{12} \right)} \right)^{\Delta} \exp\left( \xi\left(  \frac{\alpha_1'}{f_1'} + \frac{\alpha_2'}{f_2'}  - \frac{2 \cL_0}{\gamma^2} \right)\right) \nonumber \\
& \times  \exp \left(- \xi \gamma \left(\alpha_{12} - \frac{\cL_0}{\gamma^2} f_{12} \right) \cot \left( \frac{\gamma}{2}f_{12}\right) \right)\,,
\end{align}
where we remind the reader that
\begin{equation}
\gamma = \sqrt{- \cM_0} = \sqrt{1 - \frac{24 \xi_H}{c_2} }\,.
\end{equation}
We now compute the expectation value of this operator representing a light probe (with weights $\Delta_L, \xi_L \sim \cO(1)$) in a flat space cosmology background. The tree-level result is easily given by taking the BMS$_3$ fields $(f,\alpha(f))$ to correspond to the saddle point on the massive BMS$_3$ orbit $(f,\alpha(f)) = (\vp,u)$. We find
\begin{align}
\cF_{\unity}^{\rm HL} & = \left( \frac{\gamma}{2 \sin \left(\frac{\gamma}{2}\vp_{12} \right)} \right)^{2\Delta_L} e^{- \xi_L \gamma (u_{12} - \frac{\cL_0}{\gamma^2} \vp_{12} ) \cot \left( \frac{\gamma}{2}\vp_{12} \right) -  \frac{2 \xi_L \cL_0}{\gamma^2}} + \cO\left(\tfrac{1}{c_2}\right) \\ \nonumber
& \equiv \vev{\cO_L \cO_L}_{\rm FSC} + \cO\left(\tfrac{1}{c_2}\right)\,.
\end{align}

We continue to compute the $ \cO(1/c_2)$ corrections to this expression by expanding \eqref{bilocalHL} to second order in perturbation theory and computing its expectation value using the propagators \eqref{eq:Fluctuations2PointFunctions}. In expanding the bilocal \eqref{bilocalHL} around the saddle point we can safely ignore linear terms and terms quadratic in $\epsilon$, as we have $\vev{\epsilon \epsilon} = 0$. The terms that do contribute are:
\begin{equation}\label{eq:HLexpansion}
    \cF_{\unity}^{\rm HL} = \vev{\cO_L \cO_L}_{\rm FSC} \left( 1 + \xi_L V_{\xi} + \xi_L^2 V_{\xi\xi} + \Delta_L \xi_L V_{\Delta \xi} + \ldots \right),
\end{equation}
with
\begin{subequations}
\begin{align}
    V_{\xi} & = \frac{\gamma^2 \tilde \alpha_{12} \epsilon_{12}}{2 \sin^2  \left( \tfrac{\gamma}{2} \vp_{12} \right)} -  \tilde \alpha_1'\epsilon_1' - \tilde \alpha_2'\epsilon_2', \\
    V_{\Delta\xi} & = \left( \tilde \alpha'_1 + \tilde \alpha_2' - \gamma \cot  \left( \tfrac{\gamma}{2} \vp_{12} \right) \tilde \alpha_{12} \right) \left( \epsilon'_1 + \epsilon_2' - \gamma \cot  \left( \tfrac{\gamma}{2} \vp_{12} \right) \epsilon_{12} \right), \\
    V_{\xi\xi} & = \left( \tilde \alpha'_1 + \tilde \alpha_2' - \gamma \cot  \left( \tfrac{\gamma}{2} \vp_{12} \right) \tilde \alpha_{12} \right) \bigg[\frac12 \left( \tilde \alpha'_1 + \tilde \alpha_2' - \gamma \cot  \left( \tfrac{\gamma}{2} \vp_{12} \right) \tilde \alpha_{12} \right)  \\
    & \qquad  +  \frac{\gamma^2 \epsilon_{12}}{2 \sin^2 \left( \tfrac{\gamma}{2} \vp_{12} \right)} \left( u_{12} - \frac{\cL_0}{\gamma^2} \left( \vp_{12} - \frac{1}{\gamma} \sin \left( \gamma \vp_{12} \right) \right) \right) \bigg]. \nonumber
\end{align}
\end{subequations}
To continue, we need the coincidence point limit of the correlation functions. Like before, we regulate the correlators taken at the same points by displacing them by an infinitesimal imaginary amount. This gives
\begin{subequations}\label{coincidencemassive}
\begin{align}
    \vev{\tilde \alpha_1 \epsilon_1 } & = - \frac{6}{c_2 \gamma^2} \left( 2 \gamma_E + \psi(\gamma) + \psi (- \gamma)  \right)\,, \\
    \vev{\tilde \alpha_1 \tilde \alpha_1 } & =  - \frac{12 \cL_0}{c_2 \gamma^4} \left(2 \gamma_E +  \psi(\gamma) + \psi (- \gamma)  + \frac{\gamma}{2} \zeta (2, -\gamma) - \frac{\gamma}{2} \zeta(2,\gamma) \right)\,, \\
    \vev{\tilde \alpha_1' \epsilon_1' } & = - \frac{6}{c_2 } \left( 2 \gamma_E + 2 \log(\delta_\vp-\mu_L\delta_u) + \psi(\gamma) + \psi (- \gamma)  \right) \,, \\
    \vev{\tilde \alpha_1' \tilde\alpha'_1 } & =  \frac{6 \cL_0}{c_2 \gamma}\Big(\zeta(2,\gamma) - \zeta(2,-\gamma) \Big) - \frac{12 \delta_u}{c_2 (\delta_\vp-\mu_L\delta_u)}  \,,
\end{align}
\end{subequations}
where here $\gamma_E$ is the Euler–Mascheroni constant, $\psi(x) = \frac{\Gamma'(x)}{\Gamma(x)}$ is the digamma function and $\zeta(s,q)$ is the Hurwitz zeta function.

We now have all the ingredients to compute the $1/c_2$ corrections to the heavy-light identity block. The result is, taking $(\vp_1 ,u_1) \to (\vp ,u) $ and $(\vp_2 ,u_2) \to (0,0)$ 
\begin{subequations}\label{eq:heavyquantum}
\begin{align}
    V_{\xi} & =  - \frac{3}{c_2} \csc^2 \left( \frac{\gamma \vp}{2} \right) \bigg[ \Phi(e^{i \vp},1,\gamma) +\Phi(e^{-i \vp},1,\gamma) + \Phi(e^{i \vp},1,-\gamma) + \Phi(e^{-i \vp},1,-\gamma) \nonumber \\ 
     & \quad + 2 \cos(\gamma \vp) ( 2 \gamma_E + \psi(\gamma) + \psi (- \gamma)  ) + 4 \log \left(2 \sin \left(\frac{\vp}{2} \right) \right) \bigg] + \frac{24}{c_2} \log \delta_\vp \,, \\
     V_{\Delta \xi} & =  - \frac{6}{c_2}  \csc^2 \left( \frac{\gamma \vp}{2} \right) \bigg[ B(e^{i\vp},\gamma,0) +B(e^{-i\vp},\gamma,0) + B(e^{i\vp},-\gamma,0) + B(e^{-i\vp},-\gamma,0) \nonumber \\ 
     & + 2  ( 2 \gamma_E + \psi(\gamma) + \psi (- \gamma)  ) + 4 \log \left(2 \sin \left(\frac{\vp}{2} \right) \right) \bigg] + \frac{24}{c_2} \log \left( \frac{2}{\delta_\vp} \sin \left( \frac{\vp}{2} \right) \right)\,,
\end{align}
and
\begin{align}
     V_{\xi \xi} & = - \frac{6 \cL_0}{c_2 \gamma} \csc^2 \left( \frac{\gamma \vp}{2} \right) \bigg[ \frac{ i \vp}{1 - e^{i\gamma \vp}} \left(B(e^{i \vp},\gamma,0) + B(e^{-i \vp},-\gamma,0) + \Phi( e^{i \vp},1, -\gamma) + \Phi( e^{ - i \vp},1, \gamma) \right) \nonumber \\
     & - \frac12 e^{i \gamma \vp} \Phi(e^{i\vp},2, \gamma) + \frac12 e^{- i \gamma \vp} \Phi(e^{i\vp},2, -\gamma) - \frac12 e^{-i \gamma \vp} \Phi(e^{-i\vp},2, \gamma) + \frac12 e^{i \gamma \vp} \Phi(e^{-i\vp},2, -\gamma) \nonumber \\
     & + \zeta(2,\gamma) - \zeta(2, - \gamma) - \vp \cot\left(\frac{\gamma \vp}{2} \right) \left( 2 \gamma_E + \psi(\gamma) + \psi (- \gamma)  + 2 \log \left(2 \sin \left(\frac{\vp}{2} \right) \right) \right) \nonumber
    \bigg]      
     \\
    & + \frac{u}{2} \partial_\vp V_{\Delta \xi} - \frac{12 \delta_u}{c_2 \delta_\vp},
\end{align}
\end{subequations}
where we have used the identity relating the Lerch transcendant $\Phi(z,1,a)$ to the incomplete Beta function $B(z,a,0)$ as
\begin{equation}
B(z,a,0) = z^a \Phi(z,1,a).
\end{equation}
This concludes our computation of the $1/c_2$ corrections to the identity BMS$_3$ block in the haevy-light limit. We now use this result to obtain the quantum corrections to the entanglement entropy in a BMS$_3$ invariant field theory dual to a flat space cosmology.

\subsubsection{Quantum corrections to the FSC Entanglement Entropy}
In analogy to the computation performed in section \ref{sec:quantum}, we are now able to use the results of the last subsection to compute the $n \to 1$ limit of \eqref{eq:EEReplicaTrace} for correlation functions on the massive orbits of BMS$_3$ that are dual to flat space cosmologies. We do this by taking $n$-derivative of the $n$-th power of \eqref{eq:HLexpansion} with light weights given by \eqref{eq:BMSWeights} and then taking the $n\to 1$ limit. The result is
\begin{equation}
    S_{\rm EE}^{\rm FSC} = S_{\rm EE}^{\rm FSC, tree}  - \frac{c_2}{12} V_{\xi} \,,
\end{equation}
where $S_{\rm EE}^{\rm FSC, tree}$ is given in equation \eqref{eq:EEFSC} and $V_\xi$ is given in equation \eqref{eq:heavyquantum} above (but one should take $\vp \to \vp_{12})$. The answer is again exact in the perturbative expansion in $1/c_2$ by applying the same arguments of section \ref{sec:quantum}.


\section{Discussion}\label{sec:discussion}

In this work we have refined and further developed flat space holographic methods by using the geometric action on the coadjoint orbit of the BMS$_3$ group. We have performed the Hamiltonian reduction of the classical gravity action in Chern-Simons form and obtained exactly the geometric action on the coadjoint orbits of the BMS$_3$ group of \cite{Barnich:2017jgw}. The orbit representatives correspond to the zero modes of the gravitational charges as we have shown by explicitly by taking into account the bulk holonomy, generalizing earlier work by \cite{Barnich:2013yka}. 

This makes the relationship between the different gravitational saddles obeying Barnich-Comp\`ere \cite{Barnich:2006av} boundary conditions and the coadjoint orbits of the BMS$_3$ group explicit and provides an action principle for `flat space boundary gravitons'. These are the excited states generated by boundary condition preserving diffeomorphisms of a given gravitational saddle and correspond to descendants of primary operators in the boundary BMS$_3$ field theory. The action can be used to compute the contribution of stress-tensor descendants to the one-loop partition function of a given classical saddle on the torus and we show that this corresponds exactly to the BMS$_3$ character obtained earlier in \cite{Oblak:2015sea,Bagchi:2019unf}. The comparison with the BMS$_3$ characters gives us a relation between the weights of primaries in a BMS$_3$ invariant field theory and the mass and angular momentum of the flat space cosmologies. We have further noticed that zeta function regularization of the partition function induces a shift in the BMS$_3$ central charge $c_1$ by 26 in the case of the vacuum orbits and by 2 for the generic orbits.

We have also shown how to construct bilocal operators whose vacuum expectation value corresponds to two-point functions of primary operators in the BMS$_3$ invariant quantum field theory. By expanding these operators in $1/c_2$ and using perturbation theory in the geometric action, we have obtained the leading order quantum correction to these bilocal operators, coming from stress tensor descendants of the BMS$_3$ primary fields. We have used this result to compute the quantum corrections to entanglement entropy in BMS$_3$ invariant quantum field theories and to compute the $1/c_2$ corrections to the BMS$_3$ identity blocks, both for light external operators and in the heavy-light limit. The quantum corrections to the entanglement entropy also induce a shift of $c_1$ (albeit by a different number than the partition function), while keeping $c_2 = 3/G_N$ fixed. This indicates that in pure three-dimensional flat space quantum gravity, Newtons constant is not renormalized by quantum corrections. Instead quantum corrections lead to a non-zero $c_1$, indicative of a quantum gravitational anomaly.\footnote{A shift in $c_1$ can also be obtained from a deformation of the BMS$_3$ algebra, as explained in \cite{Parsa:2018kys}. In this sense one could view the quantum corrections as leading to a deformed algebra in the field theory. We thank M.~M.~Sheikh-Jabbari and H.~R.~Safari for pointing this out.}

Several subtleties and extensions of the work presented here deserve a comment. In this paper we have concerned ourselves exclusively with a single boundary, leaving aside the interesting and relevant question on how to connect $\mathscr{I}^+$ and $\mathscr{I}^-$ through the boundary theory/symmetries (see \cite{Prohazka:2017equ,Compere:2017knf} for a 3D discussion on linking $\mathscr{I}^+$ and $\mathscr{I}^-$ a la \cite{Strominger:2013jfa}). In the present setup, we have supposed the manifold to be a disk times a null line, with the boundary of the disk being either $\mathscr{I}^+$ or $\mathscr{I}^-$. If this was indeed the whole setup, we could not allow for a non-trivial holonomy, as any cycle on the disk is contractible. In fact, the presence of a holonomy assumes that we are dealing with a non-trivial topology in the interior, for instance, another boundary where we could define another set of boundary conditions. One could imagine topologically deforming the interior of 3D flat spacetime to form an annulus times a null direction, where the inner boundary of the annulus corresponds to $\mathscr{I}^-$ and the outer boundary is $\mathscr{I}^+$. In that case, following the work of \cite{Elitzur:1989nr}, the two boundaries can have separate dynamics, but are coupled though the holonomy that has the effect of coupling the zero modes of the fields on both boundaries. The present work describes a single boundary in this situation, ignoring the dynamics at the inner boundary, or at the past/future of $\mathscr{I}^{+/-}$. The complete setup with two boundaries has recently been explored in AdS$_3$ \cite{Grumiller:2019ygj,Henneaux:2019sjx} and it would be interesting to see how in flat spacetime this could lead to a linking of the theories at $\mathscr{I}^+$ and $\mathscr{I}^-$.

Besides the topology, we have also assumed the holonomy to be non-dynamical. This was sufficient to obtain an effective theory of BMS$_3$ transformations around a gravitational saddle with constant mass $\cM_0$ and angular momentum $\cL_0$. The complete setup would, however, treat the holonomy as a dynamical variable. Then one would need to extend the phase space of the boundary theory from a single BMS$_3$ coadjoint orbit to the collection of all orbits and include a canonical conjugate to the holonomy to obtain a non-degenerate symplectic form on this extended phase space \cite{Henneaux:2019sjx}. In addition, the orbit representatives in the BMS$_3$ coadjoint action need not be constant and an interesting question is what gravitational solutions (if any) correspond to BMS$_3$ orbits with non-constant representatives. To the best of our knowledge this question is also an open question in the AdS$_3$/CFT$_2$ setup, where it is not known what kind of locally AdS$_3$ solutions correspond to Virasoro coadjoint orbits with non-constant representatives.

Another open question is whether the one-loop partition function we have computed in section \ref{sec:partitionfun} is exact. We suspect that it is, by virtue of the arguments made in the AdS$_3$/CFT$_2$ setup discussed in \cite{Cotler:2018zff}. In that case, the boundary action is the Alekseev-Shatashvili action on the coadjoint orbit of the Virasoro group and a suitable generalization of the Duistermaat-Heckman theorem \cite{Duistermaat:1982vw} can be made to show that the partition function is one-loop exact (following closely the argument for the Schwarzian action of \cite{Stanford:2017thb}). We suspect that similar arguments can be made for the case described in this work, supported by the fact that the Duistermaat-Heckman argument can be generalized to the geometric action on the coadjoint orbit of any semi-simple Lie group. All that remains to be done is to generalize this to (centrally extended) semi-direct product groups, such as the BMS$_3$ group. If true, it could also be quite rewarding to investigate how the one-loop exactness can be utilized to compute other observables exactly.

The Wilson line setup we used in section \ref{sec:BMSblocks} only allowed to compute the identity BMS$_3$ block. In order to be able to compute other BMS$_3$ blocks, such as the ones considered in \cite{Bagchi:2016geg,Bagchi:2017cpu,Hijano:2017eii,Hijano:2018nhq}, one would have to work out the open Wilson line networks of \cite{Bhatta:2016hpz} for flat spacetimes, by gluing the Wilson lines in the interior using the $\mathfrak{isl}(2,\bR)$ Clebsch-Gordan coefficients. In the present context, due to our radial gauge choice, this entire Wilson line network could be reduced to the boundary completely and computed using the expectation values of the fields in the geometric theory. One could hope that the one-loop exactness of the theory would aid in giving exact results, but this is a very speculative statement at this point.

Generalizations of the present work to the supersymmetric case are also of interest. The reduction of the boundary of $\cN=1$ flat space supergravity was already performed in \cite{Barnich:2014cwa} and $\cN=2$ BMS$_3$ invariant WZW-models were considered recently in \cite{Banerjee:2019lrv,Banerjee:2019epe}, however the relation to the geometric action on the coadjoint orbits of supersymmetric extensions of the BMS$_3$ group was not yet exposed. It seems that the techniques and methods discussed here can straightforwardly be applied to the supersymmetric extensions of the BMS$_3$ algebra and flat space supergravities of \cite{Bagchi:2009ke,Lodato:2016alv,Banerjee:2016nio}. 

Another interesting connection was made recently between the geometric action on the coadjoint orbits of the Virasoro group and complexity growth in two-dimensional CFTs \cite{Caputa:2018kdj}. In that work a suitable definition for Nielsen complexity for 2D CFTs was introduced and led to the Alekseev-Shatashvili action as complexity functional for the CFT. Since the Alekseev-Shatashvili action also arises from the Hamiltonian reduction of AdS$_3$ gravity with Brown-Henneaux boundary conditions, one could view this as an explicit realization of the ``complexity equals bulk action'' proposal of \cite{Brown:2015bva}. It would be rewarding to see if these arguments can also be applied to BMS$_3$ invariant field theories and gravity in asymptotically flat spacetimes.

Finally, a flurry of activity (see for instance \cite{Kapec:2014opa,Kapec:2016jld,Pasterski:2016qvg,Donnay:2018neh,Puhm:2019zbl,Adamo:2019ipt,Ball:2019atb} among others) has recently been devoted to understanding the link between the $\mathcal{S}$-matrix in 4d Minkowski spacetimes and correlation functions on the celestial sphere, after methods pioneered in \cite{deBoer:2003vf}. The four-dimensional Lorentz group SL$(2,\mathbb{C})$ acts as the two-dimensional global conformal group on the celestial sphere at infinity that is extended to the full conformal group of quantum gravitational scattering in 4D Minkowski spacetimes \cite{Kapec:2014opa,Kapec:2016jld}. This conformal group is contained within the extended BMS$_4$ group \cite{Barnich:2010eb} as superrotations. It would be very interesting to see whether similar connections as exposed in this paper can be made between the geometric quantization methods of Kirrilov and Konstant applied to the BMS$_4$ group and gravitational scattering in 4D Minkowski spacetime.


\subsection*{Acknowledgements}
We like to thank A.~Bagchi, G.~Barnich, A.~Belin, P.~Caputa, O.~Fuentealba, D.~Grumiller, H.~Gonz\'alez, S.~Hadar, M.~Henneaux, D.~Klein Kolchmeyer, B.~Oblak, A.~Ranjbar and J.~Salzer for useful discussions and valuable comments.
WM is supported by the ERC Advanced Grant {\it High-Spin-Grav} and by FNRS-Belgium (convention
FRFC PDR T.1025.14 and convention IISN 4.4503.15). The research of MR is supported by the European Union’s Horizon 2020 research and innovation programme under the Marie Skłodowska-Curie grant agreement No 832542.


\begin{appendix}

\section{Identity BMS$_3$ block by direct computation}\label{sec:BMSblockdirect}

As a cross check of the Wilson line computation performed in Section~\ref{sec:BMSblocks}, we proceed here to compute the identity BMS$_3$ block directly, following the approach of computing the Virasoro identity block directly in the large $c$ limit presented in appendix B of \cite{Fitzpatrick:2014vua}.

\subsection{BMS$_3$ field theories}

We start by fixing notation and conventions and we review briefly the necessary elements of BMS$_3$ invariant field theory. We mostly use the same conventions as \cite{Lodato:2018gyp}, with $x \leftrightarrow t$. 
The $\mathfrak{bms}_3$ algebra is given by \eqref{BMS3}. Primary fields transform under $\mathfrak{bms}_3$ transformations as
\begin{subequations}
	\label{GCAcommutators}
	\begin{align}
	[\Lt_n, \phi_p(x,t)] & = \big[x^{n+1} \partial_x + (n+1)t x^n \partial_t + \xi_p n(n+1)t x^{n-1} \\
	& \qquad \nonumber + \Delta_p (n+1)x^n \big] \phi_p(x,t) \; , \\
	[\Mt_n, \phi_p(x,t)] & = \big[ x^{n+1}\partial_t + \xi_p (n+1) x^n \big] \phi_p(x,t)\,. 
	\end{align}
\end{subequations}
The vacuum state in the highest-weight representation is defined as being annihilated by the global Poincar\'e subalgebra of $\mathfrak{bms}_3$ and all lowering operators $\Lt_{n},\Mt_{n}$ with $n>0$.
\begin{equation}
\Lt_n|0\rangle = \Mt_n|0\rangle=0 \,,\qquad \forall\, n \geq -1\,.
\end{equation}
Inserting primary field $\phi_p(x,t)$ at the origin of the Carrollian plane $\bR^{1,1}$ generates primary states $\ket{p}$ in the highest-weight representations
\begin{equation}
\phi_p(0,0)\ket{0} \equiv \ket{p}\,,
\end{equation} 
where $\ket{p}$ is defined such that
\begin{align}
& \Lt_0 |p\rangle  = \Delta_p | p \rangle\,, &  \Mt_0 |p \rangle = \xi_p | p \rangle \,,
&& \Lt_n|0\rangle = \Mt_n|0\rangle=0 \,,&&  \forall\, n \geq 1\,.
\end{align}
The BMS modules (analogue to the Verma modules in CFT) are defined by acting with raising operators $\Lt_n, \Mt_n$, with $n<0$ on the primary states, defining the BMS descendant states at level $N$
\begin{equation}\label{descendants}
\ket{p,\{N\}} = \Lt_{-\{k\}} \Mt_{-\{l\}} \ket{p} \equiv \Lt_{-k_1} \ldots \Lt_{-k_i} \Mt_{-l_1} \ldots \Mt_{-l_j} \ket{p}\,,
\end{equation}
where $\{N\}$ denotes two sets of integers $\{ k\} $ and $\{l\}$, whose total level $N$ is the sum of all elements in the sets and we organize each of them in descending order ($k_1 \geq k_2$ and $l_1 \geq l_2$, \ldots).

The Hermitian conjugate states are
\begin{equation}\label{out}
\bra{p} = \lim_{x \to \infty} x^{2\Delta_p} \bra{0} \phi_p(x,0)\,.
\end{equation}
Hermitian conjugation inverts the order of the descendant operators and takes
\begin{equation}
\Lt_{k}^\dagger = \Lt_{-k} \,, \qquad \Mt_l^\dagger = \Mt_{-l}\,.
\end{equation}
Hence
\begin{equation}
\bra{p, \{N\}} = \bra{p}\Mt_{l_j} \ldots \Mt_{l_1}\Lt_{k_i}\ldots \Lt_{k_1} \,.
\end{equation}
The out states \eqref{out} are annihilated by the raising operators $\Lt_n,\Mt_n$ with $n < 0$. 

The correlation functions between primaries are invariant under the global Poincar\'e subalgebra. This fixes the functional form of the normalized two-point function completely 
\begin{equation}\label{twopt}
\langle \phi_m(x_1,t_1) \phi_n(x_2,t_2) \rangle = \frac{\delta_{\Delta_m,\Delta_n} \delta_{\xi_m,\xi_n}}{x_{12}^{\Delta_m+\Delta_n}}e^{ - (\xi_m + \xi_n) \frac{t_{12} }{x_{12}}}\,,
\end{equation}
where $t_{12} = t_2 - t_1$ and likewise for $x_{12}$. The three-point function between primaries depend on a single coefficient $c_{imn}$
\begin{equation}\label{threept}
\langle \phi_i(x_1,t_1) \phi_m(x_2,x_2) \phi_n(x_3,t_3) \rangle = \frac{c_{imn}}{x_{12}^{\Delta_{imn}} x_{23}^{\Delta_{mni}} x_{13}^{\Delta_{inm}}}e^{- \xi_{imn} \frac{t_{12}}{x_{12}} - \xi_{mni}\frac{t_{23}}{x_{23}} - \xi_{inm}\frac{t_{13} }{x_{13}} }\,,
\end{equation}
where $\Delta_{imn} = \Delta_i + \Delta_m - \Delta_n$ and likewise for $\xi_{imn}$. 
The four-point function can depend on a general function of the invariant cross ratios $X$ and $T$. We will write it as
\begin{align}\label{fourpt}
& \langle \phi_m(x_1) \phi_m(x_2,t_2) \phi_n(x_3,t_3) \phi_n(x_4,t_4) \rangle = x_{12}^{-2\Delta_{m}} x_{34}^{-2\Delta_{n}} e^{-2 \xi_m \frac{ t_{ij} }{x_{12}} - 2\xi_n \frac{t_{34}}{x_{34}} } F_{\rm BMS}(X,T)\,, \nonumber
\end{align}
where $F_{\rm BMS}$ is an arbitrary function of the BMS$_3$ cross ratios
\begin{equation}\label{GCAcrossratios}
X = \frac{x_{12} x_{34}}{x_{13}x_{24}}\,, \qquad \frac{T}{X} = \frac{t_{12}}{x_{12}} + \frac{t_{34}}{x_{34}} - \frac{t_{13}}{x_{13}} - \frac{t_{24}}{x_{24}} \,.
\end{equation}
The functions $F_{\rm BMS}$ can be decomposed into BMS$_3$ blocks, similarly as one would do for conformal blocks in CFTs.

\subsection{BMS$_3$ blocks}
To get a handle on the BMS$_3$ blocks we define an identity operator as a sum over a complete set of states in the theory. To this end we first consider the Gram matrix of inner products of descendents of a primary $p$ at a given level $N$. We denote it by $\mathfrak{M}^{p}_{\{N\},\{N'\}}$ and it is defined as
\begin{equation}
\mathfrak{M}_{\{N\}, \{N'\}}^{p} = \vev{p,\{N\}| p, \{N'\}}.
\end{equation}
The Gram matrix is orthogonal in the sense that it vanishes for $N\neq N'$. Within a given level $N = N'$ it is not orthogonal but instead can be brought into triangular form, with only non-zero entries in the upper left corner, extending to the anti-diagonal. The inverse of the Gram matrix $\mathfrak{M}^{\{N\},\{N'\}}_{p}$ can be used to define a projection operator
\begin{equation}
\cP = \sum_{p, \{N\}, \{N'\} } \ket{p, \{N\}} \mathfrak{M}^{\{N\},\{N'\}}_{p} \bra{p,\{N'\} }.
\end{equation}
Even though the states $\ket{p,\{N\}}$ are not orthogonal, by including the inverse of the Gram matrix this projection operator is. This can be easily checked by noting that $\cP \ket{q, \{M\}} = \ket{q, \{M\}} $ and $ \bra{q, \{M\}} \cP = \bra{q, \{M\}} $. This implies that one can always insert $\cP$ into any correlation function. In particular, inserting this into a four-point function of BMS$_3$ primary operators gives
\begin{align}
& \vev{ \phi_m(x_1,t_1) \phi_m(x_2,t_2) \cP \phi_n(x_3,t_3) \phi_n(x_4,t_4)} \\ & \quad 
 = \sum_{p,\{N\}, \{N'\}} \vev{\phi_m(x_1,t_1) \phi_m(x_2,t_2)| p, \{N\} } \mathfrak{M}^{\{N\},\{N'\}}_{p} \vev{ p, \{N'\}| \phi_n(x_3,t_3) \phi_n(x_4,t_4)} \nonumber \\ & \nonumber
 \quad \equiv   \vev{\phi_m(x_1,t_1) \phi_m(x_2,t_2)} \vev{ \phi_n(x_3,t_3) \phi_n(x_4,t_4)} \sum_p c_{mmp}c_{nnp} \cF_p(X,T;\Delta_i,\xi_i)\,.
\end{align}
Here $i = m,n$ and $c_{iip}$ are theory dependent structure constants (the coefficients of the three-point functions). The final equality defines the BMS$_3$ blocks $\cF_p(X,T; \Delta_i,\xi_i)$.
It contains all theory-independent information. They include the sum over all BMS$_3$ descendants, but exclude the sum over primaries. Hence there is a BMS$_3$ block associated to each BMS primary exchange in the four-point function and this correlator decomposes into a sum over the blocks for all primaries in the theory. 

Using a global Poincar\'e transformation we can always take the coordinates to the special values
\begin{equation}\label{points}
\{ (x_i,t_i), (x_j,t_j), (x_m,t_m), (x_n,t_n)\} = \{(\infty,0), (1,0), (x,t), (0,0) \}\,.
\end{equation}
In that case $T =t$ and $X=x$ and the BMS$_3$ blocks can be computed as
\begin{align}
 \cF_p(x,t;\Delta_i,\xi_i) & = \frac{\vev{ \phi_m(\infty,0) \phi_m(1,0) \cP \phi_n(x,t) \phi_n(0,0)}}{\vev{\phi_m(\infty,0) \phi_m(1,0)} \vev{ \phi_n(x,t) \phi_n(0,0)}}\nonumber\\
 &=\sum_{\{N\}, \{N'\}} \frac{ \vev{m | \phi_m(1,0)| p, \{N\} } \mathfrak{M}^{\{N\},\{N'\}}_{p} \vev{ p, \{N'\}| \phi_n(x,t)| n } }{ c_{mmp} c_{nnp} x^{-2 \Delta_n} \exp(-2\xi_n \frac{t}{x}) }.
\end{align}
The structure constants $c_{iip}$ in the denominator are only there to cancel the three-point function coefficients in the nominator and hence from now on we omit both in any explicit computation.

We are interested here in the case where the primary $p$ is the identity operator (that has BMS weights $(\Delta ,\xi ) = (0,0)$ ). This implies that $\phi_{\unity}(0,0)|0\rangle = |0\rangle$ and hence $\ket{\unity, \{N\}}$ does not contain any descendants generated by $\Lt_{-1}$ and $\Mt_{-1}$. This in turn implies that the inverse Gram matrix for any descendant is of order $1/c_2$, as all order one contributions to the inverse Gram matrix are coming from the descendants generated by the global subalgebra \cite{Bagchi:2016geg}.

Let us focus now on the $1/c_2$ contribution to the identity BMS$_3$ block. The only contributions to the inverse Gram matrix at this order are coming from the single descendant states $\vev{0|\Lt_{m} \Mt_{-m}|0}$ and $\vev{0|\Mt_{m} \Lt_{-m}|0}$. This, together with the triangular structure of the Gram matrix, allows us to write
\begin{align}\label{idblockO1}
\cF_{\unity}(x,t;\Delta_i,\xi_i) = 1 + & \sum_{m =2}^{\infty} \frac{ \vev{m | \phi_m(1,0) \Lt_{-m}|0} \vev{0|\Mt_{m} \phi_n(x,t)| n } }{\vev{0|\Mt_{m} \Lt_{-m}|0} x^{-2 \Delta_n} \exp(-2\xi_n \frac{t}{x}) } \\
\nonumber & + \sum_{m=2}^{\infty} \frac{ \vev{m | \phi_m(1,0) \Mt_{-m}|0} \vev{0|\Lt_{m} \phi_n(x,t)| n } }{\vev{0|\Lt_{m} \Mt_{-m}|0} x^{-2 \Delta_n} \exp(-2\xi_n \frac{t}{x}) } + \cO\left(\frac{1}{c_2^2}\right) \,.
\end{align}
One can now evaluate all the correlators in this expression and explicitly perform the sum.
The inner products $\vev{0|\Lt_{m} \Mt_{-m}|0}$ and $\vev{0|\Mt_{m} \Lt_{-m}|0}$ are easily obtained using the commutation relation \eqref{BMS3} (with conventional normalization for the central terms). They read
\begin{equation}
	\vev{0| \Lt_{m} \Mt_{-m}|0} = \vev{0|\Mt_{m} \Lt_{-m}|0} = \frac{c_2}{12} m(m^2-1).
\end{equation}
To compute the numerators we use the commutators of the $\mathfrak{bms}_3$ generators with the primaries \eqref{GCAcommutators} to turn the three-point functions into a differential operator acting on a two-point function. To be more precise, we write:
\begin{align}
	\vev{\phi_1(x_1,t_1) \phi_1(x_2,t_2) \Xt_{-m} | 0} = & - \vev{ [ \Xt_{-m}, \phi_1(x_1,t_1)] \phi_1(x_2,t_2) } \\
	&  - \vev{ \phi_1 (x_1,t_1) [\Xt_{-m}, \phi_1(x_2,t_2)]}. \nonumber
\end{align}
For $\Xt_{-m} = \Mt_{-m}$ this becomes
\begin{align}
& \vev{\phi_1(x_1,t_1) \phi_1(x_2,t_2) \Mt_{-m} | 0} = \\
& \qquad  \nonumber - \big[ x_1^{1-m} \partial_{t_1} + x_2^{1-m} \partial_{t_2} + \xi_1(1-m)(x_1^{-m} + x_2^{-m} )  \big] \vev{\phi_1(x_1,t_1) \phi_1(x_2,t_2 ) },
\end{align}
and for $\Xt_{-m} = \Lt_{-m}$ 
\begin{align}
& \vev{\phi_1(x_1,t_1) \phi_1(x_2,t_2) \Lt_{-m} | 0} =  \nonumber - \big[ x_1^{1-m} \partial_{x_1} + x_2^{1-m} \partial_{x_2} + (1-m) \left(t_1 x_1^{-m} \partial_{t_1} + t_2 x_2^{-m} \partial_{t_2} \right) \\
& + \xi_1 m (m-1)\left(t_1 x_1^{-m-1} + t_2 x_2^{-m-1} \right)   + \Delta_1 (1-m)(x_1^{-m} + x_2^{-m} )  \big] \vev{\phi_1(x_1,t_1) \phi_1(x_2,t_2 ) }.
\end{align}
Using the expression for the two-point function \eqref{twopt} this evaluates to
\begin{subequations}
\begin{align}
	\frac{\vev{\phi_1\phi_1 \Mt_{-m}}}{\vev{\phi_1 \phi_1}} & = \xi_1\left( (m-1) \left(x_1^{-m} + x_2^{-m} \right) + \frac{2}{x_{12}} (x_1^{1-m} - x_2^{1-m}) \right), \\
	\frac{\vev{\phi_1\phi_1 \Lt_{-m}}}{\vev{\phi_1 \phi_1}} & = \Delta_1 \left( (m-1) \left(x_1^{-m} + x_2^{-m} \right) + \frac{2}{x_{12}} (x_1^{1-m} - x_2^{1-m}) \right) \\
	& \nonumber  + \xi_1 \bigg( m (1-m) \left(x_1^{-1-m}t_1 + x_2^{-1-m}t_2\right) + \frac{2(1-m)}{x_{12}} \left(t_1 x_1^{-m} - t_2 x_2^{-m} \right) \\
	& \qquad  \;\;\; - \frac{2t_{12}}{x_{12}^2} \left( x_1^{1-m} - x_2^{1-m}\right)\bigg). \nonumber
\end{align}
\end{subequations}
The same technique can be used to compute the $\vev{\Xt_m \phi_2 \phi_2}$ correlators appearing in the nominator of \eqref{idblockO1}. Putting everything together we see that the $\cO(1/c_2)$ contribution to the BMS$_3$ identity block is
\begin{align}
\cF_{\unity}(x,t;\Delta_i,\xi_i) = 1 +  \frac{12}{c_2} (\Delta_1 \xi_2 + \Delta_2 \xi_1)&  \sum_{m =2}^{\infty} \frac{(m-1)^2}{m(m^2-1) } x^m  \\
\nonumber   \qquad \qquad  +  \frac{12}{c_2} \xi_1 \xi_2 & \sum_{m =2}^{\infty} \frac{(m-1)^2}{(m^2-1) } t x^{m-1}  + \cO\left(\frac{1}{c_2^2}\right) \, \\
= 1 + \frac{2}{c_2}(\Delta_1 \xi_2 + \Delta_2 \xi_1) x^2\, {}_2F_1(2,2;4,x) & + \frac{2}{c_2} t\,\xi_1 \xi_2 \partial_x \left(x^2\, {}_2F_1(2,2;4,x)\right) + \cO\left(\frac{1}{c_2^2}\right) \,. \nonumber
\end{align}
This expression matches exactly with the one computed from the Wilson lines reduced to the boundary geometric theory of the coadjoint orbits of BMS$_3$ for $c_1=0$.

\section{Propagators with $C_1\neq0$}\label{sec:NonZerocOne}
In this part of the appendix we collect the propagators \eqref{vacpropagators} and \eqref{eq:Fluctuations2PointFunctions} with $C_1\neq0$ where with $C_1$ here we mean the full central charge $C_1$ i.e. $C_1=c_1+26$ for the vacuum orbit and $C_1=c_1+2$ for the massive orbit. The vacuum orbits are given by
\begin{subequations}
\begin{align}
\vev{\epsilon(\vp, u) \epsilon(0,0)} & = 0, \\
\vev{\tilde \alpha(\vp, u) \epsilon(0,0)} & = \, \frac{3}{c_2} \bigg(  3 \zeta -2 - 2 \frac{(1-\zeta)^2}{\zeta}\log\left(1-\zeta \right)  \bigg),  \\
\vev{\tilde \alpha (\vp, u) \tilde{\alpha}(0,0)} &  =  \frac{3 i u \mu_M}{c_2} \left(  2 + \zeta - 2 \frac{(\zeta^2 -1)}{\zeta} \log (1-\zeta)  \right)-\frac{c_1+26}{c_2}\vev{\tilde \alpha(\vp, u) \epsilon(0,0)}\nonumber\\
& =  i\mu_M  u \; \zeta \partial_\zeta  \vev{\tilde \alpha(\vp, u) \epsilon(0,0)}-\frac{c_1+26}{c_2}\vev{\tilde \alpha(\vp, u) \epsilon(0,0)}.
\end{align}
\end{subequations}
For the massive orbits, the propagators are
\begin{subequations}
\begin{align}
\vev{\epsilon(\vp, u) \epsilon(0,0)} = & 0, \\
\vev{\tilde \alpha(\vp, u) \epsilon(0,0)}  = &\,  \frac{6}{c_2\gamma^2} \bigg( 2 \log (1-\zeta) + \Phi(\zeta,1,\gamma) + \Phi(\zeta, 1, - \gamma)  \bigg),  \\
\vev{\tilde \alpha(\vp, u) \tilde{\alpha}(0,0)}  = & \frac{6}{c_2 }\beta\bigg(\Phi(\zeta,2,\gamma)-\Phi(\zeta,2,-\gamma)\bigg)\\
&+\left(2\beta\gamma-\frac{c_1+2}{c_2}+i\mu_M u\, \zeta \partial_\zeta\right)\vev{\tilde \alpha(\vp, u) \epsilon(0,0)},\nonumber 
\end{align}
\end{subequations}
where $\beta = \frac{24\Delta_H+(c_1+2)\left(\gamma^2-1\right)}{2c_2\gamma^3}$. For the coincident points one has for the vacuum orbit
\begin{align}
    \vev{\tilde \alpha (\vp + i \delta_\vp, u + i \delta_u) \epsilon(\vp ,u) } & = \frac{3}{c_2}+\ldots \,, \\
    \vev{\tilde \alpha (\vp + i \delta_\vp, u + i \delta_u) \tilde \alpha(\vp ,u) } & =  -\frac{3(c_1+26)}{c_2^2} +\ldots\,, \\
    \vev{\tilde \alpha '(\vp + i \delta_\vp, u + i \delta_u) \epsilon'(\vp ,u) } & = - \frac{9 + 12 \log (\delta_\vp-\mu_L\delta_u)}{c_2} +\ldots\,, \\
    \vev{\tilde \alpha '(\vp + i \delta_\vp, u + i \delta_u) \tilde\alpha'(\vp ,u) } & = -\frac{12 \delta_u}{c_2 (\delta_\vp-\mu_L\delta_u)}+\frac{c_1+26}{c_2^2}(9 + 12 \log (\delta_\vp-\mu_L\delta_u)) +\ldots \,,
\end{align}
and for the massive orbit
\begin{subequations}\label{coincidencemassivec1}
\begin{align}
    \vev{\tilde \alpha_1 \epsilon_1 } & = - \frac{6}{c_2 \gamma^2} \left( 2 \gamma_E + \psi(\gamma) + \psi (- \gamma)  \right)\,, \\
    \vev{\tilde \alpha_1 \tilde \alpha_1 } & =  - \frac{12 \beta}{c_2 \gamma} \left(2 \gamma_E +  \psi(\gamma) + \psi (- \gamma)  + \frac{\gamma}{2} \zeta (2, -\gamma) - \frac{\gamma}{2} \zeta(2,\gamma) \right)-\frac{c_1+2}{c_2}\vev{\tilde \alpha_1 \epsilon_1 } \,, \\
    \vev{\tilde \alpha_1' \epsilon_1' } & = - \frac{6}{c_2 } \left( 2 \gamma_E + 2 \log(\delta_\vp-\mu_L\delta_u) + \psi(\gamma) + \psi (- \gamma)  \right) \,, \\
    \vev{\tilde \alpha_1' \tilde\alpha'_1 } & = \frac{6 \beta\gamma^2}{c_2 \gamma}\Big(\zeta(2,\gamma) - \zeta(2,-\gamma) \Big) - \frac{12 \delta_u}{c_2 (\delta_\vp-\mu_L\delta_u)}-\frac{c_1+2}{c_2}\vev{\tilde \alpha_1' \epsilon_1' }  \,.
\end{align}
\end{subequations}
\end{appendix}


\providecommand{\href}[2]{#2}\begingroup\raggedright\endgroup

\end{document}